\documentclass[aps,preprint,nofootinbib,preprintnumbers,eqsecnum,superscriptaddress]{revtex4-1}
\pdfoutput=1
\usepackage{graphicx}

\usepackage{color}

\newcommand{\nb}[1]{\color{blue}}

\newcommand{\HL}[1]{{\bf \textcolor{magenta}{#1}}}
\newcommand{\hl}[1]{\color{magenta}}

\usepackage[
	  pagebackref=false,
	  colorlinks=true,
      linkcolor=blue,
      urlcolor=blue,
      filecolor=black,
      citecolor=red,
      pdfstartview=FitV,
      pdftitle={},
        pdfauthor={Thomas Faulkner, Hong Liu, Mukund Rangamani},
        pdfsubject={},
        pdfkeywords={},
        pdfpagemode=None,
        bookmarksopen=true
      ]{hyperref}

\usepackage[normalem]{ulem}
\usepackage{amsmath}
\usepackage{enumerate}
\usepackage{amsfonts}
\usepackage{epsfig}
\usepackage{mathbbol}

\def\Tr{\mathop{\rm Tr}}

\def\Im{\mathop{\rm Im} }
\def\Re{\mathop{\rm Re} }

\newcommand\half{{\ensuremath{\frac{1}{2}}}}
\newcommand\p{\ensuremath{\partial}}

\newcommand\field[1]{{\ensuremath{\mathbb{{#1}}}}}

\newcommand\vev[1]{{\ensuremath{\left\langle{#1}\right\rangle}}}

\newcommand\ket[1]{\ensuremath{\lvert{#1}\rangle}}
\newcommand\bra[1]{\ensuremath{\langle{#1}\rvert}}

\newcommand{\J}{\field{J}}

\newcommand{\be}{\begin{equation}}
\newcommand{\ee}{\end{equation}}
\newcommand{\bea}{\begin{eqnarray}}
\newcommand{\eea}{\end{eqnarray}}
\newcommand{\bega}{\begin{gather}}
\newcommand{\eega}{\end{gather}}
\newcommand{\nn}{\nonumber\\}
\newcommand{\bi}{\begin{itemize}}
\newcommand{\ei}{\end{itemize}}
\newcommand{\ben}{\begin{enumerate}}
\newcommand{\een}{\end{enumerate}}
\newcommand{\bca}{\begin{cases}}
\newcommand{\eca}{\end{cases}}
\newcommand{\bln}{\begin{align}}
\newcommand{\eln}{\end{align}}
\newcommand{\bst}{\begin{split}}
\newcommand{\est}{\end{split}}
\def\ie{\begin{equation}\begin{aligned}}
\def\fe{\end{aligned}\end{equation}}
\newcommand{\bma}{\le(\begin{matrix}}
\newcommand{\ema}{\end{matrix}\ri)}

\newcommand\al{{\alpha}}
\newcommand\ep{\epsilon}
\newcommand\sig{\sigma}

\newcommand\ga{{\ensuremath{{\gamma}}}}

\newcommand\de{{\ensuremath{{\delta}}}}
\newcommand\De{{\ensuremath{{\Delta}}}}
\newcommand\vp{\varphi}

\newcommand\da{{\dagger}}

\def\th{{\theta}}

\newcommand\ra{{\rightarrow}}

\newcommand\ov{\over}
\newcommand\ha{{\half}}

\newcommand\apr{{\ensuremath{{\alpha'}}}}
\def\le{\left}
\def\ri{\right}

\newcommand\sA{{\ensuremath{{\mathcal A}}}}

\newcommand\sG{{\ensuremath{{\mathcal G}}}}
\newcommand\sH{{\ensuremath{{\mathcal H}}}}

\newcommand\sO{{\ensuremath{{\mathcal O}}}}

\newcommand\sV{{\mathcal V}}
\newcommand\sJ{{\mathcal J}}

\newcommand\vx{{\vec x}}

\newcommand{\ft}{{\mathfrak t}}

\allowdisplaybreaks
\def\nn{\nonumber}

\def\b{{\beta}}
\def\e{{\ep}}
\def\hJ{{\widehat{\J}}}

\def\del{\partial}

\global\long\def\mA{\mathcal{A}}

 \global\long\def\mG{\mathcal{G}}
 \global\long\def\mH{\mathcal{H}}
 
 \global\long\def\mJ{\mathcal{J}}
 
 \global\long\def\mL{\mathcal{L}}
 
 \global\long\def\mN{\mathcal{N}}
 \global\long\def\mO{\mathcal{O}}

 \global\long\def\mU{\mathcal{U}}
\global\long\def\mV{\mathcal{V}}

 \global\long\def\e{\epsilon}
 \global\long\def\ra{\rightarrow}
 
 \global\long\def\avg#1{\left\langle #1\right\rangle }

\global\long\def\f#1#2{\frac{#1}{#2}}
 \global\long\def\t{\theta}
 \global\long\def\a{\alpha}
 \global\long\def\b{\beta}
 \global\long\def\g{\gamma}
 \global\long\def\G{\Gamma}
 \global\long\def\s{\sigma}
 
 \global\long\def\d{\delta}
 \global\long\def\Tr{\text{Tr}}
 
 \global\long\def\ket#1{\left\langle #1\right|}
 \global\long\def\bra#1{\left|#1\right\rangle }

 \global\long\def\vp{\varphi}

 \global\long\def\D{\Delta}

 \global\long\def\app{\approx}

 \def\ge{g_{\rm eff}}

\def\rena{regenesis}
\def\Rena{Regenesis}

\begin{document}

\title{
\Rena\ and quantum traversable wormholes
}

\preprint{MIT-CTP/5067}

\author{Ping Gao}
\affiliation{Center for the Fundamental Laws of Nature,
Harvard University,
Cambridge, MA 02138}

\author{Hong Liu}
\affiliation{Center for Theoretical Physics, \\
Massachusetts
Institute of Technology,
Cambridge, MA 02139 }

\begin{abstract}

\noindent  

Recent gravity discussions of a traversable wormhole indicate that in holographic systems signals generated by a source could reappear long after they have dissipated, with the need of only performing some simple operations.
In this paper we argue the phenomenon, to which we refer as ``regenesis'', is  universal in general quantum chaotic many-body systems, and elucidate its underlying physics. The essential elements behind the phenomenon are: (i) scrambling which in a chaotic system makes  out-of-time-ordered correlation functions (OTOCs) vanish at large times; (ii) the entanglement structure of the  state of the system. The latter aspect also implies that the \rena\ phenomenon requires  fine tuning of the initial state. 
Compared to other manifestations of quantum chaos such as the initial growth of OTOCs which deals with early times, and a random matrix-type energy spectrum which reflects very large time behavior, regenesis concerns with intermediate times, of order the scrambling time of a system. 
We also study the phenomenon in detail in general two-dimensional conformal field theories in the large central charge limit, 
and highlight some interesting features including a resonant enhancement of \rena\ signals near the scrambling time and 
their oscillations in coupling. Finally, we discuss gravity implications of the phenomenon for systems with a gravity dual, arguing that there exist regimes for which  traversability of a wormhole is quantum in nature, i.e. cannot be associated with a semi-classical  spacetime causal structure.

\end{abstract}

\today

\maketitle

\tableofcontents

\section{Introduction}

When a circuit is disconnected from its battery, the electric current flowing through it quickly stops, due to dissipation.

Using modern language, treating the circuit and its environment as a single isolated quantum system, we say 
the current is scrambled among other degrees of freedom of the system. 
Once a signal is dissipated (or scrambled), it cannot be recovered in practice, as to do that one needs to have control over the full quantum state of the system, which in practice is never possible for a system of many degrees of freedom.

In this paper we discuss a new phenomenon, 
based on the recent discussion of a traversable wormhole~\cite{Gao:2016bin,Maldacena:2017axo,Susskind:2014yaa,Susskind:2016jjb,Susskind:2017nto}, where the current signal can re-appear with the need of only performing some simple operations.

\begin{figure}
\begin{centering}
\includegraphics[width=10cm]{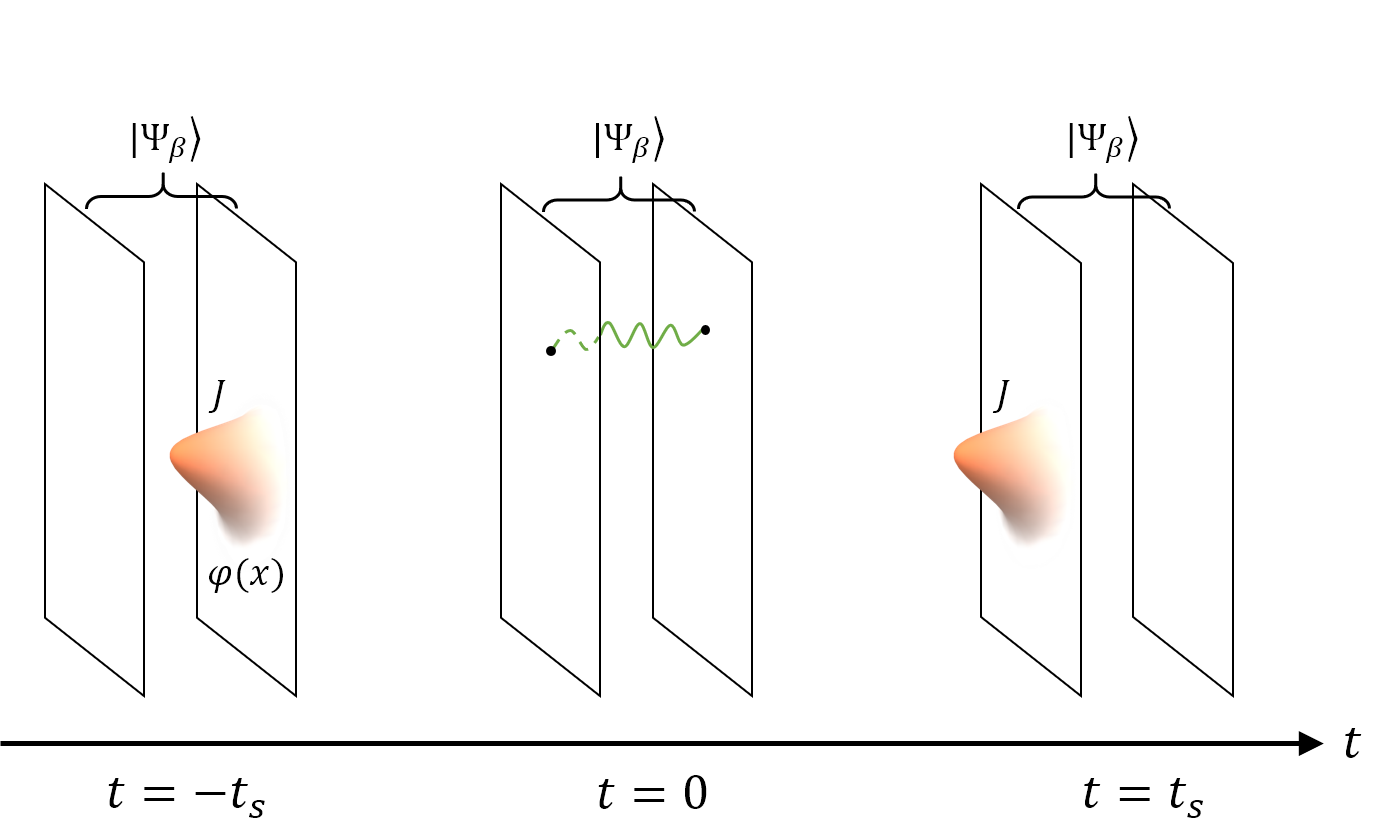}
\par\end{centering}
\caption{The setup for the \rena\ phenomenon. At $t = - t_s$, uncoupled systems $L,R$ are in an entangled state~\eqref{tfd0}. 
The signal in $J^R$ disappears shortly after we turn off the source $\vp^R$. At $t=0$, we turn on a local coupling between $L$ and $R$ for a short time, which we have approximated as a delta function in time in~\eqref{oun}. At $t=t_s$, the signal reappears in 
the $L$ system if $t_s$ is sufficiently large. The reappeared signal is not identical to the original signal, but related by a transformation. 
\label{fig:fig0-1}}
\end{figure}

Let us first describe the setup using field theory language (see also Fig.~\ref{fig:fig0-1}).  Consider two identical uncoupled quantum systems, to which we will refer as $L, R$ systems.  
The Hamiltonians for them are respectively $H^L$ and $H^R$ which by definition have the same set of eigenvalues $\{E_n\}$ with respective energy eigenstates $\bra{n}_{L,R}$.
Suppose $L,R$ systems are arranged 
in a special entangled state such that at $t =0$ it is given by a thermal field double state~\cite{Takahasi:1974zn} 
\begin{equation}\label{tfd0}
\bra{\Psi_\beta}={1 \ov Z_\beta} \sum_{n}e^{- {\beta E_{n} \ov 2}}\bra{\bar n}_{L} \bra{n}_{R}, \qquad Z_\beta = \sum_n e^{-\beta E_n}
\end{equation}
where $\bra{\bar n}$ denotes time reversal of $\bra{n}$.  $\bra{\Psi_\b}$
has the property that if one operates solely in one of the systems one finds a thermal state at inverse temperature $\b$.  Consider at some time $t = - t_s  < 0$  turning on an external source $\vp^R$ for some few-body Hermitian operator $J^R$  for a short interval.\footnote{The time interval is taken to be much smaller than $t_s$.} In the $R$ system there is an induced expectation value $\vev{J^R (t)} \equiv \vev{\Psi_\b|J^R (t)|\Psi_\b}$, but there is no response in the 
$L$ system as by definition $[J^L, J^R] =0$.  
As usual $\vev{J^R (t)}$
will dissipate and decay quickly to zero 
after $\vp^R$ is turned off. 

Now couple the two systems at $t=0$, with the total Hamiltonian 
\be \label{oun}
H = H^L + H^R - g V \delta (t=0)
\ee
where $g$ is a coupling and $V$ is an operator involving both $L$ and $R$ systems, e.g. a simplest choice is 
\be\label{pp1}
 V = \sO^L (0) \sO^R (0)  \
\ee
for some few-body operator $\sO (x)$. 
 The surprising result from the gravity analysis of~\cite{Maldacena:2017axo} is that 
a signal will re-appear in the $L$-system if $t_s$ is larger than the {\it scrambling time $t_*$} of the $L,R$ system.\footnote{Scrambling time is defined here as the time scale when  $\vev{[V(t), W(0)]^2}$ between generic few-body operators $V,W$ become $O(1)$.}
Note that here $\sO$ and $J$ are generic few-body operators which do not need to have any common degrees of freedom between them. 

The purpose of this paper is to argue that this phenomenon, to which we will refer below as ``\rena,'' 
is  universal 
for generic quantum chaotic systems, to elucidate its underlying physics, to study it in detail  
in a class of field theories, and to discuss its gravity implications. 

A general result we obtain is that in a generic chaotic system for $t, t_s \gg t_*$:
(i) $\vev{J^L(t)}_g$ is supported only for $t \approx t_s$ where $\vev{\cdots}_g$ denotes expectation value in~\eqref{tfd0} with a nonzero $g$; (ii)  as a function of $t_s$, $\vev{J^L (t=t_s, \vx)}_g$ has the following behavior
\be \label{yto}
\vev{J^L (t_s, \vx)}_g \approx C(g)  \, \vp^R (-t_s, \vx) , \qquad t_s \gg t_*
\ee
where $C (g)$ is an $O(1)$ constant depending on $g$. We thus find the ``input signal'' $\vp^R$ from the $R$ system at $t=- t_s$  regroups at $t=t_s$ in the $L$ system  long after it has dissipated!\footnote{Note that in~\eqref{yto} signals which are input earlier in the $R$ systems appear
later in the $L$ system, so in fact what one finds is the {\it time reversed} form of the input signal.}
 The result~\eqref{yto} is insensitive to the specific form of $L-R$ interaction $V$. The behavior for a system with $t_s \sim t_*$ is more complicated and will be mentioned later.

The essential elements behind the \rena\ behavior~\eqref{yto} are: (i) scrambling in a chaotic system makes  out-of-time-ordered correlation functions (OTOCs) vanish for $t \gg t_*$~\cite{Shenker:2013pqa,Maldacena:2015waa}, and (ii)  the entanglement structure of~\eqref{tfd0} which strongly correlates an operator inserted at $(-t, \vx)$ with an operator at $(t, \vx)$.
Compared to other manifestations of quantum chaos such as the initial growth of OTOCs which deals with early times, and a random matrix-type energy spectrum which reflects very large time behavior, the regenesis phenomenon concerns with intermediate times, of order the scrambling time of a system. 
 Instead of making the signal to appear at the same spatial location $\vx$, by considering a slight variation of~\eqref{tfd0} one could also make the signal from $(-t, \vx)$ to regroup at $(t, \vx + \vec a)$ for some $\vec a$. 

One may wonder what happens if we consider the same setup in a few-body or integrable system. 
In general, with $g \neq 0$, some response will be generated in the $L$-system: interactions among each subsystem
will manage to communicate the effects of $\vp^R$ to $J^L$. But there are two crucial differences: (1) it will not be ``regenesis,''
as in a few-body system (or in integrable systems) there is no dissipation, so the original signal in the $R$-system will remain there forever; (2) the signal generated in the $L$-system will depend sensitively on the specifics of an individual system and the operators used. In contrast, in chaotic systems, the behavior is universal, independent of all the details. 

At first sight, the \rena\ phenomenon appears to be miraculous: how can a dissipated signal regroup with a very simple operation like~\eqref{pp1}? If one wants to be melodramatic, we could imagine that by turning on $\vp^R$, we create a ``cat'' in the $R$ system.  The cat lives for a while, and dies. Eventually her body will be fully scrambled with the environment. 
Now it appears that we could bring her back to life in the $L$-system by simply turning on a $gV$!
There are two important catches here. Firstly, the success of the operation in fact requires extreme fine tuning in how we prepare the initial state at $-t_s$ when we turn on the external source $\vp^R$. 
The state should be prepared such that as the system evolves to $t=0$, the system is in the thermal field double state~\eqref{tfd0}. This is a highly nontrivial requirement as the scrambling time $t_*$ could be macroscopic for a macroscopic system, and as we will see explicitly the \rena\ phenomenon is somewhat fragile: various modifications could destroy the behavior~\eqref{yto}. 
A second catch is that as we will discuss later (in Sec.~\ref{sec:quan}), at least for the regime $t, t_s \gg t_*$, 
the signal~\eqref{yto} is quantum in nature, i.e.
the variance is always comparable to the expectation value itself, and one cannot cut down fluctuations using macroscopic measurements.

We also study the \rena\ phenomenon in two-dimensional conformal field theories (CFT) in the large central charge limit which is known to be chaotic~\cite{Roberts:2014ifa}. That is, we take $L$ and $R$ systems to be (1+1)-dimensional and described by a CFT.  We consider $c \gg \De_J \gg \De_\sO \sim O(1)$ 
where  $c$ is the central charge of the system, and $\De_\sO, \De_J$ are respectively the scaling dimensions of few-body operators $\sO$ and $J$. In this regime we can obtain the behavior of $\vev{J^L (t)}_g$  in detail by applying the techniques of~\cite{Fitzpatrick:2014vua,fitzpatrick2015virasoro,Fitzpatrick:2015qma}.
 In addition to~\eqref{pp1} we will also consider two other types of couplings (in~\eqref{pp2} $\al$ denotes different operator species)
\be
\label{pp2}
V = {1\ov k} \sum_{\al=1}^k \sO^L_\al (0) \sO^R_\al (0)
\ee
which were considered in~\cite{Maldacena:2017axo} in the large $k$ limit, and 
\be
V = {1 \ov L} \int_{-{L \ov 2}}^{L \ov 2} dx \, \sO^L (x) \sO^R (x)  \ .
\label{pp3}
\ee
For~\eqref{pp2} our CFT results are fully consistent with the gravity results of~\cite{Maldacena:2017axo} for a $(0+1)$-dimensional holographic system. 

We will refer to~\eqref{pp1} as a single-channel coupling, while~\eqref{pp2}--\eqref{pp3} as multiple-channel, one from multiplicity of operators, one from spatial integrations. Their local spacetime structure is chosen to maximally take advantage of the entanglement structure of~\eqref{tfd0}.
Here is a brief summary of the main features found in explicit CFT calculations (some of these features are also present in the gravity results of~\cite{Maldacena:2017axo} for~\eqref{pp2}):  

\ben 

\item 
In all cases, as a function of $t_s$, one can separate the behavior of $\vev{J^L (t_s)}_g$  
into three different regions: (1) $t_s \ll t_*$ (sub-scrambling region), where $\vev{J^L (t_s)}_g$ is exponentially small and can be considered to be zero for practical purposes; (2) $t_s \sim t_*$ (transition region), where $\vev{J^L (t_s)}_g$ becomes $O(1)$; (3)  $t_s \gg t_*$ (stable region), where $\vev{J^L (t_s)}_g$ is given by~\eqref{yto}. 
Here the scrambling time is given by  $t_* = {\beta \ov 2 \pi} \log {c \ov 6 \pi}$.

\begin{figure}
\begin{centering}
\includegraphics[width=16.5cm]{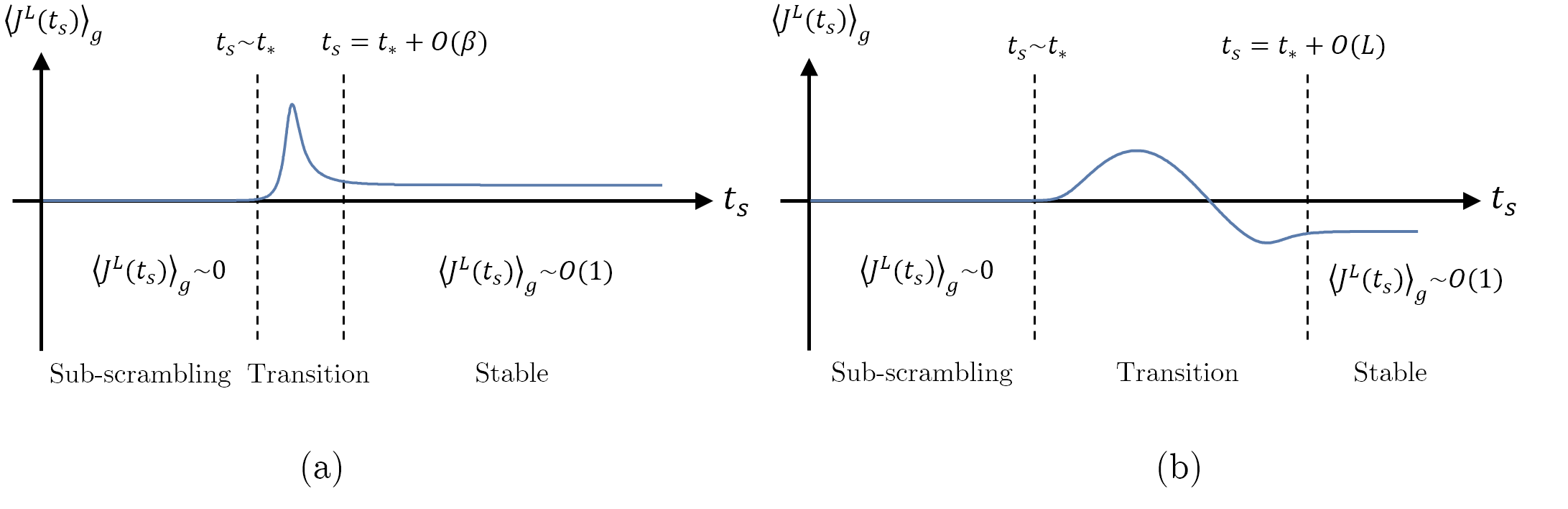}
\par\end{centering}
\caption{The behavior of $\avg{J^L(t_s)}_g$ for three regimes of $t_s$.  (a) is a cartoon for~\eqref{pp2} with transition regime of length $O(\b)$, and the resonant enhancement could be pronounced. (b) is for~\eqref{pp3} with the size of the transition region of order $O(L)$ for $L \gg \b$. 
\label{fig:fig0-2}}
\end{figure}

\item For~\eqref{pp1} and~\eqref{pp2} the transition region is very narrow, of order $O(\b)$, while for~\eqref{pp3} the transition regions is lengthened to $O(L)$ for $L \gg \b$. See Fig.~\ref{fig:fig0-2}.

\item An interesting effect for~\eqref{pp2} in the transition region is that, with a choice of a sign for $g$, the magnitude of $\vev{J^L (t_s)}_g$ could be exponentially large in $g$, to which we refer as resonant enhancement.  See Fig.~\ref{fig:fig0-2}(a) for a cartoon.

\item For~\eqref{pp2}--\eqref{pp3}, $C(g)$ in~\eqref{yto} is given by 
\be 
C(g) = 2 G_J \sin (g G)
\ee
is oscillatory in $g$. $G_J$ and $G$ are some constants. For~\eqref{pp1} there appears no oscillation in $g$, and we find for large $g$
\be 
C(g) \propto g^{-1}  \ .
\ee

\een

\begin{figure}
\begin{centering}
\includegraphics[width=9cm]{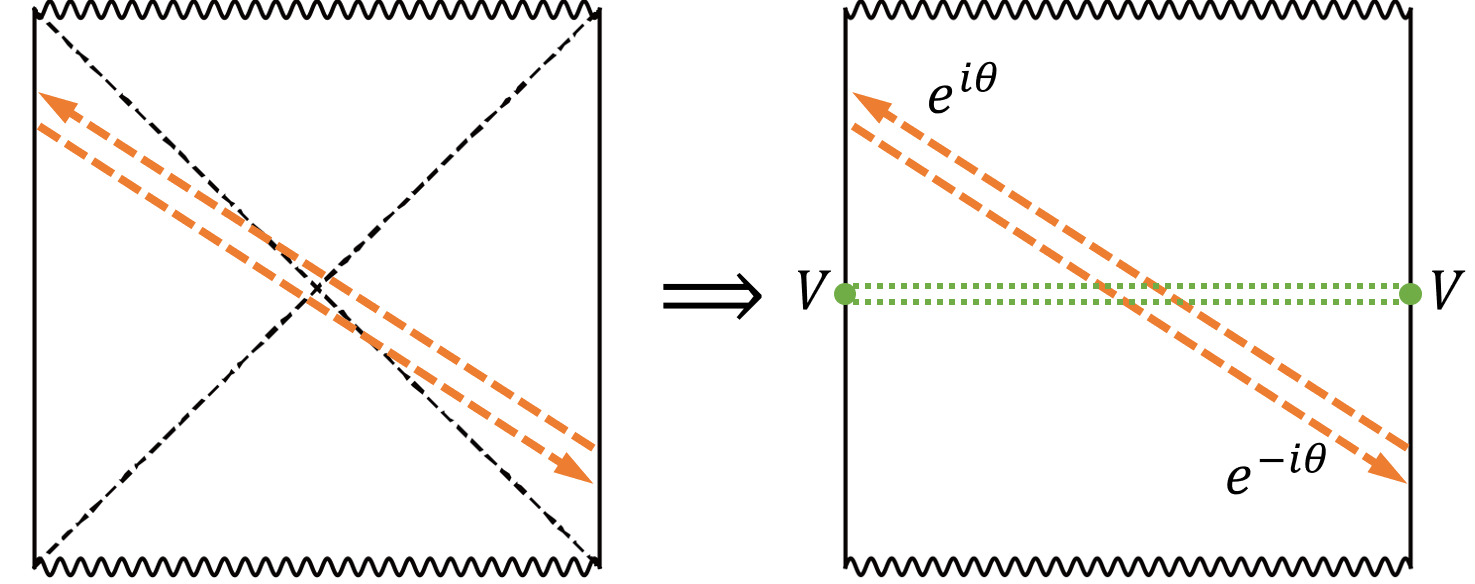}
\par\end{centering}
\caption{Left: in a wormhole described by~\eqref{tfd0}, due to entanglement between the two boundaries,  there are virtual particles which can propagate between them. The non-traversability can be understood as coming from perfect destructive interference between the process of a virtual particle traveling from $R$ to $L$, and the mirror process of traveling from $L$ to $R$. Right:  turning on interaction $V$ subtly changes the entanglement structure and gives a phase
for each propagation. Now the destructive interference is no longer perfect, resulting propagations of ``real'' particles between the boundaries. Note that the interference is not a process between ``future'' and ``past'' as the related two boundaries are actually spacelike separated. It is possible to boost the frame such that it occurs on one spatial slice.
 \label{fig:15}}
\end{figure}

Our field theory studies also have important implications for the understanding of the traversability of a wormhole on the gravity side. In~\cite{Gao:2016bin,Maldacena:2017axo}, the basic picture is that the two-sided coupling~\eqref{pp1} (or~\eqref{pp2}--\eqref{pp3}) generates negative-energy excitations which in turn deform the spacetime causal structure of the wormhole geometry, 
allowing signals to propagate  between the two boundaries of a wormhole. Combining our results and those from gravity calculations of~\cite{Maldacena:2017axo}, 
we argue that there exist physically distinct scenarios for traversability from causal propagation through the wormhole. 
For example, in the regime as described by~\eqref{yto}, the traversability appears quantum in nature, i.e. cannot be associated with a semi-classical spacetime causal structure. 
Instead it appears to involve breaking of a delicate destructive interference, see  Fig.~\ref{fig:15} for a cartoon picture (which is appropriate for $t_s \gg t_*$).

The plan of the paper is as follows. In Sec.~\ref{sec:gen} we present a general argument for the \rena\ phenomenon, explain the quantum nature of the signal, and 
discuss its robustness. We also present a simple qubit model as a contrast study of this phenomenon in a few-body system. In Sec.~\ref{sec:cft} we outline the main steps of the calculation of $\vev{J^L (t)}_g$ in a two-dimensional 
CFT in the large central charge limit, with details of the calculation given in Appendix~\ref{app:a} and Appendix~\ref{full-k}.
Details on the CFT calculation of the robustness of the phenomenon is given in Appendix~\ref{app:rob}. 
 In Sec.~\ref{sec:ana} we analyze the results obtained in Sec.~\ref{sec:cft} and various Appendices. In Sec.~\ref{sec:grav} we discuss gravity interpretation of our results, including a detailed comparison with the results of~\cite{Maldacena:2017axo}. In Sec.~\ref{sec:conc} we conclude with various discussions, including future directions and experimental realizations.

\section{A general argument for the \rena\ phenomenon} \label{sec:gen}

In this section we present a simple argument for the \rena\ behavior~\eqref{yto} for a general quantum chaotic system
and discuss the robustness of the phenomenon, i.e. how it fares against imperfections of the preparation of $\bra{\Psi_\b}$. 
The results of this section are consistent with the gravity results of~\cite{Maldacena:2017axo} for holographic systems, and will be further confirmed through explicit calculations in two-dimensional CFTs in the large central charge limit  in subsequent sections.

\subsection{More on the general setup}

As discussed in the Introduction we consider $L$ and $R$ systems in 
a thermal field double state~\eqref{tfd0}.
In this state,  the expectation value of any set of operators which act only on one of the systems
 is given by the thermal average with inverse temperature $\beta$, e.g. 
\be 
\vev{\Psi_\beta| \sO_{1}^L \cdots \sO_{n}^L  |\Psi_\beta} = {1 \ov Z_\beta} {\rm Tr} \le(e^{-\beta H} \sO_1 \cdots \sO_n \ri)
\equiv \vev{\sO_1 \cdots \sO_n}_\b
\ee
where on the right hand side the trace is performed in the left system and $\vev{\cdots}_\b$ denotes thermal average at inverse temperature $\b$. For notational simplicity we have dropped $L$ labels. We will do this for the rest of the paper: below any quantities with no explicit labels should be understood as in the $L$ system.

By definition any operators from $L$ system commute with those of  $R$ system, i.e. 
\be \label{co}
[\sO^L (x), J^R (x')] = 0, \qquad \forall \ \sO, J, x, x' \ 
\ee
where $x$ denotes spacetime coordinates, i.e. $x = x^\mu = (t, \vx)$ and $\vx$ are spatial coordinates. 
Consider turning on a source $\vp^R$ for some generic (hermitian) few-body operator $J^R$, i.e. perturbing the action $ S^R$ of the right system by 
\be \label{le}
S^R \to  S^R + \int d^d x \, \vp^ R (x) J^R (x) \ . 
\ee
We will choose $J$ such that  its thermal expectation value is zero. Then at linear order in $\vp^R$ we have 
\be \label{sop}
\vev{J^R(x)} = \int d^d x' \, G^{RR} (x, x') \vp^R (x') , \qquad G^{RR}  (x,x')= i \th (t-t') \vev{[J^R (x), J^R (x')]}_\beta  \ .
\ee
For $g =0$, there is no response in the left system 
\be 
\vev{J^L (x)} = 0
\ee
due to~\eqref{co}. On general ground one expects that the thermal response function $G^{RR}$ for a non-conserved quantity to behave for large $t-t'$ or large $|\vec x - \vec x'|$
as 
\be \label{oin}
G^{RR}  (t, \vx,t', \vx') \sim e^{-{|t-t'| \ov \tau_{r}}} , \qquad  G^{RR}  (t, \vx,t', \vx') \sim e^{-{|\vx-\vx'| \ov \ell_{r}}} 
\ee
where $\tau_r, \ell_r$ are respectively relaxation time and length, both of which will be treated as microscopic, i.e. much smaller than typical scales involved in $\vp^R$.  For a scale invariant system, they are both of  $O(\b)$, see e.g.~\eqref{co1}. 
Thus $\vev{J^R}$ will quickly decay to zero in a time scale of order $\tau_r$ after the source is turned off.

Now with $g$ nonzero in~\eqref{oun} we would like to see 
whether there is a response on the $L$ side. We take the source $\vp^R$ to be turned on for a short period around $t = - t_s < 0$ such that $\vev{J^R (t)}$ will have long decayed to zero before $V$ is turned on at $t=0$. 
From Appendix~\ref{app:lin}, we  find at full nonlinear level in $\vp^R$
\be\label{nonli-J}
\avg{J^L(t)}_g=\vev{\Phi | J^L (t) |\Phi} = 
\ket{\Psi_\b}e^{-i\int ds\, \vp^R(s) J^R(s)}e^{-igV}J^L(t)e^{igV}e^{i\int ds\, \vp^R(s) J^R(s)}\bra{\Psi_\b} 
\ee
with $\bra{\Phi}$ defined as 
\be \label{ujm}
\bra{\Phi} \equiv e^{igV}e^{i\int ds\, \vp^R(s) J^R(s)}\bra{\Psi_\b}  \ .
\ee
Expanding~\eqref{nonli-J} to linear order in $\vp^R$ we then find 
\bega \label{lresp}
\vev{J^L (x)}_g = \int d^d x' \, G^{LR} (x, x') \vp^R (x') , 
\end{gather} 
with (note we take both $J$ and $\sO$ to be Hermitian operators)
\bega \label{ma0}
G^{LR} (x,x')   = i \th (t-t')  \vev{\Psi_\b|[J^L (x), J^R (x')]|\Psi_\b}_g = i \th (t-t') (W (x, x') - W^*(x, x')) ,  \\
 W \left( x, x'  \right) = \left\langle \Psi_\b | e ^ { -i g V } J^ { L } \left(x \right) e ^ { i g V } J^{ R } \left( x' \right) | \Psi_\b \right\rangle\ .
\label{ma1}
\end{gather}

\subsection{Entanglement structure} \label{sec:ent}

The thermal field double state~\eqref{tfd0} has a rather specific entanglement structure between $L$ and $R$ 
systems. It can be checked that the state generated from a Hermitian operator $J^R$ acting on $\bra{\Psi_\beta}$ can be reproduced from the action of $J^L$ in the $L$ system with a shift in imaginary time, i.e. 
\be \label{coo1}
J^R  \bra{\Psi_\b} = \eta^* J^L ( {i \beta/2})  \bra{\Psi_\b}
\ee
where $\eta$ is the phase factor associated with time reversal on $J$ and will be dropped subsequently as it will not play any role. 
Furthermore, 
\be \label{ho2}
(H^L - H^R) \bra{\Psi_\b} = 0, \quad \to \quad e^{- i H^L t} \bra{\Psi_\b}  =  e^{- i H^R t} \bra{\Psi_\b}   \ .
\ee
The combination of~\eqref{coo1}--\eqref{ho2} implies that 
\bea 
J^R (t) \bra{\Psi_\b} &= & e^{i H^R t} J^R e^{-i H^R t} \bra{\Psi_\b}  
= e^{- i H^L t} e^{i H^R t} J^R \bra{\Psi_\b}  \cr
& =& e^{- i H^L t}  J^L (i \b/2) e^{i H^L t} \bra{\Psi_\b}  = J^L (-t + i \b/2) \bra{\Psi_\b} 
\label{co3}
\eea
where we have used $L$ and $R$ operators commute and~\eqref{ho2} repeatedly. Note that~\eqref{co3} applies to a complex $t$ with ${\rm Im} \, t \in (0, \b/2)$. By using the above equation repeatedly we 
further find that 
\be 
 \label{con}
J_1^R (t_1) J_2^R (t_2) \cdots J_n^R (t_n)  \bra{\Psi_\b} =  J^L_n (-t _n+ {i \beta/2}) \cdots  J_2^L (-t_2+ i \beta/2) 
 J_1^L (-t_1+ i \beta/2)   \bra{\Psi_\b} 
\ee
where subscripts label different operators. Due to the entanglement structure of $\bra{\Psi_\beta}$, we see from~\eqref{co3}--\eqref{con} that operators inserted at $(t, \vx)$ in the $R$ system are strongly correlated with those inserted at $(-t, \vx)$ in the $L$ system. In other words, there appears a ``time reversal symmetry'' between $L$ and $R$ systems. 
We will refer to such a pair of points as an entangled pair. This simple fact will play a key role in understanding the results of the paper. Moreover, the interactions~\eqref{pp1} and~\eqref{pp2}--\eqref{pp3} are chosen to involve couplings between operators inserted at entangled points, which as we will see makes the teleportation most efficient. 

From~\eqref{co3} we have 
\be\label{omk}
\vev{\Psi_\b|J^L (t, \vx) J^R(- t' , \vx')|\Psi_\b}  = \vev{J^L (t, \vx) J^L (t' + i \b/2, \vx')}_\b
 \sim \bca e^{- {|t-t'| \ov \tau_r}} & |t -t'| \gg \tau_r \cr 
e^{- {|\vx-\vx'| \ov \ell_r}} & |\vx- \vx'| \gg \ell_r 
\eca
\ee
where in the second equality we have again displayed the usual behavior for a thermal two-point function. 
An explicit example of~\eqref{omk} is given by~\eqref{co2} for a two-dimensional CFT, for which $\tau_r = \ell_r = {\b \ov 2 \pi}$. 
Treating $\tau_r, \ell_r$ as microscopic, we see that the two-side correlation function~\eqref{omk} is essentially nonzero only  for $t \approx t'$ and $\vx \approx \vx'$, and we will denote
\be \label{omk1}
G_J \equiv  \vev{\Psi_\b|J^L (t, \vx) J^R(- t , \vx)|\Psi_\b}
\ee
which is a constant from spacetime translational symmetries.

\subsection{\Rena\ behavior for quantum chaotic systems}  \label{sec:rena1}

Now let us examine the behavior of~\eqref{nonli-J} (or its linear version~\eqref{ma1}) for a general chaotic system. We will take $x = (t, \vx)$ and 
$x' = (-t_s, \vx')$ with $t_s > 0$.  
First consider $g=0$. Since $J^R$ commutes with $J^L$, equation~\eqref{nonli-J} then reduces to $\vev{\Psi_\b |J^L |\Psi_\b} = 0$. 
For $g \neq 0$, at small $t$, since $\sO$ and $J$ are generic few-body operators, whose degrees of freedom in general do not overlap, $[J(t), V] \approx 0$, which again leads to $\vev{J^L (t)}_g \approx 0$. 
As time increases, $J(t)$ grows and scrambles in the space of degrees of freedom. At sufficiently large $t$, $[J(t), V]$ becomes non-negligible; the time scale that this happens defines the scrambling scale $t_*$. 
Thus $\vev{J^L (t)}_g$~(thus also $G^{LR}$) can be non-negligible  only when $t$ is of the scrambling scale $t_*$.

Also due to the entanglement structure of~\eqref{tfd0}, as manifested in~\eqref{omk}, we expect $W$ defined in~\eqref{ma1}  to be non-vanishing only when $t_s$ and $t$ are close. So $t_s$ also has to be of order or larger than $t_*$ for $G^{LR}$ to be nonzero. Similarly for the full nonlinear expression~\eqref{nonli-J}. 

 When $t, t_s$ are of order the scrambling scale, even at linear order in $\vp^R$, the expression for $\vev{J^L}_g$ is complicated. 
 We will study the behavior of $W$ and $G^{LR}$ in detail for various choices of $V$ in subsequent sections in two-dimensional CFTs. Here we show that when $t, t_s \gg t_*$ 
 the behavior of full nonlinear expression~\eqref{nonli-J} is very simple .

For clarity we will illustrate the main argument using the linear expression~\eqref{ma1}. Consider expanding the exponential $e^{igV}$ between $J^L$ and $J^R$ in power series of $g$, then at $n$-th order one gets correlation functions of the form 
\be \label{demn}
M_n = \vev{\Psi_\b| e^{-i g V} J^L (t)  (\sO^L(0) \sO^R (0)  )^n J^R (-t_s) |\Psi_\b} 
\ee
where we have suppressed all the spatial dependence. 
Note that the precise form of $V$ is not important and we have only schematically indicated that it has the form of some product of $\sO^L\sO^R$ inserted at $t=0$. Now using repeatedly~\eqref{con}, we can write~\eqref{demn} as 
\be 
M_n = \vev{\Psi_\b| e^{- i gV} J^L (t) (\sO^L (0))^n J^L (t_s + i \b/2) (\sO^L (i \b/2))^n |\Psi_\b} 
\ee  
with $J$ and $(\sO^L)^n$ out-of-time-ordered. 

Now for a general quantum chaotic system, due to scrambling, we expect 
\be 
M_n \to 0, \qquad  t, t_s \gg t_* \quad {\rm for} \quad n \geq 1 
\ee
with $t_*$ the scrambling time of the system. Thus for $t, t_s \gg t_*$, equation~\eqref{ma1} should reduce to 
\be \label{ww1}
W = \vev{\Psi_\b| e^{-i g V} J^L (t, \vx) J^R (-t_s, \vx')|\Psi_\b}  \ .
\ee
Notice that expanding the exponential $e^{i gV}$ in~\eqref{ww1} in power series of $V$ will now give rise to only time-ordered correlation  functions (TOCs).  
On general grounds, one expects that such TOCs to factorize at large time separations between $J$ and $\sO$ insertions, 
we then find\footnote{The conclusion was obtained before in the large $k$ limit of~\eqref{pp2} for holographic systems using gravity calculation in~\cite{Maldacena:2017axo}.}
\be\label{oio}
W \approx \vev{e^{-i g V}} \vev{J^L (t, \vx) J^R (-t_s, \vx')}   \ .
\ee
From~\eqref{ma0},~\eqref{omk1} and the discussion below~\eqref{omk}, we thus find that $G^{LR} (t, \vx; - t_s, \vx')$ is nonzero 
only for $t \approx t_s$ and $\vx \approx \vx'$ with 
\be \label{mok}
G^{LR} (t_s, \vx; -t_s, \vx) = C (g)= {\rm const} 
\ee
and
\be \label{mok1}
C (g)= - 2 G_J \, {\rm Im} \, \vev{e^{-i g V}} \ .
\ee
For $\vp^R$  slowly varying at the scales of $\tau_r, \ell_r$, we thus see~\eqref{lresp} reduces to
\be \label{yto1}
\vev{J^L (t_s, \vx)}_g \approx C(g)  \, \vp^R (-t_s, \vx) , \qquad t_s \gg t_* \ .
 \ee
Note that generically we expect $\vev{e^{-i g V}}$ to be complex and $O(1)$ as already mentioned the operator $V$ is designed 
to couple $\sO^{L,R}$ at entangled points.

It is interesting that the sole effect of turning on the interaction $V$ between two subsystems is to generate a phase so that $W$ is no longer real, resulting a nonzero $G^{LR}$. Through the entanglement structure of $\Psi_\b$, information of the source $\vp^R$ 
is already present in the $L$ system, just as in the usual EPR story. Heuristically, the effect of turning on $V$ is to turn this information into ``real'' physical signals. 

The above discussion can be immediately generalized to the full nonlinear expression~\eqref{nonli-J}. 
In fact for any few-body operator $X^L(t)$, setting all  OTOCs to zero, we find in the limit $t\sim t_s \gg t_*$, 
\be \label{nke}
\vev{X^L}_g \equiv \vev{\Phi |X^L(t) |\Phi} = (1 -2\Re a) \vev{\Psi_\b|X^L (t)  |\Psi_\b} + \le(a \vev{\Psi_\b\le|X^L (t) U^R \ri|\Psi_\b} +h.c.\ri)
\ee
where $\bra{\Phi}$ is given by~\eqref{ujm} and we have introduced 
\be 
U^R = e^{i\int ds\, \vp^R(s) J^R(s)}, \qquad a = \avg{e^{-igV}}_\b -1  \ .
\ee
See Appendix~\ref{app:nb} for a derivation of~\eqref{nke}. 
For $X^L = J^L$, we then find 
\be\label{nonu}
\avg{J^L(t)}_g\app a \vev{\Psi_\b\le|J^L(t) U^R \ri|\Psi_\b} +h.c. \ .
\ee
Equation~\eqref{nonu} again has the form of a complex factor times correlation functions in the thermal field double state (plus its hermitian conjugate).

To summarize, equation~\eqref{yto1} and its full nonlinear version~\eqref{nonu} can be understood as due to the following three key elements: (i) vanishing of OTOCs at large times due to scrambling; (ii) factorization of TOCs at large times; (iii) entanglement structure of $\Psi_\b$. 
Note that the factorization assumption can in principle be weakened or dropped. One only needs that~\eqref{ww1} is complex and not small at large times.  

The discussion of this subsection does not apply to $t, t_s \sim t_*$ for which we will examine in two-dimensional CFTs in Sec.~\ref{sec:ana}.

\subsection{Quantum nature of the regenesis signal} \label{sec:quan}

In this section we show that the regenesis signal~\eqref{yto1} is quantum in nature.\footnote{The content of this section 
is developed from discussions with Juan Maldacena, Douglas Stanford and Zhenbin Yang. The main conclusions have also been anticipated by them.}  
We do this by comparing~\eqref{nonu} and the corresponding variance with those in a standard response setup. 

For this purpose, let us first recall the standard response story,\footnote{In a field theory we assume $J$ is suitably smeared such that both $J$ and $J^n$ are bounded operators with a finite 
norm}
\ie \label{uuj}
\tilde J \equiv \vev{\Psi_\b| U_L^\da J^L U_L|\Psi_\b}  , \quad \de \tilde J \equiv \le(\vev{\Psi_\b| U_L^\da (J^L - \tilde J)^2 U_L|\Psi_\b} \ri)^\ha ,  
\fe
where $U_L$ is the unitary operation for turning on some source $\vp^L$ in the $L$-system. $\tilde J$ is the corresponding signal 
and $\de \tilde J$ is the variance. 
Note that since both $U_L$ and $J^L$ belong to the $L$-system,~\eqref{uuj} just reduce to thermal averages. In this context we will thus suppressed index $L$.  We also denote the variance and fourth moment of $J$ in the thermal state (recall $\vev{J}_\b =0$) as 
\bega 
J_2 =\sqrt{\vev{J^2}_\b} , 
\qquad  J_4 = \le(\vev{J^4}_\b\ri)^{1 \ov 4} \ .
\end{gather}

To make the discussion explicit, let us imagine a lattice system of interacting spins. 
Suppose $J$ is given by some operator at a single site, say $\sig_z$, then clearly both $\tilde J$ and $\de \tilde J$ are order $O(1)$, and one needs multiple measurements to detect the effects of $U$.  One can make life easier by measuring the average polarization, say,  
\be \label{hhy}
J =  {1 \ov N} \sum_i Z_i  
\ee
and considering a source which acts on all spins 
\be\label{hyy1}
U = P e^{i \sum_j  \int ds \, \vp^L (s) Z_j (s)} 
\ee
where $Z_i$ is $\sig_z$ at $i$-th site and $N$ is the total number of sites.
Putting~\eqref{hhy}--\eqref{hyy1} into~\eqref{uuj}, and assuming there is no long-range spin correlation, one then finds, in the thermodynamic limit $N \to \infty$,  
the following scalings 
\be 
\tilde J \sim O(1), \qquad \de \tilde J \sim N^{-\ha} \ .
\ee
The signal $\tilde J$ is then much larger than fluctuations $\de \tilde J$, and thus it is enough to make one single measurement. In other words, the signal is macroscopic or ``classical.'' Also note pure thermal fluctuations have the scaling 
\be\label{onj}
J_2 \sim N^{-\ha}, \qquad J_4 \sim  N^{-\ha} \ .
\ee

Similar scaling behavior can also be obtained in a large $N$ matrix-type theory (including two-dimensional CFTs in the large central charge limit). In this case take $J$ to be a single-trace operator and $U \sim e^{i N J \vp}$. We then find scalings 
\be \label{njj}
\tilde J \sim O(N), \qquad \de \tilde J \sim O(1), \qquad J_2, J_4 \sim O(1)
\ee
and thus the signal $\tilde J$ is again ``classical.''

Now coming back to~\eqref{nonu},  using the Cauchy-Schwarz inequality we find that 
\be\label{nho}
\bar J_g (t) \equiv \avg{J^L(t)}_g \leq 2 |a|\, \le(\vev{\Psi_\b| J^L J^L |\Psi_\b}  \ri)^\ha   = 2 |a| J_2  \ .
\ee
In other words, up to an $O(1)$ constant $\avg{J^L(t)}_g $ is bounded from above by the variance of $J$ in the thermal ensemble. 
Now consider the variance of $J^L$ with $g \neq 0$. Using~\eqref{nke} with various choices of $X(t)$, we find
\ie  \label{nho1}
\le(\de J_g \ri)^2 & \equiv \vev{(J^L(t) - \bar J_g (t))^2}_g  \cr
&= (1 -2 \Re a) J_2^2 - \bar J_g^2 + b 
\fe
with
\ie
b \equiv a \vev{\Psi_\b\le| (J^L(t))^2 U^R \ri|\Psi_\b} + h.c.  \leq 2 |a| J_4^2  
\fe
where we have again used Cauchy-Schwarz in the last step. Given that $\bar J_g \sim J_2$, and in general $J_4 \sim J_2$, 
all terms in~\eqref{nho1} are of order $J_2^2$. We thus find that modulo miraculous cancellations  $\de J_g \sim \bar J_g$, i.e. 
the variance is always of the same of order as the signal regardless of the choice of $J$ and $U^R$. 

More explicitly, let us consider the three situations mentioned above for the standard response story. 
For $J$ to be a spin operator at a single site, again all quantities are of order $O(1)$. For $J^L$ of the form~\eqref{hhy} with $U^R$ of the form~\eqref{hyy1}, then from~\eqref{nho} and~\eqref{onj} we find that 
\be \label{kj}
\bar J_g \sim J_2 \sim \de J_g  \sim N^{-\ha} \ .
\ee
In fact in this case considering the average polarization not only does not help to reduce the fluctuations, but also reduces 
the signal itself. One might as well just measure a single spin. Note that if one considers linear order in $\vp^R$ as in~\eqref{ma0}--\eqref{ma1} one may conclude that $\bar J_g \sim O(1)$ instead of~\eqref{kj}. This suggests that the linear response analysis for $\bar J_g$ could be potentially misleading in this regime. 

Finally for $J$ given by a single-trace operator in a large $N$ matrix-type theory, the counterparts of~\eqref{njj} are 
\be 
\bar J_g \sim \de J_g \sim O(1)
\ee
with again reduced signal.  
 
To summarize, the regenesis signal $\bar J_g$ in the $L$-system due to $U^R$ and coupling $V$ is intrinsically quantum
in all situations!

\subsection{Robustness of the \rena\ phenomenon} \label{sec:rob0}

Let us now consider the robustness of the \rena\ phenomenon, i.e. how it fares against imperfections in the preparation of the initial state at the time $-t_s$ when we turn on the external field $\vp^R$.  For simplicity, we will restrict to our discussion to linear order in $\vp^R$, i.e.~\eqref{ma0}--\eqref{ma1}. The arguments presented below generalize straightforwardly to full nonlinear level. 

Here we consider two types of ``small'' perturbations. One type is that at $t=0$ instead of $\bra{\Psi_\b}$ we get 
\be \label{sp}
\bra{\Psi (t=0)} = \bra{\Psi_\b} + \ep \bra{\Phi}
\ee
where $\ep$ is a small parameter. Physically this means that in preparation of the system at $t=-t_s$ to aim for $\bra{\Psi_\b}$ at $t=0$, the aiming is not perfect, but misses a bit.  Such perturbations may result if at $-t_s$ in addition to 
$\vp^R$ there are some other ``small'' sources present (whose strengths are characterized by $\ep$). With~\eqref{sp} in~\eqref{ma1} instead of $\bra{\Psi_\b}$, the corrections are clearly controlled by $\ep$, and thus the qualitative conclusion above
should not be modified. 

Another possibility is that at $t=0$ instead of $\bra{\Psi_\b}$ we have 
\be 
 \label{sp1}
\bra{\Psi (t=0)} = \ga^L (t_0 , \vx_0) \bra{\Psi_\b} 
\ee
with some $t_0$ for some few-body operator $\ga (x)$  (suitably smeared so that~\eqref{sp1} is normalizable). 
Heuristically, this describes a state obtained adding a ``$\ga^L$-particle'' to the thermal field double at time $t_0$. 
One could consider similar states obtained by acting with some operator in the $R$-system, but from~\eqref{co3} that is equivalent to an operation in the $L$ system with inverted time, so~\eqref{sp1} covers all cases.
 Strictly speaking,~\eqref{sp1} is not a small perturbation of $\bra{\Psi_\b}$ as it is orthogonal to $\bra{\Psi_\b}$ since a generic few-body operator $\ga$ will have negligible expectation value in $\bra{\Psi_\b}$. There is, however, a physical sense that such perturbations are ``small'':  at $t=t_0$, it is hard to tell the difference between~\eqref{sp1} and $\bra{\Psi_\b}$ by making measurements using generic few-body operators, such as $J$, as they generically commute with~$\ga$.

With~\eqref{sp1}, we should replace~\eqref{ma1} by
\be\label{gga}
W_\g\equiv{\ket{\Psi}\g^L(t_0)e^{-igV}J^L(t)e^{igV}J^R(-t_s)\g^L(t_0)\bra{\Psi}\ov \ket{\Psi}\g^L(t_0)\g^L(t_0)\bra{\Psi}}
\ee
where the spatial coordinates are suppressed. Consider first $g=0$, then 
\bln
W_\g (g=0) & = {\ket{\Psi}\g^L(t_0) J^L(t) J^R(-t_s)\g^L(t_0)\bra{\Psi}\ov \ket{\Psi}\g^L(t_0)\g^L(t_0)\bra{\Psi}} \cr
& = {\ket{\Psi}\g^L(t_0) J^L(t) \g^L(t_0) J^L (t_s + i \b/2)\bra{\Psi}\ov \ket{\Psi}\g^L(t_0)\g^L(t_0) \bra{\Psi}} 
\label{jhe}
\end{align}
which is an OTOC.  Equation~\eqref{jhe} is small whenever 
\be \label{der}
|t - t_0| \gg t_*    \quad \text{(destroy correlation of $J^L$ and $J^R$)}
\ee 
in which case insertion of $\ga^L$  will destroy the correlation between points $(t, \vx)$ and $(-t_s, \vx)$ in $\bra{\Psi_\b}$, and destroy \rena\ even without worrying about possible effects of $\ga$ on the interaction between two subsystems. 
For example, for $t \sim t_s \sim t_*$, equation~\eqref{der} means any $t_0 \ll 0$ or $t_0 \gg 2 t_*$. The latter can be more intuitively understood as insertion of $\ga^R$ of order $t_*$ before we send the signal at time $- t_*$. 

\begin{figure}
\begin{centering}
\includegraphics[width=8.5cm]{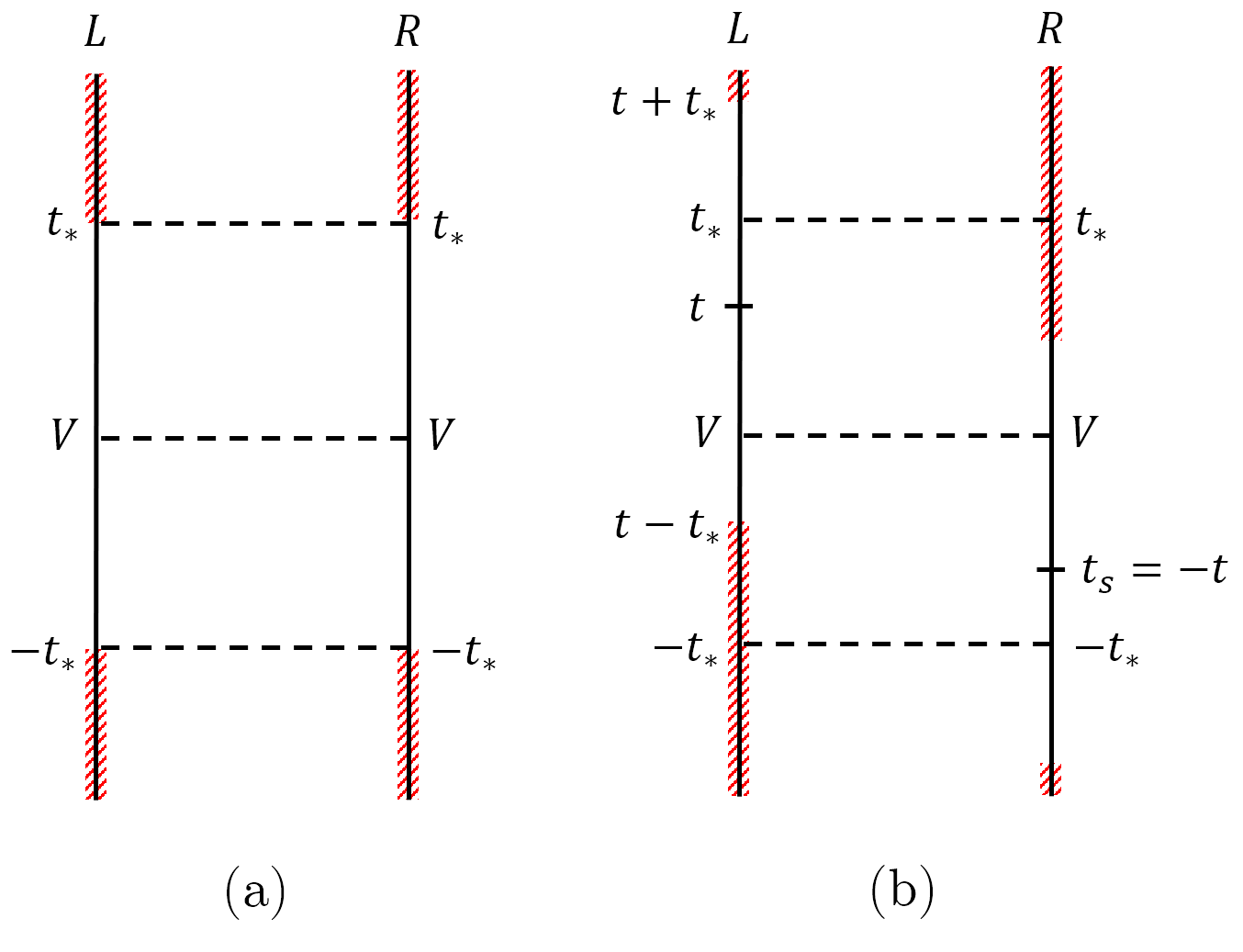}
\par\end{centering}
\caption{The \rena\ phenomenon is destroyed by insertion of $\ga$ from either side when $t_0$ is in the red shaded regions. (a) shows the destruction by diminishing effective coupling when $|t_0|\gg t_*$, and (b) shows the destruction by diminishing  $J^LJ^R$ correlation when $|t-t_0|\gg t_*$.
\label{fig:unst}}
\end{figure}

Now look at~\eqref{gga} with $g \neq 0$. Notice that the ordering between $\ga^L (t_0)$ and any $V$ insertions  are also out-of-time-ordered. From the same argument we then expect the effects of $V$ will be destroyed when 
 \be \label{der1}
 |t_0| \gg t_* \quad (\text{destroy the coupling between two systems}) \ .
 \ee 
Hence we expect \rena\ is no longer present in~\eqref{sp1} for $t_0 \gg t_*$ and $t_0 \ll 0$. 
See Fig.~\ref{fig:unst} for regions of insertion of $\ga$ which will destroy the \rena\ phenomenon. 

We will confirm  the above qualitative expectations in Sec.~\ref{sec:rob1}  by explicit calculations in two-dimensional CFTs. Our conclusion is also qualitatively consistent with 
gravity expectations discussed in~\cite{Maldacena:2017axo}, which we will elaborate more in Sec.~\ref{sec:grav}. 

\subsection{A contrast study: ``regenesis'' in a qubit  model}\label{sec:qubit}

Here we study the \rena\ phenomenon a simple qubit model to help sharpen some
essence aspects of the phenomenon in a quantum chaotic many-body system.

Consider a system consists of four qubits: $L_{1,2}$ and $R_{1,2}$, with the Hilbert space $\sH = \sH_L \otimes \sH_R = \mH_{L_1}\otimes\mH_{L_2}\otimes\mH_{R_1}\otimes\mH_{R_2}$. We will write 2 by 2 identity matrix and Pauli matrices as $\s^\mu=\{I,X,Y,Z\}$. We take the Hamiltonian to be like an Ising model
\be \label{Hamil1}
H_0=H^L+H^R,\quad H^{L}=\f{\nu}{2}(Z_{L_1}Z_{L_2})+\f{\mu}{2}(Z_{L_1}+Z_{L_2}),\quad H^{R}=\f{\nu}{2}(Z_{R_1}Z_{R_2})+\f{\mu}{2}(Z_{R_1}+Z_{R_2}) \ .
\ee

For simplicity we will consider the thermal field double~\eqref{tfd0} with $\b =0$, which is then giving by the following state of 
2 EPR pairs
\be \label{hbn}
\bra{\Psi}=\bra{EPR_{1}} \otimes \bra{EPR_{2}},\quad \bra{EPR_i}=\f {1}{ \sqrt{2}}(\bra{0_{L_i}0_{R_i}}+\bra{1_{L_i}1_{R_i}})
\ee
as one can readily check that each component $\bra{i_{L_1}j_{L_2};i_{R_1}j_{R_2}}$ $(i,j=0,1)$ in $\bra{\Psi}$ is an energy eigenstate. 

A general hermitian $J^L$ operator on $L_1$ site is $a_\mu\s^\mu$, where $a_\mu$ is a real vector. 
The corresponding operator $J^R$ acting on $R_1$ is then (note $\b =0$)
\be\label{Jr-df}
J^R\bra{\Psi} 
=J^L\bra{\Psi} \quad {\rm with} \quad J^R=a_0 I+a_1 X-a_2 Y+a_3 Z
\ee
We choose $V$ to act on site $L_2$ and $R_2$, i.e. it commutes with $J^L$ and $J^R$ as they act on different sites, so as to model the situation described in the many-body context that $V$ and $J$ are generic few-body operators whose degrees of freedom do not overlap. We will take $\sO = X$, and therefore 
\be \label{V1}
V=X_{L_2}X_{R_2} \ .
\ee

With $g=0$, we have 
\begin{align}
\ket{\Psi}J^R(-t)J^R(-t_s)\bra{\Psi}&= \ket{\Psi}J^R(t)J^L(-t_s)\bra{\Psi}\nn\\
&=a_0^2+a_3^2+(a_1^2+a_2^2)\cos\nu(t-t_s)\cos\mu(t-t_s)
\end{align}
which as expected is a function of only $t-t_s$. Since the above expression is real we have both $G^{RR} = G^{LR} = 0$. 
That even $G^{RR}=0$ is an artifact of that we are considering a $\b =0$ state. $G^{RR}$ is nonzero for other values of $\b$. 
Now turn on $V$ at $t=0$, we the have 
\begin{align}
W&\equiv\ket{\Psi}e^{-igV}J^L(t)e^{igV}J^R(-t_s)\bra{\Psi}\nn\\
&=a_0^2+a_3^2+(a_1^2+a_2^2)e^{-ig}\cos\mu(t-t_s)(\cos g \cos\nu(t-t_s)+i\sin g \cos\nu(t+t_s))
\end{align}
with 
\be 
G^{LR}(t,- t_s)=i(W-W^*)=2(a_1^2+a_2^2)\sin (2g)\sin\nu t\sin \nu t_s \cos\mu(t-t_s) \ .
\ee
$G^{RR} (t, -t_s)$ is given by the same expression as above with $t-t_s$ replaced by $t+t_s$. 

This simple example provides an interesting contrast which highlights some key elements of the \rena\ phenomenon for a chaotic many-body system: (1) for a few-body system, there is no dissipation, and thus $G^{RR}$ does not dissipate, i.e. even with $\vp^R$ turned off, the signal will remain in the $R$-system forever (turning on $g$ only modifies the signal somewhat). 
(2) With $g \neq 0$, the signal also appears in the $L$-system. The effect of $V$ is not regenesis, more like ``double genesis.''
The reason is of course trivial: interactions among degrees of freedom within each subsystem will manage to communicate $\vp^R$ to the $L$-system through $V$. 
(3) The response in the $L$-system depends sensitively on time, choice of the specific operator $J$,  the Hamiltonian $H^{L,R}$ of the subsystems, and choice of interaction $V$.

In other words, in a few-body system or an integrable many-body systems, some kind of signal in the $L$-system will be generated by turning on $V$. But it is not ``regenesis,'' and the signal will depend on all the specifics of an individual system and the operators used. In contrast, in chaotic systems, the behavior is universal, independent of all the details. 

\subsection{A generalization: \rena\ between spatially separated points}

With the understanding of the entanglement structure, the  set can be trivially generalized to be between any spatial points. 
Instead of  $\Psi_\b$ we could use a one-side spatially
shifted thermal field double state $e^{i \vec P \cdot \vec a}\bra{\Psi_\b}$ where $\vec P$
is the spatial translation operator in either left or right system.
Two choices are equivalent as $\vec P_{L}+\vec P_{R}$ is a symmetry of $\bra{\Psi_\b}$, and for definiteness we take $\vec P=\vec P_{R}$. The entangled pair of points for the shifted state are $(t, \vx)$ and $(-t, \vx - \vec a)$, and the \rena\ is now between them. We will also modify the interaction $V$ between $L$ and $R$ accordingly, e.g. replacing~\eqref{pp1} by 
\begin{equation}
V_{a}= 
\mO_{L}(0,\vec 0)\mO_{R}(0,-\vec a)  \ . \qquad 
\end{equation}
The story is then exactly same as before, with equation~\eqref{ma1} becoming
\begin{align}
W_{a}(t,\vec x; t',\vx') & =\avg{\Psi_\b|e^{-i\vec P_{R} \cdot \vec a}e^{-igV_a}J^{L}(t,\vx)e^{igV_a}J^{R}(t',\vx')e^{i\vec P_{R} \cdot \vec a}|\Psi_\b}\nonumber \\
 & =\avg{\Psi_\b|e^{-igV}J^{L}(t,\vx)e^{igV}J^{R}(t',\vx'+\vec a)|\Psi_\b} = W (t,\vec x; t',\vx' + \vec a) 
\end{align}
where we have used $e^{-i\vec P_{R} \cdot \vec a}  V_a e^{i\vec P_{R} \cdot \vec a} = V$.

\section{Explicit computations in large $c$ CFTs} \label{sec:cft}

To calculate~\eqref{nonli-J} explicitly for a general quantum many-body system is a difficult task. In~\cite{Maldacena:2017axo} it was calculated at leading order in $\vp^R$ (i.e.~\eqref{ma0}) for a $(0+1)$-dimensional holographic system by summing over scattering diagrams on gravity side.   
In this paper we will compute it in $(1+1)$-dimensional CFTs in the large central charge limit, again restricting to~\eqref{ma0}. 
This will enable us to obtain the behavior for $J^L$ for $t, t_s \sim t_*$ which one could not access using general arguments 
of Sec.~\ref{sec:gen}. In this section we will present the main technical results with the analysis of the results given in Sec.~\ref{sec:ana}. 

We will take $\sO$ to be  a scalar primary operator with conformal dimension $\De_\sO = 2 h_\sO$, and 
$J$ to be a scalar operator with dimension $\De_J = 2 h_J$. 
Furthermore, for convenience of calculation we will consider the regime 
\be\label{reg1}
O(1) \sim h_\sO \ll h_J \ll c \ .
\ee
 This regime is natural physically. We do not want the coupling $V$ to change the UV behavior of the
system, i.e. would like to take it to be a relevant operator, and thus $\De_\sO \sim O(1)$. 
$h_J$ should be much smaller than $c$ as $c$
is a measure of total number of degrees of freedom of a CFT. 
In our calculations we will neglect terms suppressed by ${1 \ov c}$ and ${h_\sO \ov c}$ 
while keeping all dependence on $h_J$ as is appropriate for~\eqref{reg1}.\footnote{This is slightly more general than the regime discussed in~\cite{fitzpatrick2015virasoro}, where the limit $h_J , c \to \infty$ with $h_J/c$ fixed was considered. See Appendix~\ref{app:a}.}  To compare with the holographic results of~\cite{Maldacena:2017axo}, we will focus on contributions from {\it the vacuum Virasoro block to~\eqref{nonli-J}}. Contributions from other primaries are discussed briefly in Sec.~\ref{sec:others} and are analyzed further in Sec. \ref{sec:eff-non-id}.


Here we will  outline the 
main steps and results, leaving technical details to Appendix~\ref{app:a}.  Readers who are only interested in the final expressions can skip this section. 

A remark on notation: below all $x$'s refer to spatial coordinate in $(1+1)$-dimension, although earlier we have used it as a shorthand for spacetime coordinates.

\subsection{Some useful expressions}

Here we first mention some standard results on two-point functions in the state $\Psi_\beta$ for a two-dimensional CFT, which we will use later. The Wightman function for two $J$'s in the same subsystem is given by 
\be\label{co1}
\vev{J^R (t_1, x_1) J^R (t_2, x_2)}_\b  = {C_J \le({2 \pi \ov \beta} \ri)^{2 \De_J}  \ov \le(2 \cosh \le({2\pi x_{12} \ov \beta} \ri) - 2 \cosh \le({2\pi (t_{12}+i\e_{12}) \ov \beta}  \ri)\ri)^{\De_J}} 
\ee
where $\e_{12}<0$ assigns the ordering of two $J^R$ operators and avoids singularity. The response function~\eqref{sop} is obtained from the imaginary part of~\eqref{co1}. 
The two-point function of $J$'s from different subsystems is given by
\bea  
\vev{\Psi_\b|J^L (t_1, x_1) J^R (- t_2,x_2)|\Psi_\b} &=&  \vev{J^L (t_1, x_1) J^L (t_2+ i \beta/2,x_2)}_\b  \cr
&=&  {C_J \le({2 \pi \ov \beta} \ri)^{2 \De_J}  \ov \le(2 \cosh \le({2\pi x_{12} \ov \beta} \ri) + 2 \cosh \le({2\pi t_{12} \ov \beta} \ri)\ri)^{\De_J}}  \
\label{co2}
\eea
where $C_J$ is a constant and $x_{12} = x_1 - x_2$. 
Note that in~\eqref{co1}--\eqref{co2}, the correlators decay exponentially for $(t_2, x_2)$ lying outside the region $(t_1 \pm {\b \ov 2\pi}, x_1 \pm {\b \ov 2 \pi})$, as indicated earlier in~\eqref{oin} and~\eqref{omk}. 
The form of~\eqref{co2} is a manifestation of the entanglement structure of~\eqref{tfd0} discussed in Sec.~\ref{sec:ent}:  the two systems 
are entangled in such a way that an operator inserted at point $(-t, x)$ in $R$ system is highly correlated with the same operator inserted in a region of size ${\b \ov 2 \pi}$ around $(t,x)$ in $L$ system.

Similarly we have 
\bega\label{oo1}
\left\langle \mathcal { O } \left( 0, x _ { i } \right) \mathcal { O } \left(0,  x _ { j } \right) \right\rangle_\b = \frac { C_\sO  ( 2 \pi / \beta ) ^ { 2 \Delta _ { O } } } { \left( 2 \cosh \frac { 2 \pi } { \beta } x _ { i j } - 2 \right) ^ { \Delta _ { O } } } , \\
 \left\langle \Psi_\b| \mathcal { O }^L \left( 0, x _ { i } \right) \mathcal { O }^R \left( 0, x _ { j } \right)  |\Psi_\b\right\rangle = \frac { C_\sO  ( 2 \pi / \beta ) ^ { 2 \Delta _ { O } } } { \left( 2 \cosh \frac { 2 \pi } { \beta } x _ { i j } + 2 \right) ^ { \Delta _ { O } } } \ .
 \label{oo2}
\end{gather}
In our discussion below we will also use the following notations 
\be \label{cc1}
\left\langle \Psi_\b |\mathcal { O }^L \left( 0 \right)  \sO^R (0) |\Psi_\b  \right\rangle = C_\mO \le({\pi \ov \beta} \ri)^{2 \De_\sO} \equiv G\ . 
\ee

\subsection{More elaborations on $W$} 

Equation~\eqref{ma1} is the central object that we would like to calculate and analyze. 
Here we elaborate a bit further on its structure. We can expand it in an infinite series (for definiteness using~\eqref{pp3} as an example)
\be \label{ma00}
W (t, x; - t_s,  x_s) =\sum_{n=0}^\infty  {(-ig)^n \ov L^n n!}  \int_{-{L \ov 2}}^{L \ov 2} \left( \prod _ { k = 1} ^ { n } d x _ { k } \ri)  W_n
\ee
with
\bega \label{eq:Wn-fomal}
W_n =  \vev{\Psi_\b|[v_n,[v_{n-1},  \cdots [v_1, J (t,x)]\cdots]  J^R (-t_s,x_s)|\Psi_\b }, \\
v_i \equiv \sO (0,x_i)  \sO^R (0,x_i) \equiv \sO_i  \sO_i^R, \quad \sO_i \equiv \sO(0, x_i) 
\end{gather}
where we have suppressed superscripts $L$ for operators in $L$ system. 
More explicitly, 
\bega
\label{jh0}
W_0 =\vev{\Psi_\b|J  J^R|\Psi_\b } = \vev{J \tilde J}_\b , \qquad  
W_1 =\vev{\Psi_\b|[\sO_1, J]  \sO_1^R J^R|\Psi_\b } =  \vev{[\sO_1, J] \tilde J \tilde \sO_1 }_\b, \\
W_2 =\vev{\Psi_\b|[\sO_2,[\sO_1, J]] \sO_1^R \sO_2^R   J^R|\Psi_\b } =  \vev{[\sO_2,[\sO_1, J]] \tilde J  \tilde \sO_2 \tilde \sO_1 }_\b \\
W_n = \vev{[\sO_n,[\sO_{n-1},  \cdots [\sO_1, J]\cdots]\ \tilde J \tilde \sO_n \cdots \tilde \sO_1}_\b
\label{jh}
\end{gather}
 where we have used~\eqref{con} repeatedly and introduced short-hand notations
 \be\label{jh1}
 J \equiv J(t, x), \qquad\sO_i \equiv \sO(0, x_i), \qquad  \tilde \sO_i \equiv \sO ( i \beta /2, x_i), \qquad \tilde J \equiv J (t_s  + i \beta /2, x_s) \ .
 \ee
Note that all $\tilde \sO$'s commute with one another.

\subsection{Evaluating $W$: part I}

We will proceed by first evaluating~\eqref{jh} and then performing the sum~\eqref{ma00}. 
The thermal correlation functions~\eqref{jh} are in turn obtained by analytic continuation from those in the Euclidean signature.
Let us first describe how to compute a multiple-point function of the form 
\be \label{onm}
w_n = \vev{J  (t_a, x_a) J (t_b, x_b) \sO (t_1, x_1)  \cdots \sO (t_{2n}, x_{2n})}_\b 
\ee
in the Euclidean signature, i.e. with all the $t = - i \tau$ understood as being pure imaginary. 
Following the standard procedure, 
we first perform 
a conformal transformation 
\be 
z = e^{{2 \pi \ov \beta} (x +t )}, \qquad \bar z = e^{{2 \pi \ov \beta}(x-t)} \  
\ee
to map the cylinder $(\tau, x)$ ($\tau$ is periodic in $\b$) to the full complex $z$ plane.  
Note that for pure imaginary $t$, $z, \bar z$ are complex conjugates of each other, but are independent variables for general complex $t$. 
The calculation of~\eqref{onm} on the $z$-plane is still nontrivial. Fortunately, in the regime $h_\sO \ll h_J \ll c$, one could do it 
by applying techniques developed recently in~\cite{fitzpatrick2015virasoro}. 

For example, at the level of 4-point function we find that 
\bega \label{cv1}
 \frac { \left\langle J _ { a } J_ { b } \mathcal { O } _ { 1}  \sO_2  \right\rangle  } { \left\langle J_ { a }  J_ { b } \right\rangle  \left\langle \mathcal { O } _ { 1} \sO_2  \right\rangle  } = \sV (u) \sV (\bar u) + O(1/c, h_\sO/c) 
 \end{gather}
 where for notational simplicity we have used the subscripts to denote the positions of operators, and ($\bar u$ is defined as $u$ with $z$'s replaced by $\bar z$'s)
 \bega \label{sv0}
 \mathcal { V } ( u ) = \left( \frac { \alpha ^ { 2} u ^ { 2} ( 1- u ) ^ { \alpha - 1} } { \left( 1- ( 1- u ) ^ { \alpha } \right) ^ { 2} } \right) ^ { h_\sO }, \quad u = \frac { z _ { 12} z _ { a  b } } { z _ { 1a } z _ { 2  b } }, \quad z_{12} = z_1 - z_2, \\
  \al = \sqrt{1 - { 24 h_J \ov c}}  \ .
   \label{borp}
 \end{gather}
More generally,  for~\eqref{onm} we have 
\be \label{bn}
\frac { \left\langle J _ { a }  J_ { b } \mathcal { O } _ { 1 }  \cdots \sO_{2n} 
\right\rangle } { \left\langle J _ { a }  J_ { b } \right\rangle  } 
 =   \sum _ {\rm all \, pairings} \prod _ { i = 1 } ^ { n } \left[ \mathcal { V } \left( u _ { i } \right) \sV (\bar u_i) \left\langle \mathcal { O } _ { i1 } \mathcal { O } _ { i2} \right\rangle \right]  + O(1/c, h_\sO/c) , \quad  u _ { i } \equiv \frac { z _{i1,i2}  z _ { a b } } { z _ { i1,  a } z _ { i2, b } }
\ee
where the sum is over all possible pairings of $\sO$'s with $(\sO_{i1}, \sO_{i2})$ denoting the $i$-th pair. 
See Appendix~\ref{app:a} for details. 

We now analytically continue the above expressions to Lorentzian signature to obtain~\eqref{jh}. 
Correlation function of 
Lorentzian operators with a specific ordering can be obtained from continuation of the corresponding Euclidean correlation function
by assigning appropriate $i \ep$'s~\cite{streater2016pct}.  
For example, 
\begin{equation}
\vev{\sO(t_{1})\cdots\sO(t_{n})}=\lim_{\{\e_{j}\}\ra0}\vev{\sO(t_{1}+ i\e_{1})\cdots\sO(t_{n}+ i\e_{n})}, \quad \e_{1}< \cdots < \e_{n}
\end{equation}
where the left hand side denotes Lorentzian correlation function of a specified order, while the right hand side denotes Euclidean correlation function with the time argument $ t= - i \tau_i$ for each operator replaced by $ t = t_i + i \ep_i$, and 
$\ep_i$ ordered as indicated. This $i \ep$-prescription instructs how one continues through possible branch cuts encountered 
when analytically continuing from imaginary to real times. 

Therefore for each term in~\eqref{jh0}--\eqref{jh} we just need to continue~\eqref{bn} by assigning different orderings of $\e_{i}$'s. For example, from~\eqref{cv1}, we find that  (recall~\eqref{jh1})
\be \label{bnm}
W_1 = \vev{\sO_1 J \tilde J \tilde \sO_1} - \vev{J \sO_1 \tilde J \tilde \sO_1} = G \vev{J \tilde J} A (u_1, \bar u_1)
\ee
where we have used~\eqref{cc1} and  
\be \label{defa}
A (u_1, \bar u_1) = \sV^+ (u_1) \sV^+ (\bar u_1)- \sV^- (u_1) \sV^- (\bar u_1), 
\ee
with
\ie \label{cv2}
u_1 &= {e^{{2 \pi \ov \b} x_1}  (e^{i \ep_1} + e^{ i \tilde \ep_1}) (e^{{2 \pi \ov \b} (x + t) + i \ep_J} +  e^{{2 \pi \ov \b} (x_s + t_s) + i \tilde \ep_J})  \ov (e^{{2 \pi \ov \b} x_1 + i \ep_1} - e^{{2 \pi \ov \b} (x + t )+ i \ep_J} )(e^{{2 \pi \ov \b} (x_s + t_s) + i \tilde \ep_J}  - e^{{2 \pi \ov \b} x_1 + i \tilde \ep_1})}, \cr \bar u_1 & = {e^{{2 \pi \ov \b} x_1}  (e^{- i \ep_1} + e^{- i \tilde \ep_1}) (e^{{2 \pi \ov \b} (x - t) - i \ep_J} +  e^{{2 \pi \ov \b} (x_s - t_s)- i \tilde \ep_J})  \ov (e^{{2 \pi \ov \b} x_1- i \ep_1} - e^{{2 \pi \ov \b} (x - t)- i \ep_J} )(e^{{2 \pi \ov \b}(x_s - t_s)- i \tilde \ep_J}  - e^{{2 \pi \ov \b} x_1-i \tilde \ep_1})} \ .
\fe
In~\eqref{defa} $\sV^+$ denotes~\eqref{sv0} with ordering  $\ep_1 < \ep_J < \tilde \ep_J < \tilde \ep_1$, while $\sV^-$ denotes~\eqref{sv0}  with ordering $\ep_J < \ep_1 < \tilde \ep_J < \tilde \ep_1$.  

For simplicity we will take $t_s = t$ and $x_s = x$, which as discussed earlier is the most relevant case. 
By tracking the motions of $u_1, \bar u_1$ as one varies $t$, we can write $A(u_1, \bar u_1)$ more explicitly as 
\be \label{aa1}
A (u_1, \bar u_1) =  (\sV_1  ( u_1 ) - \sV_{2} ( u_1 )) \sV_1 (\bar u_1) 
\ee
where  $\sV_1 (u)$  and $\sV_2 (u)$ denote respectively the values of~\eqref{sv0} along the negative real axis on its first and second sheet ($\sV (u)$ has a branch cut along $(1, +\infty)$)
\bega\label{sv1}
\sV_1 (u) =  \left( \frac { \alpha ( - u ) } { \sqrt { 1- u } } \frac { 1} {- ( 1- u ) ^ { - \alpha / 2}  + ( 1- u ) ^ { \alpha / 2} } \right) ^ { 2h_\sO } ,  \\
\label{sv2}
\sV_2 (u) = \left( \frac { \alpha ( - u ) } { \sqrt { 1- u } } \frac { 1} { ( 1- u ) ^ { - \alpha / 2} e ^ { i \pi \alpha } - ( 1- u ) ^ { \alpha / 2} e ^ { - i \pi \alpha } } \right) ^ { 2h_\sO }  
\end{gather}
and for convenience we have slightly redefined $u_1, \bar u_1$ as  
\be \label{aa2}
u_1 \equiv - {4 e^{{2 \pi \ov \b} ( t- |x - x_1|)}   \ov (1 - e^{{2 \pi \ov \b} ( t- |x - x_1|)} )^2}, \qquad  \bar u_1  \equiv - {4 e^{{2 \pi \ov \b} ( t + |x - x_1|)}   \ov (1- e^{{2 \pi \ov \b} ( t + |x - x_1|)} )^2}   \ .  
\ee
The explicit evaluation of~\eqref{aa1} is given in Appendix~\ref{app:aexp}. 

\subsection{Evaluating $W$: part II}

For general $W_n$, let us first look at the case of $V$ given by~\eqref{pp2}, for which  
\be\label{lagtj}
W_n = {1 \ov k^n} \le (\prod_{i=1}^n \sum_{\al_i=1}^k \ri) \vev{[\sO_{\al_n},[\sO_{\al_{n-1}},  \cdots [\sO_{\al_1}, J]\cdots]\ \tilde J \tilde \sO_{\al_1} \cdots \tilde \sO_{\al_n}}   
\ee
where subscripts denote different types of operators all inserted at $t, x =0$. 
Applying~\eqref{bn} to a term obtained by expanding commutators in~\eqref{lagtj}, we see that there are two types of contractions among $\sO$'s: two-sided contractions between a $\sO_{\al_i}$ and a $\tilde \sO_{\al_j}$ which are given by
$\vev{\sO_{\al_i} \tilde \sO_{\al_j}}_\b = G  \de_{\al_i \al_j}$ (recall~\eqref{cc1}), and same-sided contractions between $\sO$'s (or between $\tilde \sO$'s) which are in fact divergent. We will assume that $\sO$ and $\tilde \sO$ are smeared such that same-sided contractions are finite. 
The two-sided contractions can be further separated into  contractions among operators in the same sums or different sums. Note there is an enhancement factor $k$ if in a sum each $\sO_{\al_i}$ is contracted with the corresponding $\tilde \sO_{\al_i}$ from the same sum \cite{Maldacena:2017axo}. 
Thus in the large $k$ limit, this type of contractions will dominate over all others, including same-sided contractions. Also note that for various terms obtained by expanding commutators of~\eqref{lagtj} only orderings between $\sO$ and $J$ matter (all the $\sO$ and $\tilde\sO$ commute with one another). We then conclude that to leading order in large $k$
\be\label{wn}
W_n =  G^n A^n_0 \vev{J \tilde J} + O(1/k) \quad \to \quad W = \vev{J \tilde J}  e^{-i g G A_0}  + O(1/k) \ 
\ee
where $A_0 \equiv A (u_0, \bar u_0)$ with $u_0, \bar u_0$ obtained by setting $x_1 =0$ in $u_1, \bar u_1$. 

At finite $k$, which includes~\eqref{pp1} as a special case with $k=1$, one has to keep track of all other contractions, which is very complicated. The detailed derivations are given in Appendix~\ref{full-k}. The final result can be written in a form 
\begin{equation}
W=\avg{J\tilde{J}}\left(1+\f{i g G (A_{0}  + \eta B_{0} )}k\right)^{-k/2}\left(1+\f{i g G  (A_{0} - \eta B_{0} )}k\right)^{-k/2}
\label{gen-k}
\end{equation}
where $\eta = H/G$ and $H$ is defined as $\vev{\sO_{\al_i} \sO_{\al_j}} = \vev{\tilde \sO_{\al_i} \tilde \sO_{\al_j}} = H  \de_{\al_i \al_j}$.\footnote{We are assuming that the smearing is such that same-sided contractions of $\sO$ and $\tilde \sO$ are the same.
There is no qualitative change if one takes them to be different.} 
In~\eqref{gen-k} $B_0$ is given by 
\be \label{ooj}
B_0^2 = \left[(\mV_{1}(\mu_0)-\mV_{-1}(\mu_0))\mV_{1}(\bar{\mu}_0)+(\mV_{1}(-\mu_0)-\mV_{2}(-\mu_0))\mV_{1}(-\bar{\mu}_0)\right]\mV_{1}(\mu_0)\mV_{1}(\bar{\mu}_0)
\ee
where $\sV_1,\sV_2$ were given before in~\eqref{sv1}--\eqref{sv2}, $\sV_{-1}$ is the corresponding value on $-1$ sheet, given by  
\begin{equation}
\mV_{-1}(u)=\left(\f{\a(-u)}{\sqrt{1-u}}\f 1{(1-u)^{-\a/2}e^{-i\pi\a}-(1-u)^{\a/2}e^{i\pi\a}}\right)^{2h_{\mO}} \ .
\end{equation}
In~\eqref{ooj}, the arguments of $\sV$-functions are defined as
\be
\mu_0=\f{2i\sin \f{\pi\e}{\b}}{\sinh\f{2\pi}{\b}(t-|x|)+2i\sin \f{\pi\e}{\b}},\quad\bar{\mu}_0=\f{2i\sin \f{\pi\e}{\b}}{\sinh\f{2\pi}{\b}(t+|x|)+2i\sin \f{\pi\e}{\b}},
\ee
where $0<\e<\b$ is a regulator which makes same-sided contractions finite.
A couple of further comments on~\eqref{gen-k}. In the limit $\eta \to \infty$, $W$ becomes real and thus $G^{LR}$ is zero in that limit.  In large $k$ limit, $B_0$ terms cancel out in the exponential and recovers~\eqref{wn}.

Now finally consider~\eqref{pp3}, which we will take $L$ to be much larger than $\b$.\footnote{For $L$ comparable 
 or smaller than $\b$, it is not that different from~\eqref{pp1}.} The discussion here is similar to the large $k$ story described above, with the sums over indices $\al$ replaced by integrations over $x$. The counterpart of $k$ is ${L \ov \b}$. In the large ${L \ov \b}$ limit we will need to include contractions between $\sO_i$ and $\tilde \sO_i$ which belong to the same integral. 
 Parallel discussion as~\eqref{wn} then leads to leading order in ${\b \ov L}$ 
 \be\label{wn1}
 W 
  = \vev{J \tilde J} \exp \left( -\frac { i g G  } { L } \int _ { - L / 2} ^ { L / 2} d x_1 \, A (u_1, \bar u_1)
   \right) \ .
 \ee

\subsection{Contributions from other primaries} \label{sec:others}

Let us now consider contributions from a primary operator with weights $(h, \bar h) \sim O(1)$. For simplicity, we only consider the case~\eqref{pp2}. Using results from~\ref{app:non-id}, we can check that $W_n$ becomes 
\begin{equation}
W_{n}=nB(u_{1},\bar{u}_{1})A(u_{1},\bar{u}_{1})^{n-1}
\end{equation}
where $A(u_{1},\bar{u}_{1})$ is given by \eqref{aa1} and $B$ is
\begin{equation}
B(u_{1},\bar{u}_{1})=(\mV_{h,1}(u_{1})-\mV_{h,2}(u_{1}))\mV_{\bar{h},1}(\bar{u}_{1})
\end{equation}
where 
\begin{align}
\mV_{h,1}(u) =& \mV_{1}(u)\left[\f{1-(1-u)^{\a}}{\a}\right]^{h}{}_{2}F_{1}(h,h,2h,1-(1-u)^{\a})\\
\mV_{h,2}(u) = & \mV_{2}(u)\left[\f{1-e^{-2\pi i\a}(1-u)^{\a}}{\a}\right]^{h}\nn \\
&\times\left[_{2}F_{1}(h,h,2h,1-e^{-2\pi i\a}(1-u)^{\a})+\f{2\pi i\Gamma(2h)}{\G(h)^{2}}{}_{2}F_{1}(h,h,1,e^{-2\pi i\a}(1-u)^{\a})\right]
\end{align}
are the Virasoro block on first and second sheet. Similar for $\mV_{\bar{h},1}(\bar{u}_{1})$. Note that here we should also include the monodromy of hypergeometric
function when moving the coordinate to second sheet. We then find 
\begin{equation}
W=-igGC_{JJh}C_{h\mO\mO}C_{JJ\bar h}C_{\bar h\mO\mO}\avg{J\tilde{J}}B_{0}e^{-igGA_{0}}\label{eq:w-ni-9}
\end{equation}
where $A_{0},B_{0}\equiv A(u_{0},\bar{u}_{0}),B(u_{0},\bar{u}_{0})$
with $u_{0}$, $\bar{u}_{0}$ obtained by setting $x_{1}=0$ in $u_{1}$,
$\bar{u}_{1}$, and $C_{JJh}C_{h\mO\mO}C_{JJ\bar h}C_{\bar h\mO\mO}$ are OPE coefficients.

\section{Analysis of the results} \label{sec:ana}

In this section we analyze the expression for $W$ obtained in last section. The main expressions are~\eqref{wn} for~\eqref{pp2} in the large $k$ limit,~\eqref{gen-k} for~\eqref{pp2} at any finite $k$, which includes~\eqref{pp1} as a special case ($k=1$),
and~\eqref{wn1} for~\eqref{pp3} in the limit $L \gg \b$. 
 $A(u_1, \bar u_1)$  in those expressions are given by~\eqref{aa1}--\eqref{aa2} with $\al$ given by~\eqref{borp}, and $A_0$ is obtained from $A(u_1, \bar u_1)$ by setting $x_1 =0$. $B_0$ is given by~\eqref{ooj}.

\subsection{General remarks} \label{sec:genr}

We first note that in all cases (i.e.~\eqref{wn},~\eqref{gen-k} and~\eqref{wn1}) $W$ is proportional to 
\be
\vev{J \tilde J}_\b = \vev{\Psi_\b|J^L (t, x) J^R (- t_s,x_s)|\Psi_\b}
\ee
which from~\eqref{co2} is supported for $|t-t_s| \lesssim {\b \ov 2 \pi}$ and $|x-x_s|\lesssim {\b \ov 2 \pi}$. 
As commented earlier that the form of~\eqref{co2} is in turn determined by the entanglement structure of thermal field double state $\bra{\Psi_\b}$. Thus 
possible spacetime points $(t,x)$ to which one could send signal from $(-t_s, x_s)$ is determined by the entanglement structure
with ${\b \ov 2 \pi}$ characterizing the size of the window for possible nonzero signal. 
Now consider~\eqref{lresp} which we copy here for convenience,   
\be \label{lhre}
\vev{J^L (t,x)}_\vp = \int dt_s d x_s \, G^{LR} (t,x; -t_s, x_s) \vp ^R(-t_s, x_s) \ .
\ee
Since $G^{LR} (t,x; t_s, x_s) \propto {\rm Im} W$ falls off rapidly outside the window $|t-t_s| \lesssim {\b \ov 2 \pi}$ and $|x-x_s|\lesssim {\b \ov 2 \pi}$, for sources $\vp ^R(t,x)$ which are slowly varying in spacetime at the scale of ${\b \ov 2 \pi}$ we can approximate~\eqref{lhre} as 
\be \label{hhe}
\vev{J^L (t,x)}_\vp \approx  \bar G^{LR} (t,x; -t, x) \vp^R (-t, x) 
\ee  
where $\bar G^{LR} (t,x; -t, x)$ is obtained by averaging $G^{LR} (t,x; -t_s, x_s)$ in $(t_s, x_s)$ over the region defined by 
$|t-t_s| \lesssim {\b \ov 2 \pi}$ and $|x-x_s|\lesssim {\b \ov 2 \pi}$. Below without loss of qualitative features, instead of considering the averaged $\bar G^{LR} (t,x; -t, x)$, we will simply examine the behavior of $G^{LR} (t,x; -t_s, x_s)$ for 
$t_s=t$ and $x_s=x$. In this case we then have 
\be
\vev{J \tilde J}_\b  = C_J \le({\pi \ov \beta} \ri)^{2 \De_J} \equiv G_J\ . 
\ee
One interesting feature of~\eqref{hhe} 
is that signals which are sent earlier in the $R$ systems appear
later in the $L$ system, which is of course a direct consequence of the entanglement structure discussed in Sec.~\ref{sec:ent}.

Also notice that in~\eqref{wn},~\eqref{gen-k} and~\eqref{wn1}, the coupling $g$ always comes with 
$G$, which is the (maximal) correlation between $\sO^L$ and $\sO^R$ as following from~\eqref{oo2}. This is due to interactions~\eqref{pp1},~\eqref{pp2}--\eqref{pp3} are between $\sO^L$ and $\sO^R$ inserted at entangled spacetime points. Were we to couple $\sO^L$ and $\sO^R$ at general spatial locations or times, then one effectively diminishes the value of $G$, and weakens the effects of $V$. Below we will use
\be
g_{\rm eff} \equiv g G 
\ee
which may be interpreted as the effective coupling between two systems.

For notational simplicity we will  write $G^{LR}  (t, x; -t, x) $ as $G^{LR} (t, x)$, and then from~\eqref{ma0}
we find for~\eqref{wn} 
\be \label{glr0}
G^{LR} (t, x)   
= 2 G_J e^{\ge \, {\rm Im A_0}} \sin \le(\ge \, {\rm Re A_0}\ri) \ 
\ee
and for~\eqref{wn1}
\be \label{glr1}
G^{LR} (t, x)   
= 2 G_J e^{\ge \, {\rm Im \sA}} \sin \le(\ge \, {\rm Re \sA}\ri) 
\ee
with 
\be \label{glr2}
\sA \equiv \frac {1} { L } \int _ { - {L \ov 2}} ^ { L \ov 2} d x_1 \, A (u_1, \bar u_1) \ .
\ee
The expression of $G^{LR}$ for~\eqref{pp1} can be straightforward obtained by taking the imaginary part of~\eqref{gen-k} with $k=1$, but the formal expression is not very illuminating, so we will not write it explicitly.

Now recall that $A$ is the normalized four-point function 
\be\label{nhk}
A(u_1, \bar u_1)=  {\vev{[\sO (0, x_1), J (t, x) ] J (t+ {i \beta/2}, x)  \sO (i\b/2,x_1)}_\b \ov 
\vev{\sO (0, x_1) \sO (i\b/2,x_1)}_\b \vev{J (t, x)  J (t+ {i \beta/2}, x) }_\b }
\ee
with $A_0$ are given by setting $x_1$ to zero. The commutator upstairs is the difference between $\vev{\sO J \tilde J \tilde O}_\b$ 
and $\vev{J \sO \tilde J \tilde O}_\b$, with the latter one being an OTOC.
It has been known from previous discussion in~\cite{Roberts:2014ifa} that as $ t \to \infty$  
the OTOC $\vev{J \sO \tilde J \tilde O}_\b$ goes to zero, while $\vev{\sO J \tilde J \tilde O}_\b$ factorizes into $\vev{\sO \tilde \sO}_\b \vev{J \tilde J}_\b$ and thus 
\be 
A \to 1, \quad t \to \infty \ 
\ee
which can also be checked explicitly from~\eqref{aa1}.\footnote{Note as $t \to \infty$, $u_1, \bar u_1 \to 0$ and we find $\sV_1 (u) \to 1, \sV_2 (u) \to 0$.} 
We then find that $G^{LR}$ goes a constant, e.g. for~\eqref{glr0}--\eqref{glr1} 
\be \label{nno}
G^{LR} (t \to \infty, x) =  2 G_J \sin \ge \ .                       
\ee

Similarly for~\eqref{gen-k} we find as $t \to \infty$, $B_0 \to \sqrt{2}$, and thus
\be 
W (t \to \infty, x) = G_J \le(1 + {i \ge (1+\sqrt{2} \eta) \ov k}\ri)^{-{k \ov 2}}  \le(1 + {i \ge (1- \sqrt{2} \eta) \ov k}\ri)^{-{k \ov 2}} \ .
\ee
Note that while~\eqref{nno} is oscillatory in $\ge$ with period $2 \pi$, for $k=1$ it is not,  and goes to zero as
 $\ge^{-1}$ for large $\ge$. 

From~\eqref{hhe} we then find that in all cases 
\be \label{nno1}
\vev{J^L (t,x)}_\vp \approx C(g)  \,  \vp^R (-t, x), \qquad t \to \infty \ , 
\ee
with $C(g)$ a constant, which is consistent with~\eqref{mok}--\eqref{mok1} deduced on general ground.

We will now proceed to understand the behavior of $G^{LR}$ at general times in more detail. 

\subsection{A scaling limit}

Based on general discussion of Sec.~\ref{sec:gen} we expect the nontrivial behavior of $G^{LR}$ to arise when $t$ 
becomes of order of the scrambling time $t_*$ which for a CFT at large $c$ is proportional to $\log c$~\cite{Roberts:2014ifa}. 
Since in general we do not expect $x-x_1$ to scale with $c$, thus for values of $t$ of interest we should have $t \gg |x-x_1|$  for which 
\be \label{eq:largets}
u_1 \approx - 4 e^{{2 \pi \ov \b} (- t + |x-x_1|)} \ll 1, \qquad \bar u_1 \approx - 4 e^{- {2 \pi \ov \b} (t + |x-x_1|) } \ll 1 \ .
\ee
Denoting $\sV_{1,2} (u)= U_{1,2} (u)^{2 h_\sO}$ and expanding in small $u$ we find from~\eqref{sv1}
\be
U_1 (u) = 1 + O(u^2) \ .
\ee
We are interested in $h_J \ll c$, i.e. $\al \approx 1$. It can be readily checked all the coefficients of higher powers of $u$ in $U_1$ going to zero as $\al \to 1$ ($U_1 =1$ for $\al=1$). Expanding for small $u_1 < 0$ in $U_2$ we find from~\eqref{sv2} that
\be \label{kmn}
U_2 (u_1) = {\al |u_1| \ov 2 i \sin \pi \al} - {\al |u_1|^2 (-i + \al \cot \pi \al ) \ov 4 \sin \pi \al}  + \cdots  \ .
\ee 
Now we notice that as $ \al \to 1$, each coefficient becomes singular. There is a scaling regime
\be 
u_1 \to 0, \qquad \al \to 1, \qquad Q \equiv {2 \pi (1-\al) \ov  |u_1|} = {\rm finite}
\ee
in which one can resum the whole series~\eqref{kmn} 
\be 
U_2 (u_1) = {1 \ov 1 + i Q}  + O(1/c) 
\ee
$Q$ can be written more explicitly as 
\be \label{eq:df-Q}
Q =  {6 \pi h_J \ov c} \exp \le({2 \pi \ov \b} (t - |x-x_1|) \ri)  \equiv  \exp \le({2 \pi \ov \b} (\ft- |x-x_1|) \ri)
\ee
where we have introduced 
\be \label{bro}
 \ft \equiv t + t_J - t_*  \qquad 
t_* \equiv {\beta \ov 2 \pi} \log {c \ov 6 \pi}, \qquad t_J \equiv {\beta \ov 2 \pi} \log h_J \ .
\ee
We then find 
\be \label{mui}
A (u_1, \bar u_1)= 1 - \le({1 \ov 1 + i Q} \ri)^{2 h_\sO}    + O(1/c)
 \ .
\ee 

Similarly, in the scaling limit, from~\eqref{ooj} we find $B_0$ can be written as
\be \label{mui1}
B_{0}^{2} = 2 \left[1-\left(1+\f{1}{\sin\f{\pi\e}{\b}}Q_0 \right)^{-2h_{\mO}} \right] \le(1 + O(1/c) \ri) , \quad Q_0 \equiv Q (x_1 =0)\ .
\end{equation}
Note that in contrast to $A (u_1, \bar u_1)$ and $A_0$, $B_0$ is always real. 

In the large $c$ limit, $t_* \gg t_J, |x-x_1|$, the scaling limit helps us to focus on the time scale $t \sim t_*$  during which $Q$ is $O(1)$ (in terms of large $c$ scaling)
and the commutator~\eqref{nhk} between generic few-body operators $\sO$ and $J$ becomes sizable. This defines $t_*$ as the scrambling time. 

Note that it is curious that at leading order in~\eqref{mui}--\eqref{mui1} the only dependence on $h_J$ is through a time shift $t_J$. 

\subsection{Three regimes of $G^{LR}$}

Let us now look at the behavior of $G^{LR}$ more closely, using~\eqref{glr0} for~\eqref{pp2} as the main example, and will comment on the differences at finite $k$.  Equation~\eqref{glr1} for~\eqref{pp3} will be discussed in next subsection.

\begin{figure}
\begin{centering}
\includegraphics[width=15.5cm]{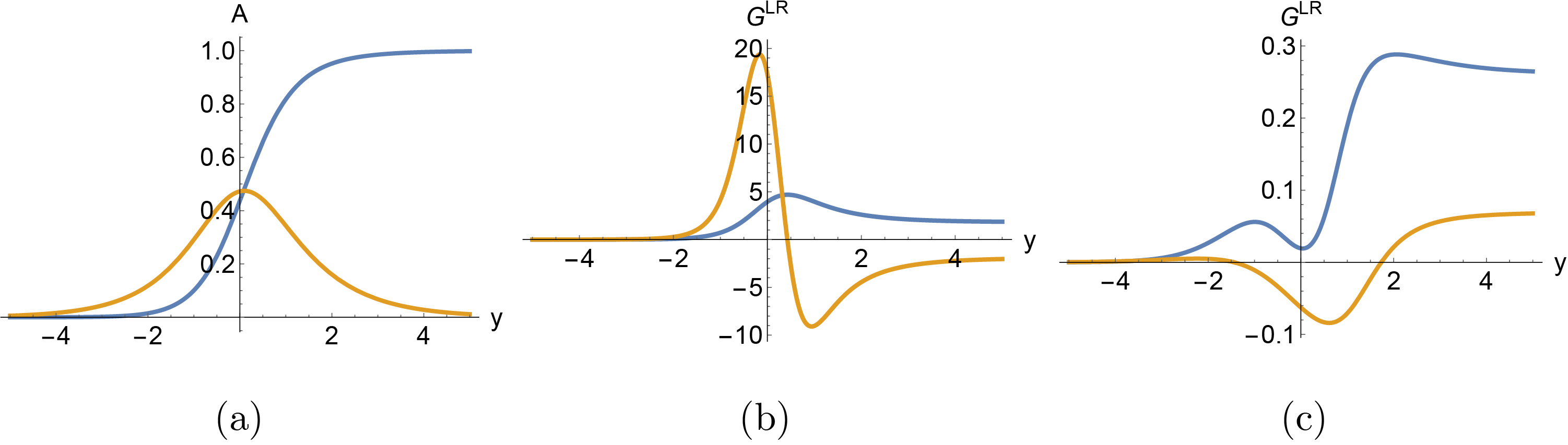}
\par\end{centering}
\caption{(a) Plots of real (blue) and imaginary (yellow) part of $A_0$ for $\De_\sO=8/9$. The horizontal axis is $y= {2 \pi \ov \b} (\ft-|x|)$. 
(b) Plots $G^{LR}$ for~\eqref{glr0} for two values of $\ge$: blue for $\ge =2$ and yellow for $\ge =5$. 
(c) Plots of $G^{LR}$ for $k=1$ in \eqref{gen-k} with $\e=\b/2$ in~\eqref{mui1}. Blue is for $\ge=2$ and yellow is for $\ge=5$.
 \label{fig:AG1}}
\end{figure}

At leading order in $1/c$, $A$ has simple dependence on $Q$ which in turn is given by a simple exponential. So the behavior of~\eqref{glr0} is straightforward to obtain. One immediate thing to notice is that $G^{LR}$ is a function of $t - |x|$ only. From~\eqref{hhe}, points with the same $t - |x|$ then get multiplied by the same factor in going from source to signal. 
The behavior of $A$ and $G^{LR}$ can be separated into three distinct 
regimes: 

\ben 

\item Sub-scrambling regime: for $t - |x| \ll t_* - t_J  $ (i.e. $\ft - |x|$ large and negative) we have $Q \ll1 $ and\footnote{Note that the first term is not pure imaginary when including $h_J/c$ corrections.}
\be \label{a1}
A_0 = h_\sO \le(2 i  +  {24 \pi  h_J \ov c}  \le(1 + {i \ov \pi} \ri) \ri) e^{{2 \pi \ov \b} (\ft - |x)}  + h_\sO (1 + 2 h_\sO)  e^{{4 \pi \ov \b} (\ft - |x)} +\cdots  
\ee
and
\be 
G^{LR} = 2 G_J \ge h_\sO \le( {24 \pi  h_J \ov c}  e^{{2 \pi \ov \b} (\ft - |x|)} + (1 + 2 h_\sO)  e^{{4 \pi \ov \b} (\ft - |x|)}\ri) +\cdots  \ .
\ee
Note that~\eqref{a1} is exponentially increasing with time with Lyapunov exponent ${2 \pi \ov \b}$ and butterfly velocity $v_B =1$~\cite{Roberts:2014ifa}, but the leading behavior is pure imaginary and does not contribute to $G^{LR}$. In this regime, the signal one sent in at time $-t$ just started getting scrambled before we set up the communication channel $V$ at time $0$. 
The signal is very weak in this regime and can be considered as being approximately zero for practical purpose.  
$G^{LR}$ for a finite $k$ has very similar behavior. 

\item Transition regime:  for 
a narrow window of size ${\beta \ov 2 \pi}$ around $t - |x| = t_* - t_J $, both real and imaginary parts of $A_0$ are $O(1)$, and $G^{LR}$ also becomes $O(1)$. 
In this window, the exponential factor $e^{\ge {\rm Im} A_0}$ in~\eqref{glr0} can enhance the magnitude of $G^{LR}$ significantly when $\ge$ has the right sign and not too small. Note as can be explicitly checked from the expression of $A_0$ (see Fig.~\ref{fig:AG1}(a) for an example), for $\De_\sO \sim O(1)$, ${\rm Im} A_0$ is always positive and smaller than $1$.\footnote{From \eqref{mui}, $\Im A_0$ is proportional to $\sin (2h_\mO \arctan Q)$. Since the value of $Q$ ranges between 0 and $\infty$, 
$\Im A_0$ is always positive for $h_\mO<1$. For any relevant $V$, $h_\mO$  is within this range.} Thus enhancement requires $\ge$ to be positive. See Fig.~\ref{fig:AG1}(b) for some examples.

In contrast, at a finite $k$ including the case for~\eqref{pp1},  from the imaginary part of~\eqref{gen-k} we find there is no 
enhancement in the transition region for generic values of regulation parameter $\ep\sim O(\b)$. See Fig.~\ref{fig:AG1}(c) for some examples. 

One may understand the exponential enhancement in the large $k$ limit as coming from constructive interference of different channels.  

\item Stable regime: as we further increase $t$ beyond the transition regime, i.e. for $t  - |x| \gg t_*-t_J$ ($\ft - |x|$ large and positive), $Q$ quickly grows to be $Q \gg 1$, for which $A_0$ approaches to $1$ exponentially (quasi-normal behavior)
\be \label{a2}
A_0= 1 - e^{- i h_\sO \pi} e^{- {4 \pi h_\sO \ov \b} (\ft -|x|)} + \cdots 
\ee
and for $G^{LR}$ we have 
\be 
G^{LR} (t, x) = G^{LR} (t=\infty) 
+ O (e^{- {4 \pi h_\sO \ov \b} (\ft -|x|)}) 
  \ .
\ee

\een 
To conclude this subsection we should emphasize that the regimes described above are not evolutions; they correspond 
to different types of behavior when we vary 
 the time separation between the time of turning on the source and the time we turn on interaction $V$ 
between $L$ and $R$ systems.

\subsection{Multiple channel from integration}

\begin{figure}
\begin{centering}
\includegraphics[width=15cm]{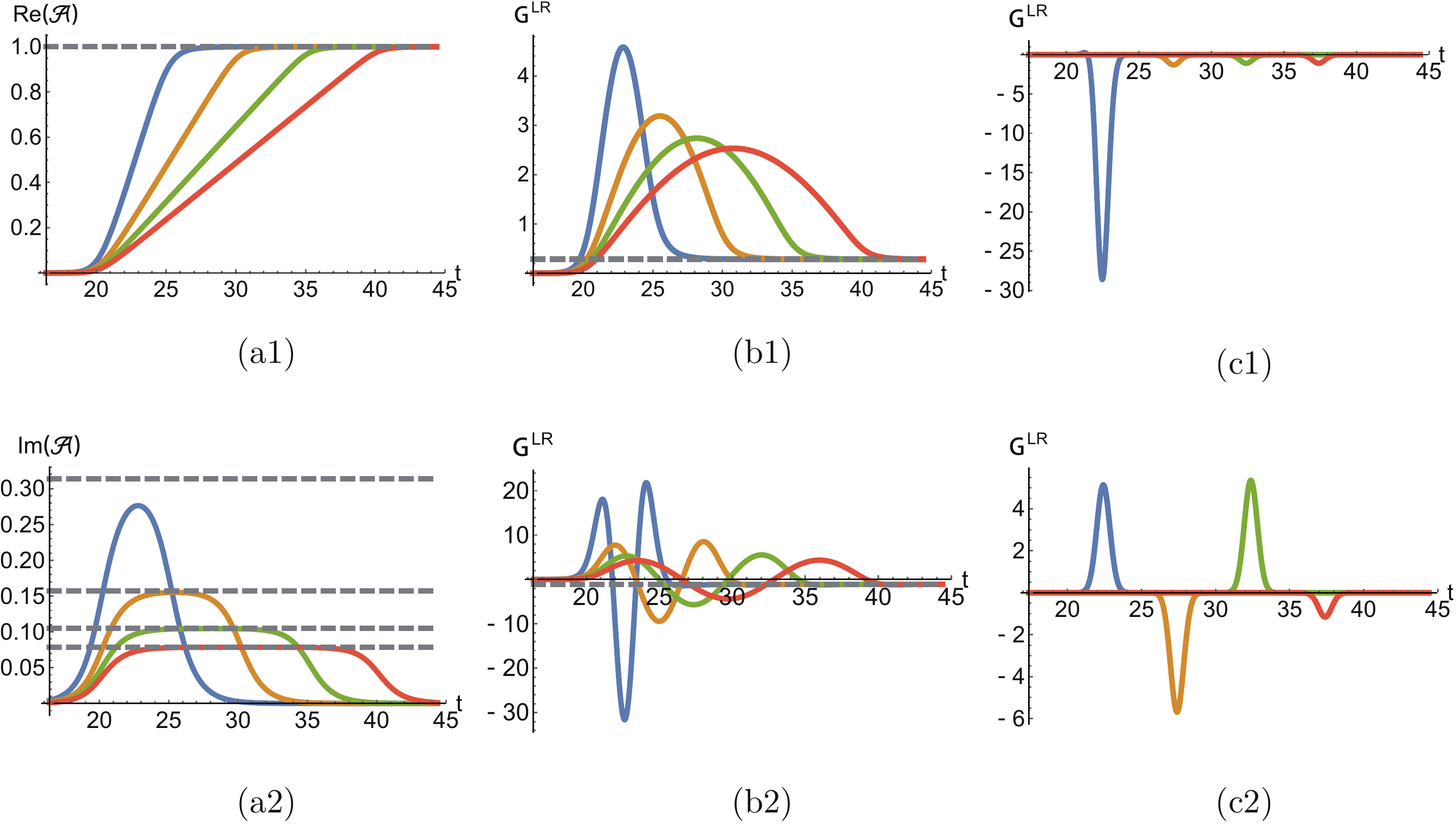}
\par\end{centering}
\caption{Plots of $\mA$ and $G^{LR}$ in various cases. We take $h_{\mO}=4/9$, $\b=2\pi$, $\D_{ J}=2h_ J=10$, $x=x_s=0$, $h_{ J}/c=10^{-10}$. Plots (a1) and (a2) are respectively real and imaginary parts of $\mA(t)$ with $\ge=3$. Blue, yellow, green, red curves are for $L/2=5,~10,~15,~20$ respectively, and the gray dashed lines are $\b/(2L)=\pi/L$. Plots (b1) and (b2) are respectively $G^{LR}(t,t_s=t)$ with $\ge=3$ and $\ge=10$.  Blue, yellow, green, red curves are for $L/2=5,~10,~15,~20$ and gray lines are asymptotic values $2\sin(\ge)$ for large $t$. In (c1) and (c2) we plot $G^{LR} (t, t_s)$ as a function of $t$ for different $t_s$, with 
$\ge =10$. For (c1) $L/2=5$ and for (c2) $L/2=15$. Blue, yellow, green and red curves are for $t_s=t_*+3,~t_*+8,~t_*+13,~t_*+18$ respectively. 
 \label{fig:fig2}}
\end{figure}

Let us now examine the behavior of~\eqref{glr1}--\eqref{glr2}.
We will consider $t_* \gg L \gg \b$ as for $L \lesssim \b$ the story is essentially the same as that of single-channel. 
There are some new elements in~\eqref{glr1} compared with~\eqref{glr0}.
Firstly due to the integration over $x_1$, $G^{LR}$ is no longer a function of $t - |x|$ only. 
Secondly,  as we will see the transition regime can be significantly lengthened.

For illustration let us consider $x=0$ 
for which we have to leading order in $1/c$
\be\label{nip}
\sA = {2 \ov L} \int _ {0} ^{{L \ov 2}}  d x_1 \, \le[1 -  \le({1 \ov 1 + i Q (x_1)}  \ri)^{2 h_\sO}   \ri]  , \quad Q (x_1)= e^{{2 \pi \ov \b} (\ft- x_1)}\ .
\ee 
The sub-scrambling regime is for  $\ft \ll 0$ such that $Q (x_1)$ is exponentially small for the whole integration range. 
The stable regime is for $\ft \gg {L \ov 2}$, for which $Q (x_1)$ is exponentially large for the whole integration range. 
The behavior of $\sA$ for these regimes  is completely parallel to the corresponding regimes of $A_0$ discussed in last subsection (with only differences in some constant prefactors), and thus the behavior of $G^{LR}$ is also parallel to those of~\eqref{glr0}. 
 Things  are more interesting for $\ft$ in the window $\ft \in (0, {L \ov 2})$ (i.e. $t \in (t_* - t_J , t_* - t_J + {L \ov 2})$)
 for which as $x_1$ changes from $0$ to $L$, $Q (x_1)$ varies from exponentially large to exponentially small.\footnote{We will be  concerned about  $\ft$'s in the middle of the window $(0, {L \ov 2})$, i.e. not close to either edges. } 
To find $\sA$ for such values of $\ft$ we note that~\eqref{nip} can in fact be exactly integrated, yielding 
\be\label{eq:AintL}
\sA (\ft)= 1-\f{\b e^{-2\pi i h_\mO}}{\pi L} \left[ B\le(\f{i} {Q( {L \ov 2})},2h_\mO,1-2h_\mO\ri)- B\le(\f{i} {Q(0)},2h_\mO,1-2h_\mO\ri) \right] 
\ee
where $B(x,a,b)$ is the incomplete beta function. Using that $Q(0)$ is exponentially large and $Q ({L \ov 2})$ exponentially small 
we find that 
\be \label{Acasei}
\mA\ra \f{2}{L}\le(\ft+c_0 {\b \ov 2 \pi}
\ri)+i\f{\b}{2L}+ \cdots 
\ee
where $c_0$ is a numerical constant. This behavior is extremely simple with linear dependence on $\ft$ and a constant imaginary part, leading to 
\be \label{kgj}
G^{LR} (t, x=0) = 2 G_J e^{\ge \b \ov 2L} \sin \le({ 2 \ge\ov L} \le( \ft + c_0 {\b \ov 2 \pi}\ri)\ri)  , \quad \ft \in (0, {L / 2}) \ .
\ee
Thus we see that for~\eqref{glr1}, the size of transition regime is extended to a region of size ${L \ov 2}$ (in contrast to ${\b \ov 2 \pi}$ for~\eqref{glr0}), but the imaginary part of $\sA$ is down by an order ${\b \ov L}$ compared with $A_0$. Thus while the resonant enhancement is extended to a much larger range of time period, the enhancement effect is more moderated. 
See Fig.~\ref{fig:fig2} for various numerical plots of $\sA$ and the corresponding $G^{LR}$. 

\begin{figure}
\begin{centering}
\includegraphics[width=8.5cm]{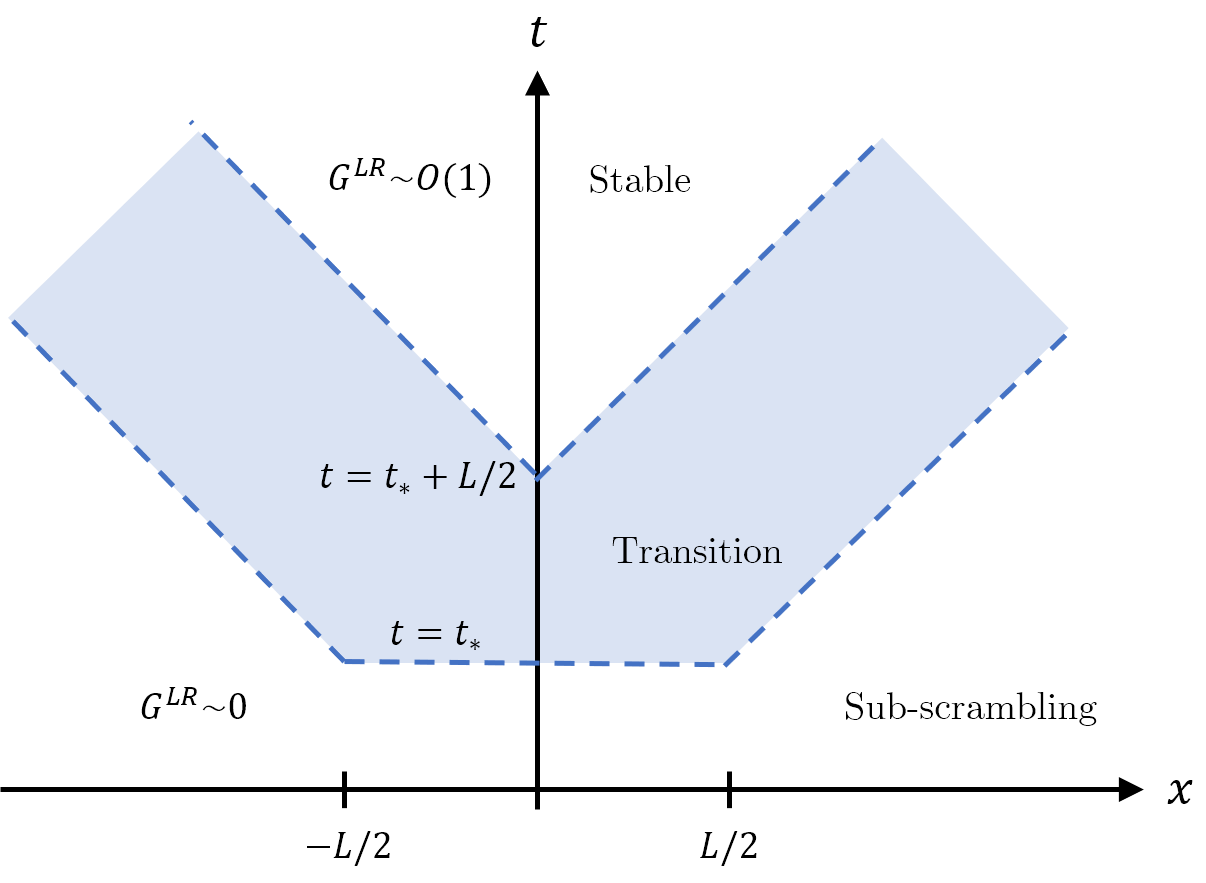}
\par\end{centering}
\caption{The sub-scrambling, transition, and stable regions for different values of $x$ for~\eqref{wn1}. 
\label{fig13}}
\end{figure}

The behavior for general $x$ is qualitatively similar. The only difference is that the transition regime is now from $t_*+\theta(|x|-L/2)(|x|-L/2)$ to $t_*+L/2+|x|$ , which has maximal length of $L$ when $|x|\geq L/2$. Here $\t(x)$ is step function. See Fig. \ref{fig13}.
Finally let us note that when $L \to \infty$, the behavior~\eqref{Acasei}--\eqref{kgj} will last forever and one never reaches the stable regime.

\subsection{Robustness of \rena\  from CFT calculations} \label{sec:rob1}

We now turn to the explicit calculation of~\eqref{gga} for CFTs in the large $c$ limit.  For simplicity we will take $t_s = t$ and consider the regime $c \gg \De_\ga \gg \De_J  \gg \De_\sO \sim O(1)$, for which we will be able to confirm explicitly the conclusion
of Sec.~\ref{sec:rob0}. 
We will present only the results here leaving details to Appendix~\ref{app:rob}. 

For simplicity we will consider only~\eqref{pp2} in the large $k$ limit and~\eqref{pp3} when $L \gg \b$. For these cases we find 
\be 
\f{W_\g (t;t_0)}{\avg{J  \tilde J}}= \sJ (t,t_0) 
\times
\begin{cases}
\exp\left(-i \ge  \sG_0 (t,t_0) \tilde A_0 (t, t_0)  \right)& \text{large \;} k \\
\exp\left(-\f{i\ge}{L} \int_{-L/2}^{L/2} dx_1\, \sG (t,t_0; x_1) \tilde A (t,t_0; x_1) \right)& \text{large } L \\
\end{cases}\label{Wg3cases-m}
\ee
where for notational simplicity we have suppressed $x, x_0$ in the arguments of various functions. 
Function $\tilde A_0 (t,t_0)$ is obtained from $ \tilde A (t,t_0; x_1)$ by setting $x_1 =0$ and 
$\sG_0$ is obtained from $\sG$ in the same way. 

By comparing~\eqref{Wg3cases-m}
with~\eqref{wn} and~\eqref{wn1}, we see that the following three differences between the corresponding expressions for~\eqref{sp1} and $\Psi_\b$ which reflect three distinct aspects how an insertion of $\ga$ in $\Psi_\b$ affects the \rena\ phenomenon : 

\ben 

\item The prefact factor $\avg{J  \tilde J}$  is multiplied by another function $\sJ$, which modifies correlation between $J^L (t, x)$ and $J^R (-t, x)$. 

\item  the effective coupling $\ge$ is multiplied by a function $\sG$, which modifies correlation between $\sO^L(0)$ and $\sO^R (0)$ and thus the effective coupling between $L$ and $R$ systems. 

\item the function $A$ is replaced by another function $\tilde A$ which reflects how ``interactions'' between $\sO$ and $J$ operators are modified 
due to presence of $\ga$. 

\een
Note that item (1) and (2) can be interpreted as coming from modification of the entanglement structure 
of $\Psi_\b$. 

Now let us look at the explicit expressions of $\sJ, \sG$ and $\tilde A$. Note that if the spatial location $x_0$ is sufficiently far away, e.g. if $|x_0 - x| \gg t-t_0$, clearly from causality $\ga$ cannot have any effect on $J$. Similar statement applies to $\sO$. 
Now from Appendix~\ref{app:rob} we find that (assuming $x-x_0$ is much smaller than $t_*$)
\be\label{ij}
\sJ (t, x; t_0, x_0) = \left(1+\f{2 h_\g }{
\ep_\ga}Q_J \right)^{-2h_J}, \qquad Q_J=e^{\f{2\pi}{\b}(|t-t_0|-t_*-|x_0-x|)}
\ee
where $\ep_\ga$ is a UV regulator need to make~\eqref{sp1} normalizable. Note that $\sJ \leq 1$, and 
$\sJ \to 0$ when $|t-t_0| \gg t_*$ as anticipated in~\eqref{der}. 
From Appendix~\ref{app:rob}, $\sG$ has the form 
\be\label{ij1}
\sG = \left(1+\f{2 h_\g }{
\e_\ga}Q_1 \right)^{-2h_\mO}, \qquad Q_1=e^{\f{2\pi}{\b}(|t_0|-t_*-|x_0-x_1|)} \ .
\ee
Again $\sG \leq 1$ and we see that $\sG \to 0$, and thus the effective coupling is destroyed, when $|t_0| \gg t_*$, which confirms the expectation~\eqref{der1}. 

The behavior of $\tilde A$ is much more difficult to work out explicitly. This is also easy to understand physically: to see how the presence of $\ga$ modifies ``interactions'' between $\sO$ and $J$ is warranted to be complicated in a strongly interacting system. 
Fortunately the conclusion does not depend on the detailed form $\tilde A$. We expect $\tilde A \approx A $ if $|t_0| \ll t_*$ and 
$|t-t_0| \ll t_*$, i.e. the effect of $\ga$ on $J-\sO$ correlation functions will be small if $\ga$ excitation does not have enough time to grow. Outside this region, the form of $\tilde A$ is expected to be complicated, but we do not really care as 
from~\eqref{der}--\eqref{der1} and~\eqref{ij}--\eqref{ij1},  outside this region, the correlations between two systems already become too weak to have the \rena\ phenomenon. 

\subsection{Effects of other primaries} \label{sec:eff-non-id}

In this subsection, we briefly discuss implications of contributions from other primaries, equation~\eqref{eq:w-ni-9},
in the limit of large $c$ with  $Q=1/(cu_0)\sim - e^{2\pi (t-|x|)/\b }/(4c)$ fixed. Compared with the identity contribution, the main difference
lies in the prefactor $B_0$, which we focus on.  Note that in the aforementioned limit various expressions in $B_0$ behaves as 
\begin{align}
\f{1-(1-u)^{\a}}{\a}\ra\f 1c Q^{-1},&\quad \f{1-e^{-2\pi i\a}(1-u)^{\a}}{\a}\ra\f 1c(-24\pi ih_{J}\pi+Q^{-1})\\
_{2}F_{1}(h,h,2h,1-(1-u)^{\a}),&\; {}_{2}F_{1}(h,h,2h,1-e^{-2\pi i\a}(1-u)^{\a})\ra1 \\
_{2}F_{1}(h,h,1,e^{-2\pi i\a}(1-u)^{\a})&\ra \f {\G(2h-1)}{\G(h)^2 c^{1-2h}}(-24\pi ih_{J}\pi+Q^{-1})^{1-2h}
\end{align}
Therefore, in this limit $\mV_{h,2}(u_0)$ and $\mV_{h,1}(u_0)$  scale as
\begin{equation}
\mV_{h,2}(u_0)\ra \f{2\pi i \G(2h)\G(2h-1)}{\G(h)^4} (cQ)^{h-1}(1-24\pi h_J Q)^{1-h-2h_\mO}, \quad \mV_{h,1}(u_0)\ra (cQ)^{-h} \ 
\end{equation}
which leads to 
\be \label{fing}
B_0 \ra -\f{2\pi i \G(2h)\G(2h-1)}{\G(h)^4 c^{1+\bar{h}-h}} Q^{h-1}\bar{Q}^{-\bar{h}}(1-24\pi i h_J Q)^{1-h-2h_\mO}
\ee
where $\bar{Q}=1/(c\bar{u}_0)\sim - e^{2\pi (t+|x|)/\b}/(4c)$.


Let us discuss the case for different spins $s = h- \bar h$. If $h$ and $\bar h $ are non-integer, as $Q$ and $\bar Q$ are negative numbers, $B_0$ is complex even in the beginning. This will give $W$ a nonzero value from beginning. If they are integers, we see only the real part of $B_0$ will contribute to $W$. The relevant part in $B_0$ is  \cite{Roberts:2014ifa, cornalba2007eikonal1, cornalba2007eikonal2, cornalba2007eikonal}
\be\label{eq:cqq}
c^{h-\bar h -1} Q^{h-1}\bar{Q}^{-\bar h}\sim c^{s-1} e^{2\pi (s-1) t/\b} e^{-2\pi (\D-1)|x|/\b}
\ee
where $\D=h+\bar h$. For a CFT flowed from some large $N$ theory, all connected 4-pt correlation functions should scale as $1/c$ for generic spacetime coordinates in large $c$ limit. This implies that the OPE coefficient should scale as
\be
C_{JJh}C_{h\mO\mO}C_{JJ\bar h}C_{\bar h\mO\mO}\sim 1/c
\ee 
In \eqref{eq:cqq}, for $s=0$, $B_0 \sim {1 \ov c}$ and the contribution to $W$ scales as $1/c^2$, whose effects can be neglected as far as the density of states does scale with $c^2$. For $s=1$,  $B_0$ scales as $O(1)$ and contributes to $W$ as $1/c$, which is also negligible. For higher spins $s\geq 2$, their contributions to $W$ are $O(c^{s-2})$ that are never smaller than the identity block. Such higher spin contributions to OTOCs violate the chaos bound individually, but 
are expected to lead to slower exponential growth when resummed~\cite{Maldacena:2015waa,shenker2015stringy}.\footnote{
Note that a theory with finite number of higher spin operators will have faster scrambling and non-trivial contribution to $B_0$ and $W$. However, these theories must be excluded as they are shown to violate causality \cite{camanho2016causality} and unitarity \cite{perlmutter2016bounding}.}
For a CFT with a string theory dual, there exists an additional parameter which is dual to $\apr/R^2$ with $\apr$ the string scale. 
In the limit of $\apr \to 0$ (in addition to the large $c$ limit), all higher spin contributions are suppressed and 
the vacuum Virasoro block gives the leading contribution (the density of states for operators with dimensions which are $c$-independent and $s \leq 1$ are always $O(c^0)$). 
 Lastly, at very late time, $t\gg \log c$, $Q,\;\bar{Q}\ra \infty$, $B_0\ra 0$ exponentially. This means that the late time interference behavior of identity block is not changed by any non-identity channels.

\section{Gravity interpretation} \label{sec:grav} 

In this section we compare the results from the vacuum Virasoro block with the gravity discussion of~\cite{Gao:2016bin,Maldacena:2017axo} and discuss implications of our results  for wormhole physics in the context of holography. Other recent papers on traversable wormholes from gravity perspective include~\cite{Maldacena:2018lmt,Bak:2018txn,Maldacena:2018gjk,Caceres:2018ehr,Fu:2018oaq}.
Note that our two-dimensional CFT calculation in the large $c$ limit may be considered as describing  a BTZ black hole. 

\begin{figure}
\begin{centering}
\includegraphics[width=8cm]{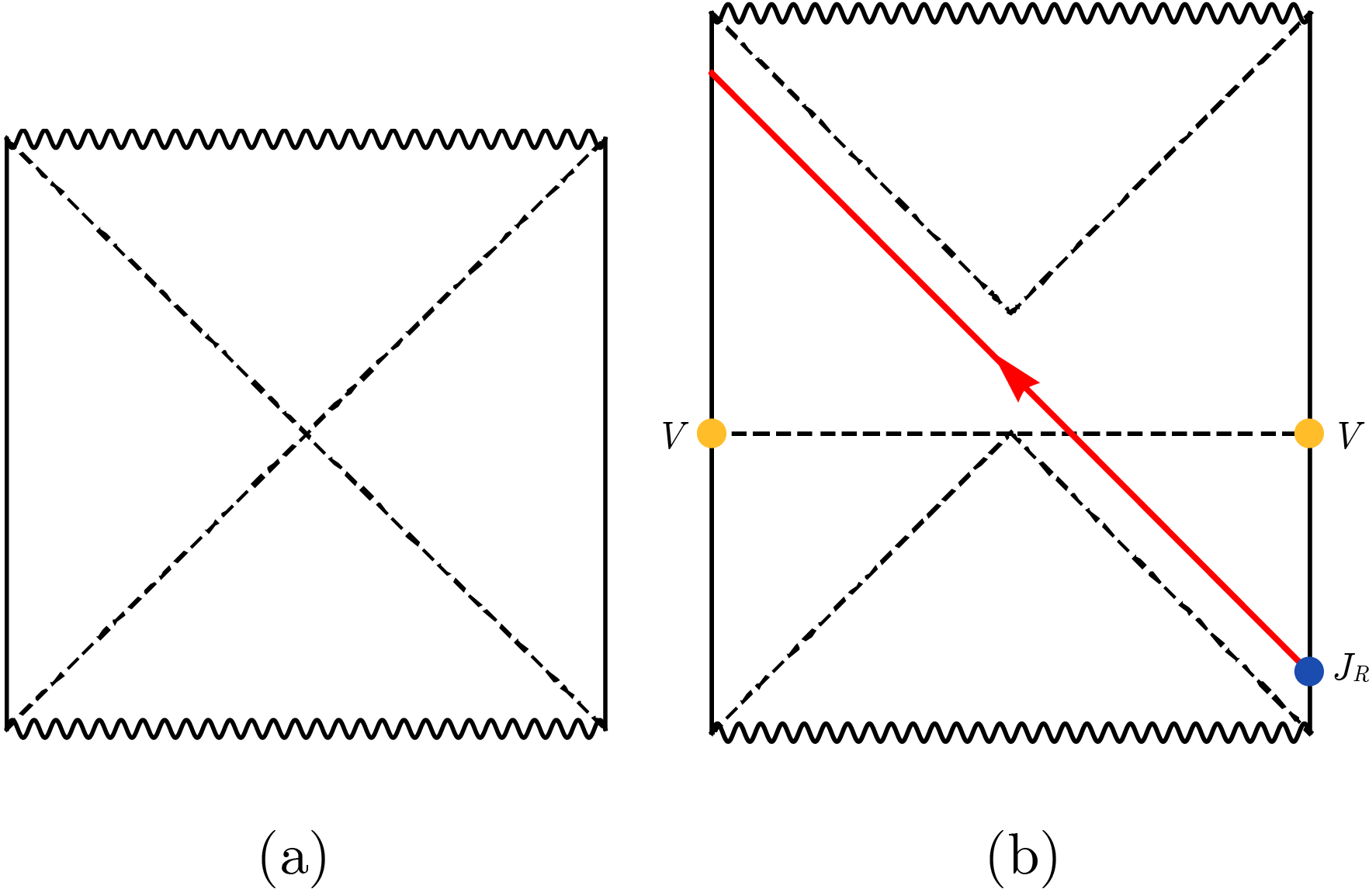}
\par\end{centering}
\caption{(a): Penrose diagram of an eternal black hole. The two boundaries are causally disconnected. 
(b): in the picture of~\cite{Gao:2016bin,Maldacena:2017axo}, presence of $V$ deforms the bulk geometry, especially the causal structure, allowing signals to pass from right to left, now following a timelike geodesic. 
 \label{fig:worm}}
\end{figure}

 In the gravity description, thermal field double~\eqref{tfd0} 
is described by an eternal black hole which has two asymptotic boundaries connected by a non-traversable wormhole (see Fig.~\ref{fig:worm}(a)).
The \rena\ phenomenon corresponds to the statement that with a coupling like~\eqref{pp1}, one could send signals between two boundaries, i.e. the wormhole becomes traversable.

Turning on the source $\vp^R$ for a short time on the right boundary generates  bulk excitations dual to $J$. These excitations fall toward and are absorbed by the black hole membrane, which corresponds to the dissipation of $\vev{J^R}$. 
The process happens very fast, with time of order $O(\b)$, as also seen in our earlier CFT calculation. 
When we briefly couple the two boundaries at $t=0$, the physical picture of~\cite{Gao:2016bin,Maldacena:2017axo} is that, the interaction $V$ deforms the bulk geometry, especially the causal structure, making the wormhole traversable, as indicated in Fig.~\ref{fig:worm}(b). 
 It is important in the discussion of~\cite{Gao:2016bin} that only one sign of the coupling $g$, i.e. $g > 0$ which generates negative bulk energy, allows for the traversability. 

Below we will first compare our CFT results with that of~\cite{Maldacena:2017axo} obtained from gravity scatterings. 
We then discuss implications of the regenesis phenomenon on gravity side. In particular, we will argue that there are other scenarios for wormhole traversability in addition to that suggested by Fig.~\ref{fig:worm}(b). For example, the regime for $t_s \gg t_*$ should correspond to a ``quantum traversable wormhole.''

\subsection{Explicit comparison with gravity results} 

\begin{figure}
\begin{centering}
\includegraphics[width=10cm]{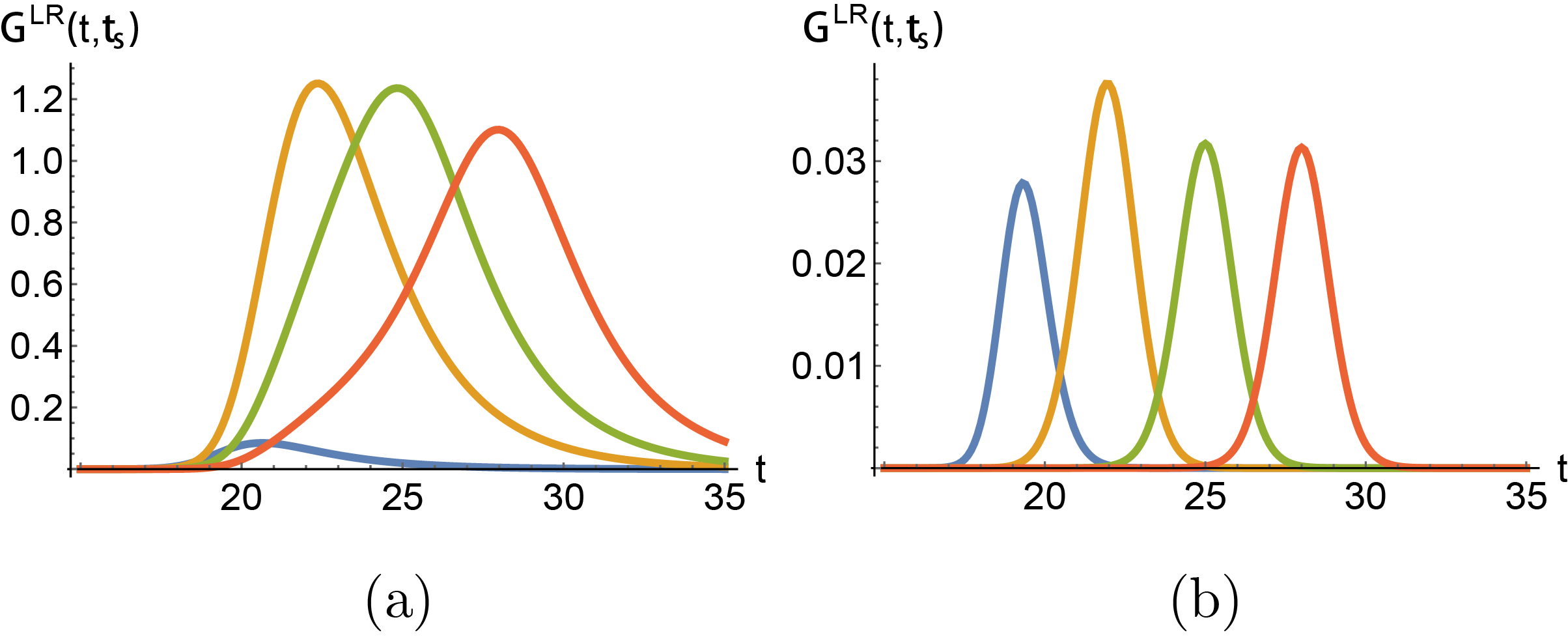}
\par\end{centering}
\caption{Plot of $G^{LR}(t,t_s)$ as a function of $t$ for fixed $t_s$. In (a) we choose $\D_J=\D_\mO=4/9$, in (b) we enlarge $\D_J$ as $\D_J=3$. Blue, yellow, green and red curves are for $t_s=t_*+0.5,~t_*+3.5,~t_*+6.5,~t_*+9.5$.
 \label{fig:fig14}}
\end{figure}

Here we compare our explicit expression for $W$ calculated from  two-dimensional CFTs with that of~\cite{Maldacena:2017axo}
calculated for a $(0+1)$-dimensional boundary theory from gravity. 
By using graviton scattering amplitudes between $J\tilde{J}$ and $\mO\tilde{\mO}$ near horizon, for $V$  given by~\eqref{pp2}  they derived the following expression for $W$ (below $\b$ has been set to $2 \pi$, and we have rescaled $p$ and used notations 
introduced in Sec.~\ref{sec:genr})  
\be\label{med} 
W(t, -t_s)=e^{-i \ge } \vev{J \tilde J}_\b   \f{(-i)^{2\D_J}}{\G(2\D_J)}\int_{0}^\infty dp \, p^{2\D_J-1}e^{ip}\exp\left[\f{i \ge}{\left(1+\f{p \hat Q}{2\De_J}\right)^{2\D_\mO}}\right] , 
\ee
where $p$ is a bulk momentum of the $J$-quantum, and 
\be 
\vev{J \tilde J}_\b = \le({1 \ov 2 \cosh {t-t_s \ov 2}} \ri)^{2 \De_J} , \qquad \hat Q =  {\De_J G_N\ov 8} {e^{t+t_s \ov 2} \ov \cosh {t-t_s \ov 2}}  \ . 
\ee

To compare~\eqref{med} with our results, it is convenient to deform the integral contour to be along the imaginary $p$ axis from $0$ to $i\infty$ (note the integral along the arc from $+i \infty$ to $+\infty$ vanishes). We then find (after a scaling $p \to {i} p$) for $t_s = t$
\be\label{med1} 
W(t)= G_J \f{e^{-i \ge } }{\G(2\D_J)}\int_{0}^\infty dp\, p^{2\D_J-1}e^{-p}\exp\left[\f{i \ge}{\left(1+ {i p Q \ov 2 \De_J} \right)^{2\D_\mO}} \right] \ 
\ee
where 
\be 
G_J = 2^{- 2 \De_J}, \qquad Q=  \hat Q |_{t=t_s}  = {\De_J G_N\ov 8} e^{t}    = e^{t + t_J - t_*} , \quad t_J \equiv \log \De_J , \quad t_* \equiv \log {G_N \ov 8} \ .
\ee

Let us now consider $\De_J$ large.  
Assuming the exponential factor involving $\ge$ in~\eqref{med1} is slowly varying in $p$ we can approximate the factor $p^{2\D_J-1}e^{-p}$
in the integrand of~\eqref{med1}, which is a Poisson distribution by Gaussian distribution with center value
\be\label{p-sad}
p_{J}={2\D_J-1}\app 2 \D_J  
\ee
and variance $\sqrt{2\De_J}$.
Evaluated at~\eqref{p-sad}, equation~\eqref{med1} becomes  
\be \label{ghw}
W (t) = G_J \exp \le[-i \ge \le(1 - \le({1 \ov 1 + i Q} \ri)^{2 \De_\sO} \ri) \ri],
\ee
which has exactly the same form as \eqref{wn} with $A_0$ given by~\eqref{mui} (with $x=x_1=0$).\footnote{$h_\sO$ and $h_J$ 
in our (1+1)-dimensional expressions are replaced respectively here in (0+1)-dimension by $\De_\sO$ and $\De_J$.}  
To make sure our Gaussian approximation is valid, we need the exponential term involving $\ge$ in~\eqref{med}
to be slowly varying within the variance of the Gaussian distribution, i.e. 
\be  
 {\ge \De_\sO \ov \sqrt{\De_J}} Q \, |1 + i  Q |^{- 2 \De_\sO-1} \ll 1 \ .
\ee
The above equation is satisfied for all values of $Q$ if 
\be \label{unn1}
\ge \De_\sO \ll \sqrt{\De_J} \ .
\ee 
Our CFT calculation was performed for $\De_J \gg \De_\sO \sim O(1)$ with $g$ independent of $\De_J$, so is consistent.

For $\De_J \sim \De_\sO\sim O(1)$, while we do not have explicit counterpart from CFT calculation, equation~\eqref{med} is consistent with various general features we discussed earlier. For example, in the limit $t,t_s \to \infty$, it reduces to~\eqref{oio}. Another important aspect of our discussion is the reversed time ordering between the input signal $\vp^R$ and output signal $\vev{J^L}$ as indicated for example in~\eqref{hhe}. It can be checked that~\eqref{med} also has this property, as can be seen explicitly from the plots of the resulting $G^{LR}$ in Fig. \ref{fig:fig14}.

\subsection{A semi-classical regime} \label{sec:semi} 

We now elaborate on  a ``semi-classical'' regime of~\eqref{med} which was identified in~\cite{Maldacena:2017axo}.\footnote{See Appendix B.1 there. We than Douglas Stanford and Zhenbin Yang for clarifications.} Consider smearing $J$-operator so that its high energy component is suppressed. One can represent such a smearing by inserting a Gaussian factor $e^{- {p^2 \ov 2 \sig^2}}$ in the momentum integral of~\eqref{med}. Then one finds that, 
in the large $\ge$ limit (with $\sig$ and $\De_J, \De_\sO \sim O(1)$ fixed), there exists a regime corresponding to the picture of 
Fig.~\ref{fig:worm}(b).

 More explicitly, consider expanding
in $\hat Q$  in the exponent of~\eqref{med}
\be\label{ine}
\le(1+ {p \hat Q \ov 2 \De_J }\ri)^{- 2 \De_\sO} = 1 -  {\De_\sO \ov \De_J} p \hat Q + {\De_\sO (2 \De_\sO +1) \ov 4 \De_J^2} p^2 \hat Q^2 + \cdots 
\ee 
 and keeping only the linear term in $\hat Q$.  Equation~\eqref{med} can then be approximated by  
\be \label{oju}
W_\sig (t, -t_s) \approx \vev{J \tilde J}  \f{(-i)^{2\D_J}}{\G(2\D_J)}\int_{0}^\infty dp \, p^{2\D_J-1} e^{ i p X - {p^2 \ov 2 \sig^2}}, \quad 
X \equiv 1 - {\ge \De_\sO \ov \De_J}  \hat Q  \ .
\ee
Note that the prefactor $e^{-i \ge}$ has now been canceled. Without the Gaussian factor in~\eqref{oju}, the integral in~\eqref{oju} would yield
\be 
\vev{J \tilde J} {1 \ov (X + i \ep)^{2 \De_J}} = G_J {1 \ov ( \cosh {t-t_s \ov 2} - \ge \De_\sO  e^{{t+t_s \ov 2} - t_*}   )^{2 \De_J}}
\ee
which has a ``light-cone'' singularity at 
\be \label{uyo}
\cosh {t-t_s \ov 2} = \ge \De_\sO  e^{{t+t_s \ov 2} - t_*}  \quad \to \quad
t =  t_{l} \equiv t_* - \log \le(2 \De_\sO \ge - e^{t_* - t_s} \ri)
  \ .
\ee
Thus we can view~\eqref{oju} as the propagator of 
a smeared field in a spacetime where points on two boundaries satisfying~\eqref{uyo} are connected by light rays, as 
indicated in Fig~\ref{fig:worm} (b). 

Let us make some further remarks: 

\ben 

\item For the approximation in~\eqref{oju} to be valid, we need for the range of $p$ allowed by the Gaussian factor, the terms of $O(Q^2)$ and higher in~\eqref{ine} are suppressed, i.e. 
\be \label{oii}
{\hat Q \sig  \ov \De_J} \ll 1, \quad {\ge \De_\sO \ov \De_J}  \hat Q \sig \gg 1 , \quad  {\ge\De_\sO^2 \ov \De_J^2} (\hat Q \sig)^2 \ll 1  \quad \to \quad
{1 \ov \ge \sig} \ll {\De_\sO \ov \De_J} \hat Q \ll {1 \ov \sqrt{\ge} \sig} 
\ee 
Equation~\eqref{uyo} (which corresponds to ${\ge \De_\sO \ov \De_J} \hat Q =1$)  lies within the range~\eqref{oii} provided that (recall we have $\b = 2 \pi$ which sets the unit)
\be \label{uyo1}
\sqrt{\ge} \gg \sig \gg 1   \ .
\ee

\item Note that equation~\eqref{uyo} has a solution only for $\ge > 0$ and 
\be \label{hu} 
 t_s  \geq   t_* - \log (2 \De_\sO \ge )  \equiv t_c  \ 
 \ee
 and for $t_s$ satisfying~\eqref{hu}, a corresponding $t_l (t_s)$ always exists. Note  ${dt_l \ov d (-t_s)} > 0$ when~\eqref{hu} is satisfied, which means the light-cone structure is such that 
the earlier a signal (smaller $-t_s$) is sent, the earlier (smaller $t_l$) the arrival of the signal at the other boundary.  
In particular, as $- t_s \to - \infty$, $t_l (t_s) \to t_c $
which is the earliest possible arrival time, and as $- t_s \to  - t_c$, $t_l (t_s) \to + \infty$.

\begin{figure}
\begin{centering}
\includegraphics[width=10.5cm]{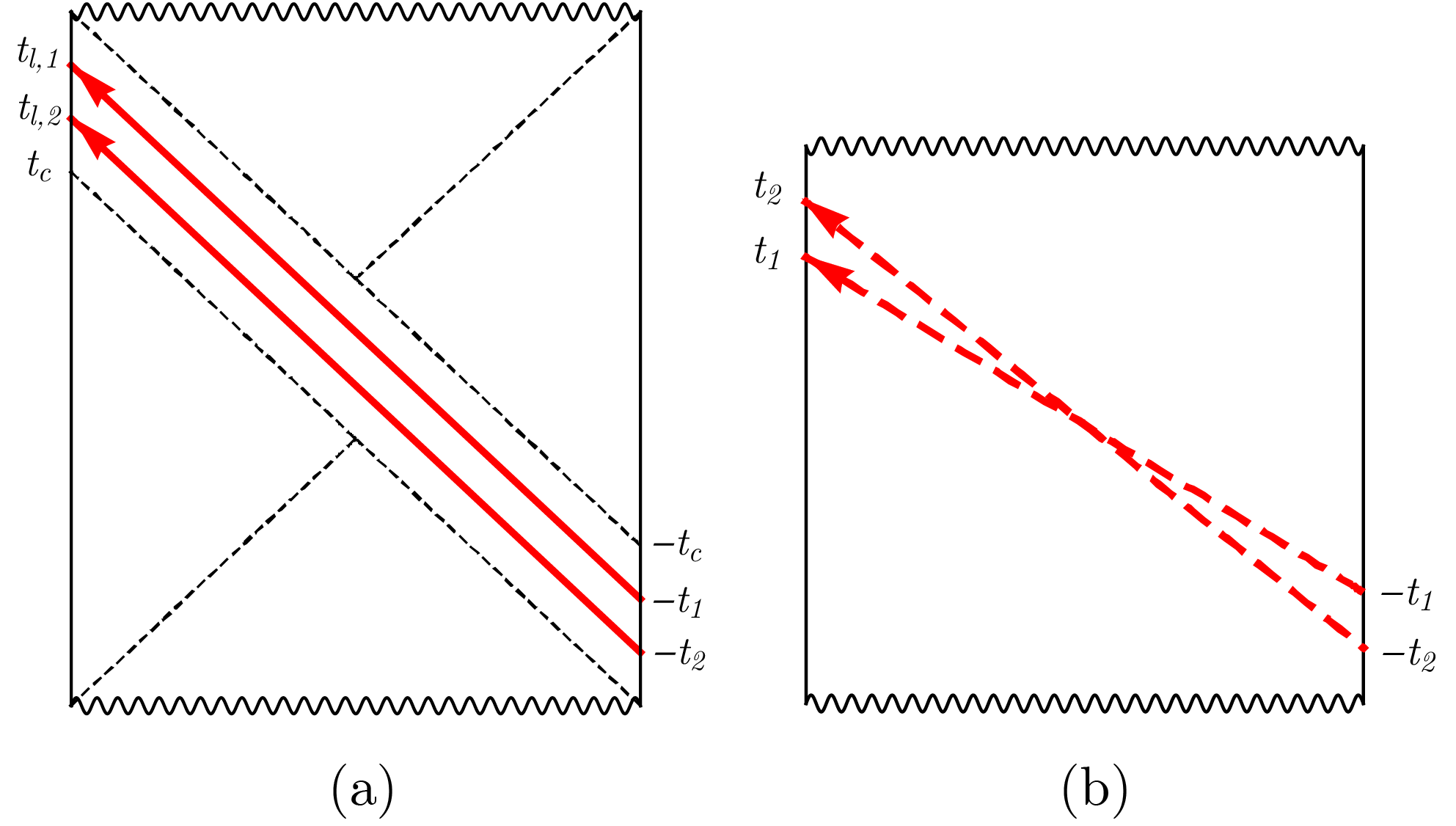}
\par\end{centering}
\caption{The relation between emission time and arrival time in two different regimes. (a) is the semi-classical regime, in which the earlier a signal (smaller $-t_s$) is sent, the earlier (smaller $t_l$) the arrival of the signal at the other boundary.  
In particular, as $- t_s \to - \infty$, $t_l (t_s) \to t_c $
which is the earliest possible arrival time, and as $- t_s \to  - t_c$, $t_l (t_s) \to + \infty$. (b) describes the situation of~\eqref{ghw} and the CFT results of Sec.~\ref{sec:ana}, where the arrival time $t$ is the negative of emission time $-t_s$.
 \label{fig:contr}}
\end{figure}

This is consistent with the heuristic picture of Fig.~\ref{fig:worm}(b), but is sharply different from the behavior exhibited in~\eqref{ghw} or for general $\De_\sO, \De_J, \ge$, for which the signal sent from $-t_s$ arrives at $t \approx t_s$. 
 We contrast these two different types of behavior in Fig.~\ref{fig:contr}.

\item For $\sig \sim \sqrt{\ge}$ or larger,  terms with quadratic and higher powers in $\hat Q$ in the expansion of $(1+ {p \hat Q \ov 2\De_J})^{- 2 \De_\sO}$  become important. Note since $\hat Q \propto G_N$, these terms may be understood as due to backreaction of $J$-quanta on the geometry. Without the momentum suppression factor $e^{- {p^2 \ov 2 \sig^2}}$ (i.e. $\sig \to \infty$), such backreactions are always important.

\een

\subsection{Old cats never die} 

Connecting the boundary and bulk pictures for the regenesis phenomenon in the semi-classical regime discussed in the last subsection raises leads to a rather amusing scenario.  

To make the contrast of the two pictures a bit sharper, let us imagine by turning on $\vp^R$, we create a ``cat,'' which 
contains only ``low energy'' constituents compared with the coupling $\ge$ which we will eventually turn on (i.e. with their bulk momenta satisfying~\eqref{uyo1}). 
From the boundary picture, the cat lives for a while, but eventually her body gets more and more scrambled with the environment. We wait until her body is fully scrambled, turn on the interaction $g V$. From normal standards, it should be safe to say by this time the cat has long died (in other words, her body should have long been decomposed). 
From the genesis phenomenon, after another scrambling time, the cat is reborn in the other universe. 
As emphasized in the Introduction this process requires extreme fine tuning of the initial state at the time we created the cat.

Now let us refer to the bulk dual of the cat as the bulk cat, which we suppose is also a living object.
Then in the bulk picture, the bulk cat travels in the deformed geometry created by the interaction $g V$.
She never ``dies''\footnote{By dying here we refer to something a bit more general, i.e. the body remains as a whole.}, just sailing through the bulk geometry. The journey should be smooth as no regions of large curvature will be encountered.
 In this picture, the reborn cat simply corresponds to the arrival of the bulk cat through a wormhole. In fact if she travels close to the light cone, the proper time that the cat experiences might not be too long. 

From the holographic duality,  these two pictures must be equivalent. In particular, since all boundary events should have 
a bulk version, the bulk cat should be able to ``see'' her own funeral on the boundary. This is the bulk way to say that 
the regenesis cat in fact contains the ``memory'' of her previous life.

\subsection{Quantum traversable wormholes} \label{sec:quant}

In Sec.~\ref{sec:semi} we discussed in detail that the semi-classical scenario of Fig~\ref{fig:worm}(b) corresponds to the parameter region of~\eqref{uyo1} and $k \to \infty$, as well as the ranges of $t, t_s$ satisfying~\eqref{oii}. 
Outside these parameter regions, how do we interpret the traversability of a wormhole? 
Here we offer some qualitative discussions by combining the general discussion of Sec.~\ref{sec:gen},  the explicit CFT results of Sec.~\ref{sec:ana}, as well as discussion of~\eqref{med}. 
One can identify the following  different physical scenarios/regimes: 

\ben 

\item A most straightforward scenario  is that the picture of Fig.~\ref{fig:worm}(b) still qualitatively applies, but with the deformed geometry interpreted as including both effects of $g V$ and the backreaction of $J$ itself. Since the deformed geometry now depends on the quanta propagating through it, strictly speaking, one can no longer talk about a background causal structure as in the case of a linear wave moving in a fixed geometry. Nevertheless, the essential physics picture remains qualitatively similar. So we will still refer to this scenario as the semi-classical scenario. 
For example, consider~\eqref{med} (which is $k \to \infty$ limit) with a large $\ge$ and a momentum suppression factor $e^{-{p^2 \ov 2 \sig^2}}$, and slowly
 increase $\sig$ beyond the range of~\eqref{uyo1}. As we increase $\sig$, the backreactions of $J$ become more and more
 important. One should be able to include them perturbatively, say first including the $\hat Q^2$ term in~\eqref{ine} but neglecting
 higher order terms, and then the $\hat Q^3$ term, and so on.\footnote{In the limit $\sig \to \infty$ (or $\sig \gg \ge$), one finds that for certain range of $t$ around $t_*$  the integral is dominated by the contribution of a (real) saddle-point with  $p_{\rm saddle} \propto \ge$ (see Appendix B of~\cite{Maldacena:2017axo}). Such a saddle point may have an interpretation in terms of bulk Einstein equations.} 

 \item When $t \sim t_s \gg t_*$, the traversability should arise from a distinct physical picture from that of Fig.~\ref{fig:worm}(b). 
In this regime, independent of the details of any theory (and for general $g, k$ and $V$), we discussed in Sec.~\ref{sec:gen} that
the sole effect of $gV$ is to  generate a complex factor in $W$ proportional to $\vev{e^{-i g V}}$ with no dependence on the quantum numbers of $J$ at all. There is no scattering between $J$ and $\sO$ quanta. For a CFT, the conclusion also does not depend on the large $c$ limit (as far as the theory is chaotic).
 
Let us also highlight some features which are sharply different from those of the semi-classical regime discussed above: 

\ben

\item Even in the $k \to \infty$ limit, the \rena\ phenomenon, and thus traversability of the wormhole, does not depend on the sign of $g$, while the semi-classical scenario of  item 1 requires $g>0$~\cite{Gao:2016bin}.

\item  Regardless of the bulk geometry, the form of $V$, and details of how the signal is sent (i.e. its energy and direction etc.), a bulk signal departing from $(-t_s, \vx)$ from the $R$ boundary can only arrive on the $L$ boundary at $(t_s, \vx)$.\footnote{Again we are treating relaxation time scales as microscopic.} See Fig.~\ref{fig:contr}. As emphasized in Sec.~\ref{sec:ent}, the pairing 
of  $(t_s, \vx)$ and $(-t_s, \vx)$ is solely determined by the entanglement structure of the thermal field double state~\eqref{tfd0}. 

\een 
Thus the traversability for this regime does not appear to be associated with any spacetime causal structure at all, and 
is in a sense driven by entanglement. In other words, one has a ``quantum traversable wormhole.''

We do not yet have a precise way to characterize the ``quantum geometry'' associate with such a traversable wormhole. 
Here we offer some preliminary thoughts. Heuristically,  the transmission of a signal from the $R$ to $L$ boundary feels like a tunneling process across the horizon mediated by $gV$ interaction. 
Directly translating the discussion of Sec.~\ref{sec:rena1}  to the bulk gives the following  picture. 
Consider first $g=0$, for which the wormhole in non-traversable. Nevertheless, the bulk Wightman and Feynman functions between $L$ and $R$ for the bulk field dual to $J$ is nonzero. We may interpret the vanishing of $\vev{[J^L, J^R]} = \vev{J^L J^R} - \vev{J^R J^L}$ as perfect
destructive interference between the process of a {\it virtual} particle traveling from $R$ to $L$, and the mirror process of traveling from $L$ to $R$, as indicated in the left plot of Fig.~\ref{fig:15}. Turning on a nonzero $g$ gives a phase shift to each propagator, and in general the destructive inference is no longer perfect, resulting propagation of {\it real} particles. See right plot of Fig. \ref{fig:15}. 
In~\eqref{nno} we saw that $G^{LR}$ is periodic in $\ge$, thus as one dials the value of $g$, perfect  destructive interference  can be again reached at various special values.

\item Now for general $k, g$ and $\sig$, and $t, t_s \sim t_*$, the picture is no longer so sharp. There is a continuous spectrum in going from the semi-classical regime of item 1 to the quantum regime of  item 2. 

For example, for large, but finite $k$, while the bulk stress tensor induced from turning on $V$ will have a finite spread, for $\ge, \sig$ satisfying~\eqref{uyo1}, we expect the physics should still be close to the semi-classical picture. 

On the other hand, for~\eqref{wn} and~\eqref{glr0} (or its $(0+1)$-dimensional counterpart~\eqref{ghw}), both feature (a) and (b) listed in item 2 also apply. So it appears reasonable to expect the traversability is governed by the mechanism of item 2 except that in general there are also scatterings between $J$ and $\sO$ involved. Note that from~\eqref{med}, equation~\eqref{ghw} 
is obtained from a pure imaginary ``saddle'' $p_{\rm saddle} \approx i 2 \De_J$, which also suggests that the underlying physics cannot be understood straightforwardly in terms of classical scatterings of $\sO$ and $J$ quanta.\footnote{Also note that once one adds the Gaussian suppression factor $e^{-{p^2 \ov 2 \sig^2}}$ to~\eqref{med}, one can no longer deform the contour 
to imaginary $p$, which also indirectly suggests that~\eqref{oju} and~\eqref{ghw} are controlled by very different physics.}

\een

\section{Discussions and future directions} \label{sec:conc}

In this paper we presented a general argument for the \rena\ phenomenon in a many-body chaotic system and studied it in detail in two-dimensional CFTs in the large central charge limit. 
We also discussed the implications of these field theory results for wormhole physics. 

Here we end with some further discussion, including future directions: 

\ben 

\item {\it Teleportation?} 

As discussed in~\cite{Gao:2016bin,Maldacena:2017axo} (see also~\cite{Susskind:2017nto}), the coupling $V$ between $L$ and $R$ system is reminiscent of the operations in a  teleportation process.
During the time evolution of the system, the effect of having the $gV\delta (t=0)$ term in the Hamiltonian can be considered as being equivalent to the process of performing some measurements in the $R$ system,
communicating the results to the $L$ system, and then performing operations on the $L$ system.
But we would like to stress that the \rena\ phenomenon is in fact 
very different from quantum teleportation in the usual sense. In teleportation one would like to send an unknown state to another party. 
Here while the signal from the $R$ system re-appears in the $L$ system, in general there is no state teleportation. 
It can be readily checked in the qubit model of Sec.~\ref{sec:qubit} that  general $H^{L,R}$ and coupling $V$, do not implement 
teleportation of a state. 
 Thus the \rena\ phenomenon can at most be considered as a ``signal teleportation''. 
 This conclusion is also supported by the discussion of~\cite{Susskind:2017nto} that the operation for a state teleportation 
 involve a much larger complexity than that of the regenesis setup. As emphasized in~\cite{Maldacena:2017axo} the regenesis setup also shares some similarities with that of Hayden-Preskill~\cite{Hayden:2007cs}. But there are also some important differences. As we emphasized in the Introduction, regenesis requires extra fine tuning in the preparation of the initial state, with essentially no decoding needed for the signal, while for Hayden-Preskill, the initial state is generic, but decoding requires much higher level of complexity~\cite{Yoshida:2017non}. Interestingly, the decoding process described in \cite{Yoshida:2017non} applies to the case much longer than the scrambling time, which corresponds to the interference regime in our paper. 
 Further study is needed to connect better these two stories.

There is another interesting question for further study, say if one wants to send some known signals (one knows the input signal $\vp^R$ one is applying) from $R$ to $L$,  whether the current protocol is an efficient one (see~\cite{Bao:2018msr} for a discussion of treating the wormhole setup as a quantum channel).

\item {\it Nature of quantum traversable wormholes} 

It is important to have a more precise bulk picture for ``quantum traversable wormholes'' which we argued in Sec.~\ref{sec:quant}.
In particular, it would be ideal to have a real-time evolution picture for it.  

Equation~\eqref{nonu} can be reproduced from~\eqref{nonli-J} by replacing $\Phi$ of~\eqref{ujm} by\footnote{We thank J.~Maldacena, D.~Stanford and Z. Yang for this observation.}
$\bra{\Phi} \approx  a^* \bra{\Psi_\b} + U^R \bra{\Psi_\b}$. Note that this cannot be a true identity as the right hand side does not have the right normalization (and it does not reproduce~\eqref{nke}). Nevertheless, this expression is suggestive as it indicates that $\Phi$ is a superposition of two macroscopic states, one from acting $U^R$ on TFD, while the other corresponding to multiplying a TFD by a complex number. Equation~\eqref{nonu}, and thus traversability, arises from interference between them. 
How should we think the bulk geometry corresponding to $\Phi$? Is there a firewall at the horizon? 

\item{\it Other systems}

It is clearly of interest to study this phenomenon in other systems like spin chains or using random unitary circuits (see e.g.~\cite{Nahum:2017yvy}) which have generated lots of insights into chaotic systems. Note that since \rena\ concerns with time scales of order the scrambling time, thus it should be insensitive to the early time behavior such as whether the system has a 
nonzero Lyapunov exponent.

\item {\it Using effective field theories (EFTs)} 

The computation of the behavior $G^{LR}$ in the transition regime (i.e. for $t_s \sim t_*$) in two-dimensional CFTs is rather complicated and technical even in the large $c$ limit. While there are reasons to believe that the qualitative behavior we obtained should apply to generic chaotic systems, it would be good to understand it in a system-independent way. 
Recently a class of EFTs which aims to capture scrambling of general operators in chaotic systems (at least for those close to being maximally chaotic) has been proposed in~\cite{Blake:2017ris}. In particular, the EFT for two-dimensional CFT in the large $c$ limit has been obtained in~\cite{Cotler:2018zff,moshe}. 
The EFT approach could provide a simpler and system-independent way to study many aspects of the \rena\ phenomenon and wormhole physics. We will leave this for future investigation.

\item {\it Experimental realizations} 

It would be interesting to observe the \rena\ phenomenon experimentally. For example, one could imagine setting up 
the protocol in  bilayer graphene or quantum hall systems. 
Experimentally realizing a thermal field double state in a many-body system appears difficult.\footnote{Perhaps discussions in~\cite{levitov,Maldacena:2018lmt} could be relevant.} If one could realize~\eqref{tfd0}, to fine tune the state at $t= - t_s$ 
is similar to realizing an OTOC, as one needs to run the system ``backward'' in time, turn on the source, and then move forward in time. Recently there has been significant progress in realizing OTOCs in the lab (see e.g.~\cite{garttner2017measuring}), so perhaps
such tuning is not that far-fetched.  In the spirit of ER = EPR~\cite{Maldacena:2013xja}, realizing \rena\ experimentally may be interpreted 
as creating a quantum traversable wormhole in the lab!

\een

\vspace{0.2in}   \centerline{\bf{Acknowledgements}} \vspace{0.2in}
We would like to thank David Gross, Yingfei Gu, Gary Horowitz, Daniel Jafferis, Kristan Jensen, Patrick Ledwith, Patrick Lee, Leonid Levitov, Juan Maldacena, Don Marolf, Krishna Rajagopal, Moshe Rozali, Subir Sachdev, Douglas Stanford, Leonard Susskind, Zhenbin Yang,  Yizhuang You, and Ying Zhao for conversations and discussions. This work is supported by the Office of High Energy Physics of U.S. Department of Energy under grant Contract Number  DE-SC0012567. H.~L. would also like to thank Galileo Galilei Institute for Theoretical Physics for the hospitality during the workshop ``Entanglement in Quantum Systems'' and the Simons Foundation for partial support during the completion of this work.

\appendix

\section{Linear responses} \label{app:lin}

Consider perturbing a system by 
\be 
H = H_0 + H'_S (t)   
= H_0 + H^{(1)}_S (t) + H^{(2)}_S (t)
\ee
where subscripts $S$ denote the Schrodinger picture operators. 
More explicitly 
\be 
H^{(1)}_S = - g \int d^3 \vx f(t, \vx) \sO^L (\vx) \sO^R (\vx) , \quad 
H^{(2)}_S = -\int d^3 \vx \, \vp^R (t, \vx)  J^R (\vx)  \ .
\ee
We will take the support of $f(t,\vx)$ to be around $t =0$ and that for $\vp^R (t, \vx)$ to be around some $- t_1 \ll 0$. 
The two functions can be considered to have no overlap.

We take the system at $t =-\infty$ to be given by some state $\rho_0$,
and consider $H_S'$ going to zero at $t \to \pm \infty$ (as well as at spatial infinity).  Now let us consider the expectation value for some operator $A$ 
\be 
\vev{A} = \Tr (\rho (t, t_0) A_S) = \Tr (\rho_0 A_H (t,t_0)) 
\ee  
where 
\be 
\rho (t,t_0) = U(t, t_0) \rho_0 U^\da (t, t_0) , \qquad A_H (t,t_0) = U^\da (t, t_0) A U (t, t_0) , \qquad t_0 \to - \infty \ 
\ee
and $U (t, t_0)$ is the evolution operator under the full Hamiltonian $H$,
\be \label{1u}
U(t, t_0) = T  \exp \le(-i \int_{t_0}^t ds \, H (s) \ri) \ .
\ee
Below we will use 
\be 
A (t,t_0) = e^{i H_0 (t-t_0)} A e^{-i H_0 (t-t_0)} 
\ee
to denote the Heisenberg operator associated with $H_0$ with reference time $t_0$,
and $A(t)$ to denote the Heisenberg operator with $t_0$ set to zero. 

Using the standard interaction picture technique we can write 
\be 
U(t, t_0) = e^{- i H_0 (t-t_0)} U_I (t,t_0), \qquad U_I (t,t_0) = T  \exp \le(-i \int_{t_0}^t ds \, H' (s,t_0) \ri)
\ee
where 
\be 
H' (t,t_0) = e^{i H_0 (t-t_0)} H_S' (t) e^{-i H_0 (t-t_0)} = H^{(1)} (t,t_0) + H^{(2)} (t, t_0)
 \ ,
\ee
where 
\be 
H^{(1)} (t, t_0) = - g \int d^3 \vx f(t, \vx) \sO^L (t,t_0) \sO^R (t,t_0) , \quad 
H^{(2)} (t, t_0) =- \int d^3 \vx \, \vp^R (t, \vx)  J^R (t,t_0)  \ .
\ee
We consider to linear order in $H_2$ while to full nonlinear order in $H_1$, i.e. 
\be 
U_I (t, t_0) = U_1 (t, t_0) \le(1- i \int_{t_0}^t ds \, H^{(2)} (s,t_0) \ri), \quad 
U_1 (t,t_0) = T  \exp \le(-i \int_{t_0}^t ds \, H^{(1)} (s,t_0) \ri)
\ee
where we have used that the support of $\vp^R (x)$ is much earlier than that of $f(x)$.
Now for simplicity we will take 
\be 
f(x) = \de (t) f (\vx)
\ee
we will the have 
\be 
U_1 (t,t_0) =  \bca 1 & t \leq 0 \cr
\exp \le(i g V \ri) & t > 0 
\eca, \quad V = \int d^3 \vx  \,  f (\vx)\sO^L (0,t_0) \sO^R (0,t_0) \ .
\ee

We thus find that 
\be 
A_H (t, t_0) = U_I^\da (t, t_0) A (t,t_0) U_I (t, t_0) = \bca  A (t,t_0)-  i \int ds [A (t,t_0), H^{(2)} (s, t_0) ] & t < 0 \cr
 A_V (t,t_0) - i \int ds [A_V (t,t_0), H^{(2)} (s, t_0) ] & t > 0 \eca 
\ee
where 
\be 
A_V (t,t_0)= e^{-i g V} A (t,t_0) e^{i g V}  \ .
\ee

\section{An identity} \label{app:nb} 

For an operator $X$ in the $L$-system, consider 
  \be \label{nke1}
\vev{X}_g \equiv \vev{\Phi |X(t) |\Phi} 
\ee
where
\be
\bra{\Phi} \equiv e^{igV} U^R \bra{\Psi_\b} , \qquad U^R = e^{i\int ds\, \vp^R(s) J^R(s)} ,
\ee
and the source $\vp^R (t)$ is supported near $t = - t_s$, while $V$ is supported at $t=0$. In the limit $t, t_s \gg t_*$, with OTOCs set to zero, $\vev{X}_g$ can
be greatly simplified. More explicitly, we have~(suppressing $\Psi_\b$)
\ie
\vev{X}_g & = \vev{U_R^\da e^{- i g V} X (t) e^{i gV} U_R} = \vev{(U_R^\da -1) e^{- i g V} X (t) e^{i gV} (U_R-1)} \cr
& +  \vev{e^{- i g V} X (t) e^{i gV} } 
+  \le[\vev{e^{- i g V} X (t) e^{i gV} (U_R-1)} + h.c. \ri] \cr
& =  \vev{ (U_R^\da -1)  X (t) (U_R-1) } +  \vev{e^{- i g V} X (t) e^{i gV} }  +  \le[\vev{e^{- i g V} X (t)  (U_R-1)} + h.c. \ri]  \ .
\fe
Note that $X$ commutes with $U_R$. Now as in~\eqref{oio} we factorize parts of a correlator which are widely separated in time (with $\vev{e^{- i g V} X (t) e^{i gV} }  = \vev{X}$), which then gives 
\ie
\vev{X}_g &  = (1 -2\Re a) \vev{\Psi_\b|X (t)  |\Psi_\b} + \le(a \vev{\Psi_\b\le|X(t) U^R \ri|\Psi_\b} +h.c.\ri)
\fe
 with
 \be
a = \avg{e^{-igV}}_\b -1   \ .
  \ee

\section{Details of CFT calculation} \label{app:a}

\subsection{Approximation of identity Virasoro block by conformal transformation}

In this Appendix we first review a method developed in~\cite{fitzpatrick2015virasoro} to calculate correlation functions in a CFT in the large $c$ limit, and then use it to calculate~\eqref{onm} in the regime of $c\gg h_{ J}\gg h_{\mO}\sim O(1)$. While their original method was considered in the limit of $c \ra\infty$ with $h_{J}/c$ fixed, we will justify the same
method in the weaker limit stated above. We will keep general dependence for $h_J$ but ignoring all $1/c$ and $h_\mO/c$
corrections. 

Consider first  the four-point function $\avg{ J_{a}(z_{a}) J_{b}(z_{b})\mO_{1}(z_{1})\mO_{2}(z_{2})}$.
Since $h_{ J}\gg h_{\mO}$, we can treat this four point function
as if the two point function of $\mO$ in the background of $ J$.
To be more precise, we are going to do the following conformal map
from $z$ plane to $w$ plane:
\begin{equation}
1-w=\left(1-\f{z_{ab}z}{z_{b}(z_{a}-z)}\right)^{\a},\;\a=\sqrt{1-24\f{h_{ J}}c}\label{eq:trans}
\end{equation}
which maps $z_{a}\ra w_{a}=\infty$, $z_{b}\ra w_{b}=1$ and $0\ra0$.
This map has a branch cut in $z$ plane from $z_{b}$ to $z_{a}$.
The Jacobian is 
\begin{equation}
\J(z)\equiv\f{\p z}{\p w}=\f{(z-z_{a})(z-z_{b})}{\a z_{ab}}\left[\f{z_{a}(z-z_{b})}{z_{b}(z-z_{a})}\right]^{-\a}
\end{equation}
The relation between 4-pt function in $z$ and $w$ planes is 
\begin{equation}
\avg{ J_{a} J_{b}\mO_{1}\mO_{2}}_{w}=\J_{a}^{h_{ J}}\J_{b}^{h_{ J}}\J_{1}^{h_{\mO}}\J_{2}^{h_{\mO}}\avg{ J_{a} J_{b}\mO_{1}\mO_{2}}_{z}
\end{equation}
where the subscript denotes the coordinate. Note that $\J_{a}$ and
$\J_{b}$ are vanishing, so the above formula should be regarded as
taken in a proper limit since the LHS is also vanishing as $w_{a}\ra\infty$.

The advantage of this conformal transformation is that in $w$ plane,
4-pt function $\avg{ J(w_{a}) J(w_{b})T(w)}$ does
not depend on $h_{ J}$ explicitly.
In large $c$ limit, this amounts to the leading order approximation
in $1/c$ expansion. To be more precise, by Ward identity, the 4-pt function in $z$ plane
is
\begin{equation}
\avg{ J(z_{a}) J(z_{b})T(z)}=\f 1{z_{ab}^{2h_{ J}}}\f{h_{ J}z_{ab}^{2}}{(z-z_{a})^{2}(z-z_{b})^{2}}
\end{equation}
Since stress tensor is not primary field, it has an extra Schwarzian
term under conformal transformations. Given the transformation (\ref{eq:trans}),
the stress tensor transformations as
\begin{equation}
T(w)=\J(z)^{2}\left(T(z)-\f{h_{ J}z_{ab}^{2}}{(z-z_{a})^{2}(z-z_{b})^{2}}\right)\label{eq:27}
\end{equation}
Hence we find that in $w$ plane
\begin{equation}
\avg{ J(w_{a}) J(w_{b})T(w)}=\J_{a}^{h_{ J}}\J_{b}^{h_{ J}}\J_{z}^{2}\cdot 0\label{eq:28}
\end{equation}
where the Schwarzian cancels the term that is proportional to
$h_{ J}$. In $w$ plane, the whole Virasoro block is summing over
all Virasoro descendents just like $z$ plane. We can Taylor expand
$T(w)$ around $w=0$:
\begin{equation}
T(w)=\sum_{n}w^{-n-2}\mL_{n}\label{eq:29}
\end{equation}
One can show that the commutation between $\mL_{n}$ and general primary
operator $X(w)$ obeys the same rule as in $z$ plane:
\begin{equation}
[\mL_{n},X(w)]=h_{X}(1+n)w^{n}X(w)+w^{1+n}\p_{w}X(w)\label{eq:30}
\end{equation}
and Virasoro algebra still holds for all $\mL_{n}$:
\begin{equation}
[\mL_{n},\mL_{m}]=(n-m)\mL_{n+m}+\f c{12}n(n^{2}-1)\d_{n,-m}
\end{equation}
Indeed, above relations always hold when the Jacobian $\J(z)$ is nonsingular
around $w$.

The relation between $L_{n}$, the Virasoro mode of $T(z)$, and $\mL_{n}$
can be solved explicitly by the conformal transformation (\ref{eq:27})
\begin{align}
\sum_{n}w^{-n-2}\mL_{n} & =J(z)^{2}\left(\sum_{n}z(w)^{-n-2}L_{n}-\f{h_{ J}z_{ab}^{2}}{(z-z_{a})^{2}(z-z_{b})^{2}}\right)\nonumber \\
 & =\f{(1-w)^{2/\a-2}z_{a}^{2}z_{b}^{2}z_{ab}^{2}}{\a^{2}(z_{a}-(1-w)^{1/\a}z_{b})^{4}}\sum_{n}\left[\f{z_{a}z_{b}(1-(1-w)^{1/\a})}{z_{a}-(1-w)^{1/\a}z_{b}}\right]^{-n-2}L_{n}-\f{h_{ J}}{\a^{2}(1-w)^{2}}\label{eq:33}
\end{align}
which implies that all $\mL_{n}$ are linear combinations of $L_{m}$
with $m\geq n$. This immediately gives an important result 
\begin{equation}\label{eq:c11}
\mL_{n}\bra h=0,\;n\geq0
\end{equation}
for any primary $\bra h$. \footnote{For $\bra{h}=\bra{0}$, \eqref{eq:c11} holds for $n\geq -1$.} However, since the expansion (\ref{eq:29})
is not convergent around infinity due to the existence of branch cut
from 1 to $\infty$, we should not expect $\mL_{n}^{\dagger}=\mL_{-n}$.
In other words, the radius of convergence of series (\ref{eq:29})
is bounded by the location of branch cut. Formally we can define a
``$w$-primary state'' $\ket{h_{w}}$ as 
\begin{equation}
\ket{h_{w}}=\lim_{w\ra\infty}\ket{0_{w}}w^{2h_{X}}X(w)\label{eq:34}
\end{equation}
with $\ket{h_{w}}\mL_{-n}=0$ for $n\geq0$ and normalization $\ket{h_{w}}h\rangle=1$.
The whole Virasoro block in $w$ plane can be calculated as insertion
of projection $P_{T^{k}}$ between $ J_{a} J_{b}$ and $\mO_{1}\mO_{2}$,
where 
\begin{equation}
P_{T^{k}}\equiv\f{\mL_{-n_{1}}\cdots\mL_{-n_{k}}\bra h\ket{h_{w}}\mL_{n_{k}}\cdots\mL_{n_{1}}}{\avg{h_{w}|\mL_{n_{k}}\cdots\mL_{n_{1}}\mL_{-n_{1}}\cdots\mL_{-n_{k}}|h}}\label{eq:Pt}
\end{equation}
Note that these $P_{T^{k}}$'s are not orthogonal, and in general
we need to take all overlaps between different projectors into account. Let us take $h=0$ for identity Virasoro block for now.

There are a few features of this construction. First, $\avg{0_{w}|\mL_{n_{1}}\cdots\mL_{n_{k}}|0}_{w}=\avg{0|L_{n_{1}}\cdots L_{n_{k}}|0}_{z}$
because $\mL_{n}$ and $L_{n}$ obey the same algebra. Therefore,
we can simply estimate the denominator of (\ref{eq:Pt}) in large
$c$ limit. For $n_{i}\geq2$, 
\begin{equation}
\avg{0_{w}|\mL_{n_{k}}\cdots\mL_{n_{1}}\mL_{-n_{1}}\cdots\mL_{-n_{k}}|0}=O(c^{k})\label{eq:37}
\end{equation}
Second, $\ket{0_{w}}\mL_{n_{k}}\cdots\mL_{n_{1}}\mO(w_{1})\cdots\mO(w_{n})|0\rangle$
is the same as that in $z$ plane because of the same algebra (\ref{eq:30})
and (\ref{eq:34}). In particular, the two point function is
\begin{equation}
\ket{0_{w}}\mO(w_{1})\mO(w_{2})|0\rangle=\f 1{w_{12}^{2h_{\mO}}}
\end{equation}
Note that this is different from conformal transformed version of
$\avg{0|\mO(z_{1})\mO(z_{2})|0}_{z}$ to $w$ plane. Physically, this
means that ignoring the neighborhood of branch cut, we regard all
other regions in $w$ plane the same as ordinary CFT on complex plane.
Third, $\ket 0 J_{a} J_{b}\mL_{-n_{1}}\cdots\mL_{-n_{k}}\bra{0}$
is not the same as $z$-plane CFT due to the branch cut, but restricted
to the conformal transformation rules from $z$-plane to $w$-plane.
In particular, the two point function obeys
\begin{equation}
\avg{0| J(w_{a}) J(w_{b})|0}=\J_{a}^{h_{ J}}\J_{b}^{h_{ J}}\avg{0| J(z_{a}) J(z_{b})|0}
\end{equation}

The advantage of this special conformal transformation is that in
$w$ plane, 
\be
\avg{ 0|J(w_{a}) J(w_{b})\mL_{-n}|0}=0
\ee 
for all $n\geq0$ due to (\ref{eq:28}). This method can be generalized to multiple
$T$ insertion in (\ref{eq:28}) and one can show that the leading
order of $h_{ J}$ vanishes in $w$-plane. Indeed, notice that $T$
is different from primary field only by a central term. Therefore,
for multiple $T$ insertion in a correlation function of primaries
$\avg X$, there is a induction relation:
\begin{align}
\avg{T(z_{1})\cdots T(z_{k})X}= & \sum_{i}\left(\f{h_{i}}{(z_{1}-z_{i})^{2}}+\f 1{z_{1}-z_{i}}\p_{i}\right)\avg{T(z_{2})\cdots T(z_{k})X}\nonumber\\
& +\sum_{i=2}^{k}\f{c/2}{(z_{1}-z_{i})^{4}}\avg{T(z_{2})\cdots\hat{T}(z_{i})\cdots T(z_{k})X}\label{eq:TTX}
\end{align}
where hat means omiting $T(z_{i})$. From above formula, the $k$
insertion should have the following expansion
\begin{equation}
\avg{T(z_{1})\cdots T(z_{k})J_{a}J_{b}}=\sum_{i=0}^{[k/2]}\sum_{j=0}^{k-2i}c^{i}h_J^{j}F_{ij}(z_{n}),\qquad F_{00}(z_{n})=0
\end{equation}
where $z_{n}$ denote coordinates collectively. In $c\gg h_{ J}$
limit, the expansion has the orders from high to low as
\begin{equation}
c^{[k/2]}h_J^{k-2[k/2]},c^{[k/2]-1}h_J^{k-2[k/2]+2},\cdots,h_{ J}^{k},\cdots
\end{equation}
Note that if we are considering $L_{-n}$ rather than $\mL_{-n}$,
the powers of $c$ does not contribute to $\ket 0 J_{a} J_{b}L_{-n_{1}}\cdots L_{-n_{k}}\bra 0$.
But transforming to $w$ plane, these powers are leading contributions
to $\ket 0 J_{a} J_{b}\mL_{-n_{1}}\cdots\mL_{-n_{k}}\bra 0$.
On the other hand, terms involving $\mO_{1}$ and $\mO_{2}$ after
inserting $P_{T^{k}}$ has scaling 
\begin{equation}
\avg{0_{w}|\mL_{n_{k}}\cdots\mL_{n_{1}}\mO_{1}\mO_{2}}\sim\sum_{n=1}^{k}O(h_{\mO}^{n})\sim O(1)
\end{equation}
Therefore, the highest order terms in first a few orders of $P_{T^{k}}$
insertion are
\begin{equation}
O(1)+O(h_{ J}/c)+O(1/c)+O(h_{ J}/c^{2})+O(1/c^{2})+O(h_{ J}/c^{2})+\cdots\label{eq:43}
\end{equation}
where above explicit terms are from $k=0$ to $k=5$. 

The purpose of the transformation (\ref{eq:trans}) is to cancel all
$O(h_{ J}/c^{[k/2]})$ order terms in odd $k$ and leave highest
orders of the expansion (\ref{eq:43}) as 
\begin{equation}
O(1)+O(1/c)+O(1/c^{2})+\cdots\label{eq:44}
\end{equation}
This can be seen by noting that the first term in (\ref{eq:TTX})
does not contribute with $c$ and only $TT$ fusion gives powers of
$c$. For $k=2s+1$, $c^{s}h_{ J}$ order only comes from the OPE
between all different $T(z_{i})$'s, which in total gives $c^{s}$:
\begin{equation}
\avg{ J_{a} J_{b}T(z_{1})\cdots T(z_{2s+1})}\supset(c/2)^{s}\sum_{k=1}^{2s+1}\sum_{\substack{\{p,q\}\\
p_{i},q_{i}\neq k
}
}\prod_{i}\f 1{z_{p_{i}q_{i}}^{4}}\avg{ J_{a} J_{b}T(z_{k})}\sim O(c^{s}h_{ J})
\end{equation}
where $\{p,q\}$ are all different choices of pairs of indices from
$1$ to $2s+1$ except $k$. Transformation to $w$ plane, due to
(\ref{eq:27}) we have 
\begin{align}
\avg{ J_{a} J_{b}T(w_{1})\cdots T(w_{2s+1})} & =\J_{a}^{h_{ J}}\J_{b}^{h_{ J}}\left(\prod_{i=1}^{2s+1}J_{z_{i}}^{2}\right)\avg{ J_{a} J_{b}\prod_{i=1}^{2s+1}\left(T(z_{i})-\f{h_{ J}z_{ab}^{2}}{z_{ia}^{2}z_{ib}^{2}}\right)}\nonumber \\
 & \supset \J_{a}^{h_{ J}}\J_{b}^{h_{ J}}\left(\prod_{i=1}^{2s+1}J_{z_{i}}^{2}\right)(c/2)^{s}\sum_{k=1}^{2s+1}\sum_{\substack{\{p,q\}\\
p_{i},q_{i}\neq k
}
}\prod_{i}\f 1{z_{p_{i}q_{i}}^{4}}\avg{ J_{a} J_{b}\left(T(z_{k})-\f{h_{ J}z_{ab}^{2}}{z_{ka}^{2}z_{kb}^{2}}\right)}\nonumber \\
 & =0\cdot O(c^{s}h_{ J})
\end{align}
This shows that, up to terms suppressed by $1/c$, the insertion of $P_{T^k}$ reduces to just insertion of vacuum $\bra{0}\ket{0_w}$.

In $h_{ J}\sim c\ra\infty$ limit, the leading order of all $P_{T^{k}}$
insertions are $h_{ J}^{k}/c^{k}$ terms. One can also see all these
terms vanishes in $w$ plane. That is the argument in \cite{fitzpatrick2015virasoro}. The reason why their method also applies to our weaker limit $O(1)\sim h_\mO\ll h_J\ll c$ is that we are using the same expansion of $h_J/c$ and $1/c$ in $w$ plane, in which only terms suppressed by $1/c$ survive due to the conformal transformation.

\subsection{Application to $W$}
Using the conformal transformation, four point function
reads
\begin{align}
\avg{ J_{a} J_{b}\mO_{1}\mO_{2}}_{z} & =\J_{a}^{-h_{ J}}\J_{b}^{-h_{ J}}\J_{1}^{-h_{\mO}}\J_{2}^{-h_{\mO}}\avg{ J_{a} J_{b}\mO_{1}\mO_{2}}_{w}\nonumber \\
 & \app \J_{a}^{-h_{ J}}\J_{b}^{-h_{ J}}\J_{1}^{-h_{\mO}}\J_{2}^{-h_{\mO}}\avg{ J_{a} J_{b}}_{w}\avg{0_{w}|\mO_{1}\mO_{2}}_{w}\nonumber \\
 & =\avg{ J_{a} J_{b}}_{z}\J_{1}^{-h_{\mO}}\J_{2}^{-h_{\mO}}\f 1{w_{12}^{2h_{\mO}}}
\end{align}
Plugin the coordinate transformation, we find that
\begin{equation}
\mV(u)\equiv\f{\avg{ J_{a} J_{b}\mO_{1}\mO_{2}}_{z}}{\avg{ J_{a} J_{b}}_{z}\avg{\mO_{1}\mO_{2}}_{z}}=\left(\f{z_{12}^{2}}{\J_{1}\J_{2}w_{12}^{2}}\right)^{h_{\mO}}=\left(\f{\a^{2}u^{2}(1-u)^{\a-1}}{(1-(1-u)^{\a})^{2}}\right)^{h_{\mO}},\qquad u=\f{z_{12}z_{ab}}{z_{1a}z_{2b}}
\end{equation}
which has branch cut of $u$ from $1$ to $+\infty$.

We are interested in cases with even number $\mO$ insertion. In leading
order, this can be calculated as
\begin{align}
\f{\avg{ J_{a} J_{b}\mO_{1}\cdots\mO_{2n}}_{z}}{\avg{ J_{a} J_{b}}_{z}} & =\f{\J_{a}^{-h_{ J}}\J_{b}^{-h_{ J}}}{\avg{ J_{a} J_{b}}_{z}}\left(\prod_{i=1}^{2n}\J_{i}^{-h_{\mO}}\right)\avg{ J_{a} J_{b}\mO_{1}\cdots\mO_{2n}}_{w}\nonumber \\
 & \app\left(\prod_{i=1}^{2n}\J_{i}^{-h_{\mO}}\right)\avg{0_{w}|\mO_{1}\cdots\mO_{2n}}_{w}\nonumber \\
 & \app\left(\prod_{i=1}^{2n}\J_{i}^{-h_{\mO}}\right)\sum_{\{(s_{2i},s_{2i+1})\}}\prod_{i=1}^{n}\avg{0_{w}|\mO_{s_{2i}}\mO_{s_{2i+1}}}_{w}\nonumber \\
 & =\sum_{\{(s_{2i},s_{2i+1})\}}\prod_{i=1}^{n}\left[\left(\f{z_{s_{2i},s_{2i+1}}^{2}}{\J_{s_{2i}}\J_{s_{2i+1}}w_{s_{2i},s_{2i+1}}^{2}}\right)^{h_{\mO}}\avg{\mO_{s_{2i}}\mO_{s_{2i+1}}}_{z}\right]\nonumber \\
 & =\sum_{\{(s_{2i},s_{2i+1})\}}\prod_{i=1}^{n}\left[\mV(u_{s,i})\avg{\mO_{s_{2i}}\mO_{s_{2i+1}}}_{z}\right],\qquad u_{s,i}\equiv\f{z_{s_{2i},s_{2i+1}}z_{ab}}{z_{s_{2i},a}z_{s_{2i+1},b}}
 \label{eq:49}
\end{align}
where $\{(s_{2i},s_{2i+1})\}$ is the collection of contractions between
$s_{2i}$-th and $s_{2i+1}$-th operators. In the second line we ignored
higher orders of $1/c$, and in third line we used large $\mN$ (large
$c$) ansatz to factorize all $\mO$'s in two point functions. Associated with antiholomorphic part, \eqref{eq:49} becomes \eqref{bn}.

\subsection{With non-identity channel} \label{app:non-id}

For a generic but light non-identity intermediate channel, the difference
with identity block is as follows. The main point is that we need
to slightly modify \eqref{eq:49}.
\begin{align}
 & \f{\avg{J_{a}J_{b}\mO_{1}\cdots\mO_{2n}}_{z}}{\avg{J_{a}J_{b}}_{z}}\nonumber \\
= & \f{\J_{a}^{-h_{J}}\J_{b}^{-h_{J}}}{\avg{J_{a}J_{b}}_{z}}\left(\prod_{i=1}^{2n}\J_{i}^{-h_{\mO}}\right)\avg{J_{a}J_{b}\mO_{1}\cdots\mO_{2n}}_{w}\nonumber \\
= & \f{\J_{a}^{-h_{J}}\J_{b}^{-h_{J}}}{\avg{J_{a}J_{b}}_{z}}\left(\prod_{i=1}^{2n}\J_{i}^{-h_{\mO}}\right)\sum_{m=0}^{\infty}\f{\avg{J_{a}J_{b}\mL_{-1}^{m}|h}_{w}\avg{h_{w}|\mL_{1}^{m}\mO_{1}\cdots\mO_{2n}}_{w}}{\avg{h_{w}|\mL_{1}^{m}\mL_{-1}^{m}|h}}+\mO(1/c)\nonumber \\
= & \f{\left(\prod_{i=1}^{2n}\J_{i}^{-h_{\mO}}\right)}{\avg{J_{a}J_{b}}_{w}}\sum_{m=0}^{\infty}\left[\sum_{i=1}^{2n}\left(2h_{\mO}w_{i}+w_{i}^{2}\del_{i}^{2}\right)\right]^{m}\f{\avg{J_{a}J_{b}\mL_{-1}^{m}|h}_{w}\avg{h_{w}|\mO_{1}\cdots\mO_{2n}}_{w}}{\avg{h_{w}|\mL_{1}^{m}\mL_{-1}^{m}|h}}\nonumber \\
\app & \f{\left(\prod_{i=1}^{2n}\J_{i}^{-h_{\mO}}\right)}{\avg{J_{a}J_{b}}_{w}}\sum_{m=0}^{\infty}\f{\avg{J_{a}J_{b}\mL_{-1}^{m}|h}_{w}}{\avg{h_{w}|\mL_{1}^{m}\mL_{-1}^{m}|h}}\left[\sum_{i=1}^{2n}\left(2h_{\mO}w_{i}+w_{i}^{2}\del_{i}^{2}\right)\right]^{m}\sum_{\{(s_{2i-1},s_{2i})\}}\avg{h_{w}|\mO_{s_{1}}\mO_{s_{2}}}_{w}\prod_{i=2}^{n}\avg{0_{w}|\mO_{s_{2i-1}}\mO_{s_{2i}}}_{w}\nonumber \\
= & \f{\left(\prod_{i=1}^{2n}\J_{i}^{-h_{\mO}}\right)}{\avg{J_{a}J_{b}}_{w}}\sum_{m=0}^{\infty}\f{\avg{J_{a}J_{b}\mL_{-1}^{m}|h}_{w}}{\avg{h_{w}|\mL_{1}^{m}\mL_{-1}^{m}|h}}\sum_{\{(s_{2i-1},s_{2i})\}}\left[\sum_{i=1}^{2}\left(2h_{\mO}w_{s_{i}}+w_{s_{i}}^{2}\del_{s_{i}}^{2}\right)\right]^{m}\avg{h_{w}|\mO_{s_{1}}\mO_{s_{2}}}_{w}\prod_{i=2}^{n}\avg{0_{w}|\mO_{s_{2i-1}}\mO_{s_{2i}}}_{w}\nonumber \\
= & \f{\left(\prod_{i=1}^{2n}\J_{i}^{-h_{\mO}}\right)}{\avg{J_{a}J_{b}}_{w}}\sum_{\{(s_{2i-1},s_{2i})\}}\sum_{m=0}^{\infty}\f{\avg{J_{a}J_{b}\mL_{-1}^{m}|h}_{w}\avg{h_{w}|\mL_{1}^{m}\mO_{s_{1}}\mO_{s_{2}}}_{w}}{\avg{h_{w}|\mL_{1}^{m}\mL_{-1}^{m}|h}}\prod_{i=2}^{n}\avg{0_{w}|\mO_{s_{2i-1}}\mO_{s_{2i}}}_{w}\nonumber \\
\app & \sum_{\{(s_{2i-1},s_{2i})\}}\f{\mV_{h}(u_{s,1})}{\mV(u_{s,1})}\prod_{i=1}^{n}\left[\mV(u_{s,i})\avg{\mO_{s_{2i-1}}\mO_{s_{2i}}}_{z}\right],\quad u_{s,i}=\f{z_{s_{2i-1},s_{2i}}z_{ab}}{z_{s_{2i-1},a}z}
\end{align}
where in the third line we used the fact that only global block contributes
in leading order, in the forth line we used identity \eqref{eq:30}, in fifth
line we factorize the $\mO$ correlation function into product of
two point functions and one three point function, in sixth line we
used the fact that
\begin{equation}
\avg{\mL_{1}^{m}\mO_{1}\mO_{2}}=\sum_{i=1}^{2}\left(2h_{\mO}w_{i}+w_{i}^{2}\del_{i}^{2}\right)\f 1{w_{12}^{2h_{\mO}}}=0
\end{equation}
and in last line we use the result in \cite{fitzpatrick2015virasoro} 
\begin{equation}
\mV_{h}(z)\equiv\f{\avg{J(\infty)J(1)\mO(z)\mO(0)}_{z}}{\avg{J(\infty)J(1)}_{z}\avg{\mO(z)\mO(0)}_{z}}=\mV(z)\left[\f{1-(1-z)^{\a}}{\a}\right]^{h}{}_{2}F_{1}(h,h,2h,1-(1-z)^{\a})
\end{equation}
and $\mV(z)$ is the vacuum block.

\subsection{Explicit expression of $A$} \label{app:aexp}

Here we give an explicit derivation of~\eqref{aa1} from~\eqref{defa}  for  $t_s = t$ and $x_s = x$. 
The discussion of this subsection has some parallel to that of~\cite{Roberts:2014ifa}
for OTOCs in a large $c$ CFT.

From~\eqref{sv0}, $\sV (u)$ is real for $u < 1$ and has a branch cut along $u \in (1, \infty)$. 
 With $t_s=t, x_s = x$ we have 
\ie \label{cv31}
u_1 = {4 e^{{2 \pi \ov \b} y_+}   \ov (e^{i \ep_1} - e^{{2 \pi \ov \b} y_+ + i \ep_J} )(e^{{2 \pi \ov \b} y_+ + i \tilde \ep_J}  - e^{ i \tilde \ep_1})}, \quad  \bar u_1  = {4 e^{{2 \pi \ov \b} y_-}   \ov (e^{- i \ep_1} - e^{{2 \pi \ov \b} y_-- i \ep_J} )(e^{{2 \pi \ov \b} y_- - i \tilde \ep_J}  - e^{-i \tilde \ep_1})} 
\fe
where we have introduced 
\be 
y_+ = x - x_1 + t, \qquad y_- = x - x_1 - t \ .
\ee
Recall that $\sV^+$ is defined with ordering  $\ep_1 < \ep_J < \tilde \ep_J < \tilde \ep_1$, while $\sV^-$  with ordering $\ep_J < \ep_1 < \tilde \ep_J < \tilde \ep_1$.  Let us look at the behavior of $A(u_1, \bar u_1)$ as we increase $t$ from $0$ while keeping $x - x_1$ fixed (assuming $x- x_1$ is not exactly zero).  For sufficiently small $t$, regardless of the sign of $x -x_1$, $(0, x_1)$ and $(t, x)$ are spacelike separated, with $u_1, \bar u_1 < 0$. 
 In this case $\ep$'s do not matter, and thus $\sV^{+} = \sV^{-}$, leading to $A =0$. This of course can be deduced from~\eqref{bnm} without doing any calculations as the commutator of spacelike separated operators must vanish. 

\begin{figure}
\begin{centering}
\includegraphics[height=8cm]{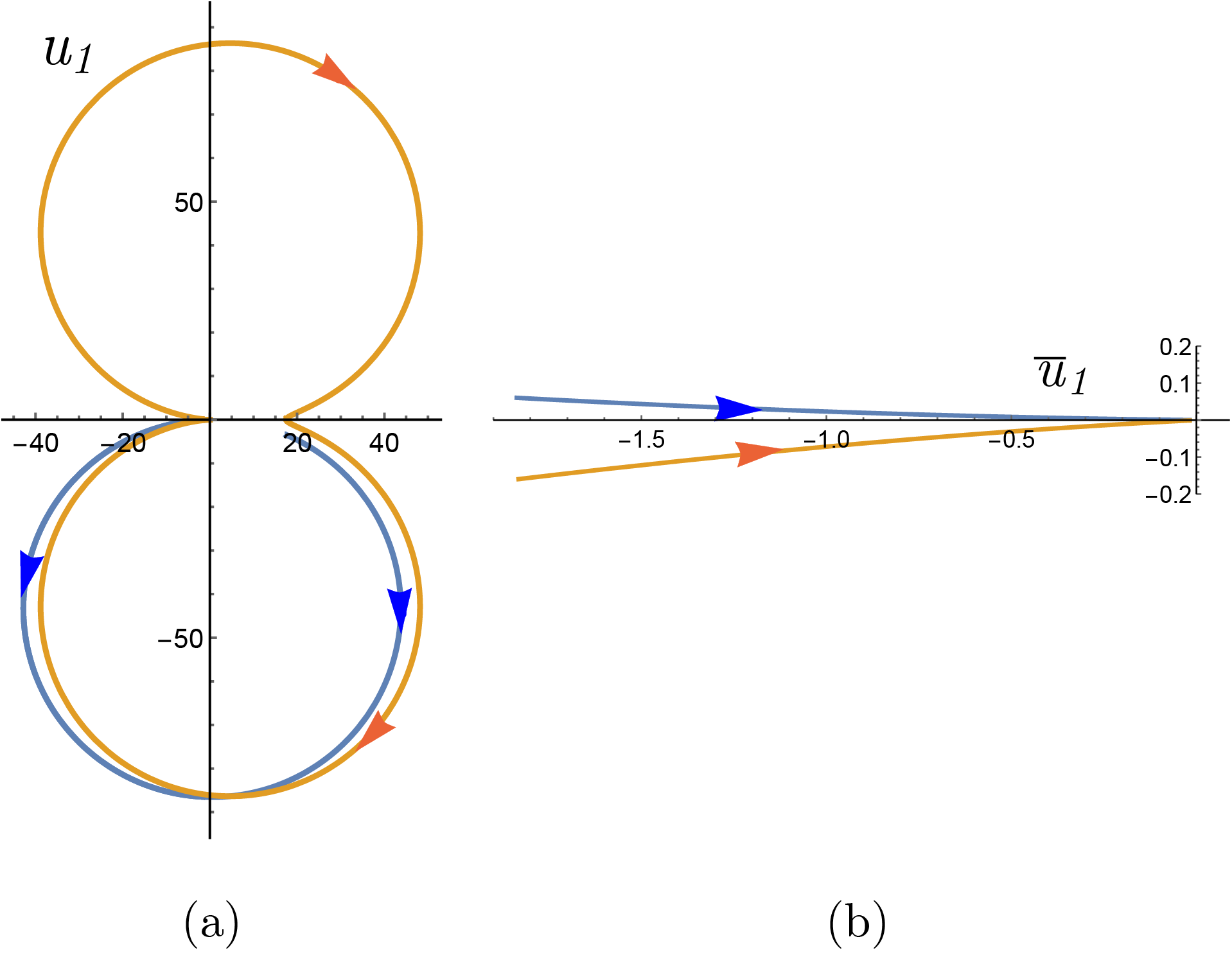}
\;\;
\par\end{centering}
\caption{Plots of $u_1$ (plot (a))  and $\bar{u}_1$ (plot (b)) as a function of $t$. Blue and yellow are for $\mV^+$ and $\mV^-$ respectively. One can see that $u_1$ stays on first sheet for $\mV^{+}(u_1)$, but moves to second sheet for $\mV^{-}(u_1)$; $\bar{u}_1$ stays on first sheet for both $\mV^{\pm}(\bar{u}_1)$. 
\label{fig:fig-uplane}}
\end{figure}

With sufficiently large $t$,  $(0, x_1)$ and $(t, x)$ become timelike separated. 
Consider, e.g.  $x - x_1 < 0$, for which $u_1$ starts being negative at $t =0$ and is again negative for large $t$, but 
in between $u_1$ undergoes nontrivial motions in the complex plane as the lightcone $y_+ =0$ is crossed. 
For  $\sV^{(+)}$, one finds that $u_1$ remains on the first sheet throughout the process, while for $\sV^{(-)}$, $u_1$ crosses the branch cut from upper half $u$-plane and moves to the second sheet. In contrast, $\bar u_1$ always remains real and negative.  See Fig. \ref{fig:fig-uplane}. We thus find that for sufficiently large $t$, 
\be\label{eq:A-1}
A (u_1, \bar u_1) = (\sV_1  ( u_1 ) - \sV_{2} ( u_1 )) \sV_1 (\bar u_1) , \quad x - x_1 < 0 
\ee
where $\sV_1 (u)$ ($\sV_2 (u)$) denotes the value along the negative real axis on the first (second) sheet,
\bega\label{sv11}
\sV_1 (u) =  \left( \frac { \alpha ( - u ) } { \sqrt { 1- u } } \frac { 1} {- ( 1- u ) ^ { - \alpha / 2}  + ( 1- u ) ^ { \alpha / 2} } \right) ^ { 2h_\sO } ,  \\
\label{sv21}
\sV_2 (u) = \left( \frac { \alpha ( - u ) } { \sqrt { 1- u } } \frac { 1} { ( 1- u ) ^ { - \alpha / 2} e ^ { i \pi \alpha } - ( 1- u ) ^ { \alpha / 2} e ^ { - i \pi \alpha } } \right) ^ { 2h_\sO }   \ .
\end{gather}

Similarly for $x_s - x_1 > 0$ we find that, $u_1$ always remains real, negative, but 
 $\bar u_1$ moves nontrivially in the complex plane as the light cone is crossed. Again one finds that 
$\sV^{(+)}$ remains on the first sheet, while $\sV^{(-)}$ moves to the second sheet from above. We thus find 
\be\label{eq:A+1}
A (u_1, \bar u_1) = (\sV_1  ( \bar u_1 ) - \sV_{2} ( \bar u_1 )) \sV_1 ( u_1), \quad x - x_1 > 0  \ .
\ee
One consistent check is that~\eqref{eq:A-1} and~\eqref{eq:A+1} agree when $x=x_1$.
We can write~\eqref{eq:A-1} and~\eqref{eq:A+1} in a unified way as 
\be \label{aa1a}
A (u_1, \bar u_1) =  (\sV_1  ( u_1 ) - \sV_{2} ( u_1 )) \sV_1 (\bar u_1) 
\ee
with now $u_1$ and $\bar u_1$ defined as
\be \label{aa2a}
u_1 \equiv - {4 e^{{2 \pi \ov \b} ( t- |x - x_1|)}   \ov (1 - e^{{2 \pi \ov \b} ( t- |x - x_1|)} )^2}, \qquad  \bar u_1  \equiv - {4 e^{{2 \pi \ov \b} ( t + |x - x_1|)}   \ov (1- e^{{2 \pi \ov \b} ( t + |x - x_1|)} )^2}   \ .
\ee

\section{Full $k$-dependence in multiple operator species} \label{full-k}

In this appendix, we will use the following notation:
\be  
W=\sum_{n=0}^\infty \f{(-ig)^n}{n!}\sum_{\{\a_i\}}W_n, \quad W_n = \f{1}{k^n}\vev{[\sO_{\a_n},[\sO_{\a_{n-1}},  \cdots [\sO_{\a_1}, J]\cdots]\ \tilde J \tilde \sO_{\a_n} \cdots \tilde \sO_{\a_1}}_\b
\ee
where we suppressed all spacetime coordinates. In order to calculate $W$ in the case of multiple operator species, we need to note that there are three types of contractions, 
\be 
G\d_{\a_i\a_j}\equiv\avg{\mO_{\a_i}\tilde{\mO}_{\a_j}}, \quad
H\d_{\a_i\a_j}\equiv\avg{\mO_{\a_i}\mO_{\a_j}}, \quad 
\tilde{H}\d_{\a_i\a_j}\equiv\avg{\tilde{\mO}_{\a_i}\tilde{\mO}_{\a_j}}
\ee
where we used the fact that all locations of $\mO$ and $\tilde{\mO}$
are the same and at origin. 

\begin{figure}
\begin{centering}
\includegraphics[width=10cm]{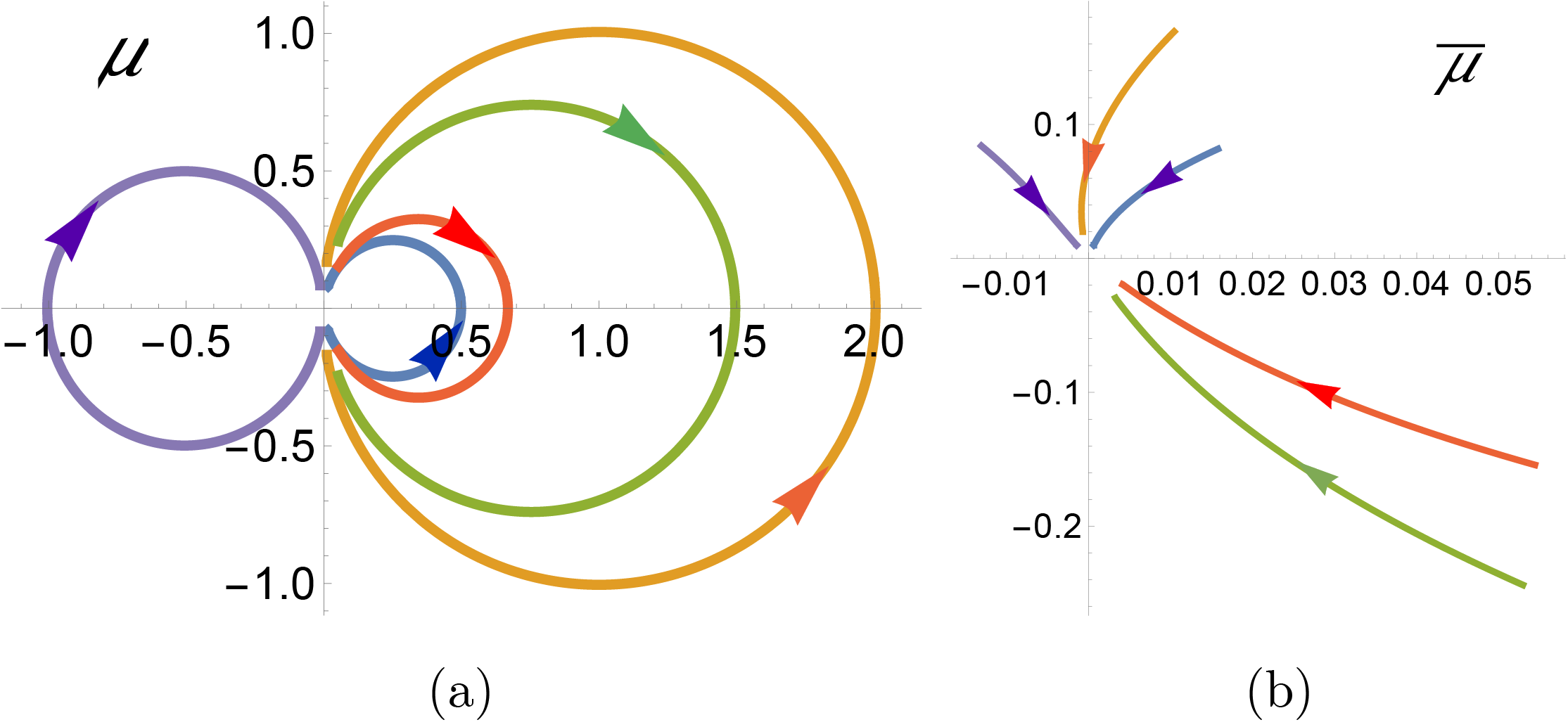}
\par\end{centering}
\caption{The plot of $\mu$ and $\bar{\mu}$. The blue, yellow, green, red
and purple are for $\protect\avg{12ab}$, $\protect\avg{1a2b}$, $\protect\avg{2a1b}$,
$\protect\avg{a21b}$ and $\protect\avg{ab\tilde{2}\tilde{1}}$ respectively.\label{fig:mu}}
\end{figure}

Since \eqref{bn} is factorized as product of 4-pt functions encoded
in $\mV(u_{i})\mV(\bar{u}_{i})$, to calculate $W_{n}$ we only need
to worry about the relative ordering of the four operators related
to each $u_{i}$ and $\bar{u}_{i}$. For $G$ type contraction, there
are two different orderings: $\avg{1ab\tilde{1}}$ and $\avg{a1b\tilde{1}}$,
which are $\mV_{1}(u)\mV_{1}(\bar{u})$ and $\mV_{2}(u)\mV_{1}(\bar{u})$
respectively. For $H$ type contraction, there are four different
orderings: $\avg{12ab}$, $\avg{1a2b}$, $\avg{2a1b}$ and $\avg{a21b}$
which are $\mV_{1}(\mu)\mV_{1}(\bar{\mu})$, $\mV_{-1}(\mu)\mV_{1}(\bar{\mu})$,
$\mV_{2}(-\mu)\mV_{1}(-\bar{\mu})$ and $\mV_{1}(-\mu)\mV_{1}(-\bar{\mu})$
respectively (see Fig. \ref{fig:mu}). For $\tilde{H}$ type contraction,
there is only one ordering: $\avg{ab\tilde{2}\tilde{1}}$, which is
$\mV_{1}(\mu)\mV_{1}(\bar{\mu})$. Here the subscript $i$ of $\mV_{i}$
means the value on $i$-th sheet given by \eqref{sv1}, \eqref{sv2} and 
\begin{equation}
\mV_{-1}(u)=\left(\f{\a(-u)}{\sqrt{1-u}}\f 1{(1-u)^{-\a/2}e^{-i\pi\a}-(1-u)^{\a/2}e^{i\pi\a}}\right)^{2h_{\mO}}
\end{equation}
In all above statements, the cross ratios are $u=u_1$ in \eqref{aa2a} with $x_1=0$ and 
\be
\mu=\f{2i\sin \f{\pi\e}{\b}}{\sinh\f{2\pi}{\b}(t-|x|)+2i\sin \f{\pi\e}{\b}},\quad\bar{\mu}=\f{2i\sin \f{\pi\e}{\b}}{\sinh\f{2\pi}{\b}(t+|x|)+2i\sin \f{\pi\e}{\b}},\quad 0<\e<\b
\ee
where $\e$ is the difference of $\e$-prescription in the time ordering of $\mO_1$ and $\mO_2$
and it now plays a role as UV regulator \footnote{In order to evaluate the Lorentzian correlation function in \eqref{bn} one has to assign $\e_i$ for each $\mO_i$ before continuation. Since all the $\sO$'s commute, the relative values of their $\ep$'s do not matter. 
Therefore we can assign $\e$ ordering for each pairing such that $\e$ for each Virasoro block is the same. Physically it is natural to take $\e$ of order $O(\b)$, and in the main body of this paper, we will choose $\e=\b/2$ for definiteness.}.

For any given choice of contractions, the number of $H$ contraction
must be the same as $\tilde{H}$. To be more precise, with $q$ contractions
of $H$, then number of $\tilde{H}$ is $q$ and the number of $G$
is $n-2q$. In such a contraction, the scaling of \eqref{bn} is $H^{q}\tilde{H}^{q}G^{n-2q}$.
If we track all commutators in $W_{n}$, we will find that the contribution
from this contraction is
\begin{align}
W_{n}/\avg{J_{a}J_{b}}\supset & C_{q}H^{q}\left[(\mV_{1}(\mu)-\mV_{-1}(\mu))\mV_{1}(\bar{\mu})+(\mV_{1}(-\mu)-\mV_{2}(-\mu))\mV_{1}(-\bar{\mu})\right]^{q}\nonumber \\
 & \times\tilde{H}^{q}\left[\mV_{1}(\mu)\mV_{1}(\bar{\mu})\right]^{q}G^{n-2q}\left[(\mV_{1}(u)-\mV_{2}(u))\mV_{1}(\bar{u})\right]^{n-2q}\label{eq:wn5}
\end{align}
where $C_{q}$ is a constant. A consistent check is that in (\ref{eq:wn5})
there are in total $4^{q}\cdot2^{n-2q}=2^{n}$ terms, which is the
same number of terms in the expansion of commutators (\ref{jh}).
In general, $C_{q}$ depends on how we choose the $q$ contractions.
However, the structure of conformal blocks does not depend on the
detail of the $q$ contraction choices. This implies a simple counting
rule for $W_{n}$: we only need to track the contractions of $H$,
$\tilde{H}$ and $G$, then replace all $H\tilde{H}$ as $H_{0}^{2}$
and $G$ as $G_{0}$, where we define 
\begin{align}
H_{0}^{2} & =H\tilde{H}\left[(\mV_{1}(\mu)-\mV_{-1}(\mu))\mV_{1}(\bar{\mu})+(\mV_{1}(-\mu)-\mV_{2}(-\mu))\mV_{1}(-\bar{\mu})\right]\mV_{1}(\mu)\mV_{1}(\bar{\mu})\\
G_{0} & =G(\mV_{1}(u)-\mV_{2}(u))\mV_{1}(\bar{u})
\end{align}
Hence, evaluation of $W_{n}$ boils down to the problem of calculating
$C_{q}$ for various contractions. In the following, since $H=\tilde{H}$
by definition, we will not distinguish them and only use $H$.

To count $C_{q}$ is somewhat tricky. Due to the factor $1/k^n$ in $W_n$, any $\mO_{\a}$ contracting with its dual $\tilde{\mO}_{\a}$ contributes with $\sum_\a \d_{\a\a}/k=1$, but contracting with any other operators contributes with only $1/k$. Keeping this in mind, we will use the following
procedure: 
\begin{enumerate}
\item For $n$ pairs of $\mO_{\a_{s}}$ and $\tilde{\mO}_{\a_{s}}$, we
define the sum of all possible contractions as $S_{n}$. Starting
from $\mO_{\a_{1}}$, there are two types of contraction: i) contracting
$\tilde{\mO}_{\a_{1}}$ gives $G$; ii) $n-1$ choices of contracting
with $\mO_{\a_{i}}$ with $i\neq1$ in total give $(n-1)H/k$; iii)
$n-1$ contracting with $\tilde{\mO}_{\a_{i}}$ with $i\neq1$ in
total give $(n-1)G/k$. For i), we call it as completing a chain.
Then redo step 1 with the next $\mO$, say $\mO_{\a_{2}}$. For ii),
do step 2. For iii), do step 3.
\item For ii) in step 1, continue with its counterpart $\tilde{\mO}_{\a_{i}}$.
We define the sum of all possible contractions in such case, namely
$n-2$ pairs of $\mO_{\a_{s}}$ and $\tilde{\mO}_{\a_{s}}$ and two
$\tilde{\mO}$'s (namely $\tilde{\mO}_{\a_{1}}$ and $\tilde{\mO}_{\a_{i}}$),
as $T_{n-2}$. We have three choices: i) contracting with $\tilde{\mO}_{\a_{1}}$
completes a chain and gives $H$; ii) $n-2$ choices of contracting
with $\mO_{\a_{j}}$ with $j\neq1,i$ in total give $(n-2)G/k$;
iii) $n-2$ contracting with $\tilde{\mO}_{\a_{j}}$ with $j\neq1,i$
in total give $(n-2)H/k$.
\item For iii) in step 1, continue with its counterpart $\mO_{\a_{i}}$.
We define the sum of all possible contractions in such case, namely
$n-2$ pairs of $\mO_{\a_{k}}$ and $\tilde{\mO}_{\a_{k}}$ plus one
pair of $\mO_{\a_{1}}$ and $\tilde{\mO}_{\a_{i}}$ with different
indicies, as $U_{n-2}$. We have three choices: i) contracting with
$\tilde{\mO}_{\a_{1}}$ completes a chain and gives $G$; ii)
$n-2$ choices of contracting with $\mO_{\a_{j}}$ with $j\neq1,i$
in total give $(n-2)H/k$; iii) $n-2$ contracting with $\tilde{\mO}_{\a_{j}}$
with $j\neq1,i$ in total give $(n-2)G/k$.
\item In step 1, completing a chain becomes $S_{n-1}$. In step 2, i) becomes
$S_{n-2}$, ii) becomes $T_{n-3}$, and iii) becomes $U_{n-3}$. In
step 3, i) becomes $S_{n-2}$, ii) becomes $T_{n-3}$, and iii) becomes
$U_{n-3}$. For any $S_{\#}$ cases, do step 1, for any $T_{\#}$
cases, do step 2, and for any $U_{\#}$ cases, do step 3. This process
ends when we finish all contractions.
\end{enumerate}
By above constructive procedure, we have the following induction relations:
\begin{align}
S_{n} & =GS_{n-1}+(n-1)G/kU_{n-2}+(n-1)H/kT_{n-2}\label{eq:Sn}\\
U_{n} & =GS_{n}+nG/kU_{n-1}+nH/kT_{n-1}\label{eq:Un}\\
T_{n} & =HS_{n}+nG/kT_{n-1}+nH/kU_{n-1}\label{eq:Tn}
\end{align}
Comparing (\ref{eq:Sn}) with (\ref{eq:Un}), we find $S_{n}=U_{n-1}$.
Plug this back and cancel $T_{n}$. The induction relation of $S_{n}$
turns out to be
\begin{equation}
S_{n+2}=G\left(1+\f{2(n+1)}k\right)S_{n+1}+(H^{2}-G^{2})\f{n+1}k\left(1+\f nk\right)S_{n}\label{eq:indS}
\end{equation}

To solve this induction relation, define $\g=H^{2}/G^{2}-1$ and $S_{n}\ra G^{n}S_{n}$.
(\ref{eq:indS}) becomes
\begin{equation}
S_{n+2}=\left(1+\f{2(n+1)}k\right)S_{n+1}+\f{n+1}k\left(1+\f nk\right)\g S_{n}\label{eq:inds}
\end{equation}
We can check first a few terms explicitly starting with $S_{1}=1$
and $S_{0}=1$, which suggests that $S_{n}$ should be a polynomial
like
\begin{equation}
S_{n}=\sum_{i=0}^{[n/2]}a_{i}(n)\g^{i}\label{eq:snseris}
\end{equation}
Taking this ansatz, the induction splits into even and odd cases of
$n$. Plug this into (\ref{eq:inds}) and compare the coefficients
of $\g^{i}$. For $n=2p$,
\begin{align}
a_{i}(2p+2) & =\f{4p+k+2}ka_{i}(2p+1)+\f{(2p+1)(2p+k)}{k^{2}}a_{i-1}(2p),\;i=1,\cdots,p\label{eq:a-1}\\
a_{p+1}(2p+2) & =\f{(2p+1)(2p+k)}{k^{2}}a_{p}(2p)\label{eq:a-2}\\
a_{0}(2p+2) & =\f{4p+k+2}ka_{0}(2p+1)\label{eq:a-3}
\end{align}
For $n=2p-1$,
\begin{align}
a_{i}(2p+1) & =\f{4p+k}ka_{i}(2p)+\f{2p(2p+k-1)}{k^{2}}a_{i-1}(2p-1),\;i=1,\cdots,p\label{eq:a-4}\\
a_{0}(2p+1) & =\f{4p+k}ka_{0}(2p)\label{eq:a-5}
\end{align}
Solving (\ref{eq:a-2}) gives
\begin{equation}\label{ap2p}
a_{p}(2p)=\f{(2p-1)!!(2p+k-2)!!}{k^{2p-1}k!!}
\end{equation}
Solving (\ref{eq:a-3}) and (\ref{eq:a-5}) gives
\begin{equation}
a_{0}(n)=\f{(2n+k-2)!!}{k^{n-1}k!!}\label{eq:a0n}
\end{equation}
Multiply (\ref{eq:a-4}) with $\f{4p+k+2}k$ and sum with (\ref{eq:a-1}).
We get
\begin{align}
a_{i}(2p+2)= & \f{(4p+k+2)(4p+k)}{k^{2}}a_{i}(2p)+\f{2p(2p+k-1)(4p+k+2)}{k^{3}}a_{i-1}(2p-1)\nonumber \\
 & +\f{(2p+1)(2p+k)}{k^{2}}a_{i-1}(2p)
\end{align}

Let us check a few $i$'s to see if any rule exists. Setting $i=1$ and using (\ref{eq:a0n}) leads to
\begin{equation}
a_{1}(2p+2)=\f{(4p+k+2)(4p+k)}{k^{2}}a_{1}(2p)+\f{(4p+k-4)!!}{k^{2p-1}k!!}\cdot\f{(4p+k)(8p^{2}+4kp+k-2)}{k^{2}}
\end{equation}
Taking the ansatz
\begin{equation}
a_{1}(2p)=\f{(4p+k-4)!!}{k^{2p-1}k!!}\bar{a}_{1}(2p)
\end{equation}
the induction is simplified as
\begin{equation}
(4p+k-2)\bar{a}_{1}(2p+2)=(k+4p+2)\bar{a}_{1}(2p)+(8p^{2}+4kp+k-2)
\end{equation}
This is very easy to solve if we assume $\bar{a}_{1}(2p)$ is a qudratic
polynomial of $p$. Using the initial condition $\bar{a}_{1}(2)=1$ from \eqref{ap2p},
we solve it as $\bar{a}_{1}(2p)=p(2p-1)$ and 
\begin{equation}
a_{1}(2p)=\f{(4p+k-4)!!}{k^{2p-1}k!!}p(2p-1)
\end{equation}
With (\ref{eq:a-4}), we also get
\begin{equation}
a_{1}(2p+1)=\f{(4p+k-2)!!}{k^{2p}k!!}p(2p+1)
\end{equation}

We can follow the induction relation again to solve $a_{2}(2p)$ and
$a_{2}(2p+1)$. It turns out that
\begin{align}
a_{2}(2p) & =\f{(4p+k-6)!!}{k^{2p-1}k!!}\cdot\f{p(p-1)}2\cdot(2p-1)(2p-3)\\
a_{2}(2p+1) & =\f{(4p+k-4)!!}{k^{2p}k!!}\cdot\f{p(p-1)}2\cdot(2p+1)(2p-1)
\end{align}
It is very tempting to guess the following general formula
\begin{align}
a_{i}(2p) & =\f{(4p+k-2i-2)!!}{k^{2p-1}k!!}\cdot\f{p!}{(p-i)!i!}\cdot\f{(2p-1)!!}{(2p-2i-1)!!}\\
a_{i}(2p+1) & =\f{(4p+k-2i)!!}{k^{2p}k!!}\cdot\f{p!}{(p-i)!i!}\cdot\f{(2p+1)!!}{(2p-2i+1)!!}
\end{align}
One can easily check that this ansatz solves the induction relations
(\ref{eq:a-1})-(\ref{eq:a-5}). Substitute the solution to (\ref{eq:snseris}),
and we get
\begin{align}
S_{n} & =\f{k^{1-n}(2n+k-2)!!}{k!!}{}_{2}F_{1}(\f 12-\f n2,-\f n2,1-\f k2-n,-\g)\nonumber \\
 & =\f{k^{1-n}(2n+k-2)!!}{k!!}(1+\g)^{n/2}{}_{2}F_{1}(-n,1-k-n,1-\f k2-n,\f{\sqrt{1+\g}-1}{2\sqrt{1+\g}})\nonumber \\
 & =(k)_{n}\left(\f{1+\sqrt{1+\g}}k\right)^{n}{}_{2}F_{1}(-n,\f k2,k,\f{2\sqrt{1+\g}}{\sqrt{1+\g}+1})\label{eq:32}
\end{align}
where in the second line we used identity \cite{Prudnikov1992}
\begin{equation}
_{2}F_{1}(a,a+\f 12,c,z)=(1-z)^{-a}{}_{2}F_{1}(2a,2c-2a-1,c,\f{\sqrt{1-z}-1}{2\sqrt{1-z}})
\end{equation}
and in the third line we used the reduction formula
\begin{equation}
_{2}F_{1}(-n,b,c,z)=\f{(b)_{n}}{(c)_{n}}(1-z)^{n}{}_{2}F_{1}(-n,c-b,1-b-n,\f 1{1-z})
\end{equation}
Indeed, given (\ref{eq:32}), (\ref{eq:inds}) is exactly the contiguous
relation of hypergeometric function
\begin{equation}
(c-a)_{2}F_{1}(a-1,b,c;z)+(2a-c+(b-a)z)_{2}F_{1}(a,b,c;z)+a(z-1){}_{2}F_{1}(a+1,b,c;z)=0
\end{equation}
for $a=-n-1$, $b=k/2$, $c=k$ and $z=\f{2\sqrt{1+\g}}{\sqrt{1+\g}+1}$.

Note that $W_{n}=G^{n}S_{n}$ with replacement $H^{2}\ra H_{0}^{2}$
and $G\ra G_{0}$. It is ready to calculate $W$ \footnote{The convergence of series requires $|g(G_0\pm H_0)/k|<1$. For large $k\ra\infty$ any $g$ is allowed, but in $k=1$ $g$ is bounded above. However, we will treat $W$ as an analytic function of $t$ and $t_s$. Note that in small $t$,$t_s$ case, $G_0$ and $H_0$ are close to zero, which releases the upper bound of $g$.}
\begin{equation}
W/\avg{J\tilde{J}}=\sum_{n=0}^{\infty}\f 1{n!}(-ig)^{n}G_{0}^{n}S_{n}=\left(1+\f{ig(G_{0}+H_{0})}k\right)^{-k/2}\left(1+\f{ig(G_{0}-H_{0})}k\right)^{-k/2}\label{eq:W-gh}
\end{equation}
where we used the fact
\begin{equation}
\sum_{n=0}^{\infty}\f{(c-a)_{n}}{n!}t^{n}{}_{2}F_{1}(a-n,b,c,x)=\f{(1-t)^{a-c}}{(1-x)^{b}}{}_{2}F_{1}(c-a,b,c,\f x{(1-x)(t-1)})
\end{equation}

\section{Robustness of regenesis} \label{app:rob}

For the deformed thermofield double state $\g^L(t_0,x_0)\bra{\Psi}$, we will calculate $G^{LR}(t,t_s)$ in this appendix to see how the regenesis is affected.

The left and right two point function in this deformed state is
\be
W_\g\equiv{\ket{\Psi}\g^L(t_0)e^{-igV}J^L(t)e^{igV}J^R(-t_s)\g^L(t_0)\bra{\Psi}\ov \ket{\Psi}\g^L(t_0)\g^L(t_0)\bra{\Psi}}
\ee
where the spatial coordinates are suppressed. Similar to \eqref{ma00}, using BCH formula, it becomes
\be 
W_\g=\sum_{n=0}^{\infty}\f{(-ig)^n}{n!L^n}\int_{-{L\ov 2}}^{L\ov 2} \left(\prod_{k=1}^n dx_k\right) W_{\g,n}
\ee
with 
\be \label{wgn}
W_{\g,n}={\avg{\g[\mO_n,[\mO_{n-1},\cdots[\mO_1,J]\cdots]\g\tilde{J}\tilde{\mO}_n\cdots\tilde{\mO}_1}_\b\ov \avg{\g\g}_\b}
\ee
where each term in the commutator is 
\be \label{a32}
w_{\g,n}\equiv\f{\avg{\g_A\g_BJ_aJ_b\mO_1\cdots\mO_{2n}}}{\avg{\g_A\g_B}}
\ee
in which the coordinate of $\g$ is $z_A$ and $z_B$. 

Assuming $c\gg h_\g\gg h_J$ , we could first do a conformal transformation with respect to $\g^L$ by introducing a branch cut from $z_A$ to $z_B$
\begin{equation}
1-v=\left(1-\f{z_{AB}z}{z_{B}(z_{A}-z)}\right)^{\eta},\;\eta=\sqrt{1-24\f{h_{ \g}}c}\label{eq:trans-v}
\end{equation}
with a Jacobian
\begin{equation}
\hJ(z)\equiv\f{\p z}{\p v}=\f{(z-z_{A})(z-z_{B})}{\eta z_{AB}}\left[\f{z_{A}(z-z_{B})}{z_{B}(z-z_{A})}\right]^{-\eta}
\end{equation}
Similar to \eqref{eq:49}, \eqref{a32} becomes
\begin{align}
w_{\g,n}&=\f{\hJ_{a}^{-h_{\g}}\hJ_{b}^{-h_{\g}}}{\avg{\g_A \g_B }_{z}}\left(\hJ_a^{-h_J}\hJ_b^{-h_J}\prod_{i=1}^{2n}\hJ_{i}^{-h_{\mO}}\right)\avg{\g_A \g_B J_{a} J_{b}\mO_{1}\cdots\mO_{2n}}_{v}\nonumber \\
 & \app\left(\hJ_a^{-h_J}\hJ_b^{-h_J}\prod_{i=1}^{2n}\hJ_{i}^{-h_{\mO}}\right) \avg{0_{v}|J_a J_b \mO_{1}\cdots\mO_{2n}}_{v}
\end{align}
For the expectation value in $v$ plane $\avg{0_{v}|J_a J_b \mO_{1}\cdots\mO_{2n}}_{v}$, since $\ket{0_v}$ behaves exactly as ordinary vacuum state, we can apply the same technique to simply it. Previous result \eqref{eq:49} applies with a simple coordinate replacement $z\ra v$. After some manipulation, it turns out that
\begin{align}\label{wgan-1}
w_{\g,n}\app\avg{J_a J_b}_z \mU_\eta(u_J)\sum_{\{(s_{2i},s_{2i+1})\}}\prod_{i=1}^{n}\left[\mV_\eta(u_{s,i})\mV_\a(r_{s,i})\avg{\mO_{s_{2i}}\mO_{s_{2i+1}}}_{z}\right]
\end{align}
where 
\be \label{a36}
\mU_\eta(u)=\left(\f{\eta^{2}u^{2}(1-u)^{\eta-1}}{(1-(1-u)^{\eta})^{2}}\right)^{h_{J}},\quad
\mV_\eta(u)=\left(\f{\eta^{2}u^{2}(1-u)^{\eta-1}}{(1-(1-u)^{\eta})^{2}}\right)^{h_{\mO}}
\ee
and
\be\label{wgan-3}
u_J\equiv\f{z_{ab}z_{AB}}{z_{aA}z_{bB}}, \quad
u_{s,i}\equiv\f{z_{s_{2i},s_{2i+1}}z_{AB}}{z_{s_{2i},A}z_{s_{2i+1},B}}, \quad
r_{s,i}\equiv\f{v_{s_{2i},s_{2i+1}}v_{ab}}{v_{s_{2i},a}v_{s_{2i+1},b}}
\ee
where the new variable $v=v(z)$ is given by \eqref{eq:trans-v}. The antiholomorphic part is parallel with above discussion.

It is clear that from \eqref{wgn}, the relative ordering among $\g,~J$ and $\tilde{J}$ is always fixed as $\avg{\g_A J\g_B\tilde{J}}$, and that among $\g,~\mO$ and $\tilde{\mO}$ is also fixed as $\avg{\g_A\mO_i\g_B\tilde{\mO}_i}$. The commutators in \eqref{wgn} will only be affected by the relative ordering among $\mO,~J$ and $\tilde{J}$, and this leads \eqref{wgn} to
\be
W_{\g,n}=\avg{J_a J_b}\mU_\eta(u_{ab})\mU_\eta(\bar{u}_{ab})\sum_{\text{all pairings}}\prod_{i=1}^{n}\left[ \mV_\eta(u_i)\mV_\eta(\bar{u}_i)A_\a(r_i,\bar{r}_i)\avg{\mO_{i1}\mO_{i2}}\right]
\ee
where $A_\a$ is the difference of two orderings $\avg{\mO J \tilde{J}\tilde{\mO}}$ and $\avg{J \mO \tilde{J}\tilde{\mO}}$
\be 
A_\a(r_i,\bar{r}_i)\equiv \mV^+_\a (r_i)\mV^+_\a(\bar{r}_i)-\mV^-_\a (r_i)\mV^-_\a(\bar{r}_i)
\ee
It follows that for large $k$ and large $L$ cases, we have
\be 
\f{W_\g}{\avg{J_a J_b}}=\mU_\eta(u_{J})\mU_\eta(\bar{u}_{J}) \times
\begin{cases}
\exp\left(-igG \mV_\eta(u_0)\mV_\eta(\bar{u}_0)A_\a(r_0,\bar{r}_0)\right)& \text{large $k$}\\
\exp\left(-\f{igG}{L} \int_{-L/2}^{L/2} dx_1\mV_\eta(u_1)\mV_\eta(\bar{u}_1)A_\a(r_1,\bar{r}_1)\right)& \text{large $L$}\\
\end{cases}\label{Wg3cases}
\ee
where subscript 0 means taking $x_1=0$. In these formula, $\mV_\eta$ can be absorbed into the definition of coefficient $g$ (at least in large $k$ case), and the prefactor $\mU_\eta$ controls the overall traversability. In the main body of this paper, we use a simpler notation $\mJ(t,t_0)\equiv\mU_\eta(u_{J})\mU_\eta(\bar{u}_{J})$ and $\mG(t,t_0;x_1)\equiv\mV_\eta(u_1)\mV_\eta(\bar{u}_1)$.

To be more precise, let us study the value of these conformal blocks in $\e$-prescription. For $\mU_\eta$ and $\mV_\eta$, we assign $\e_A<\{\e_J,\e_1\}<\e_B$ and any positive small value for $\tilde{\e}_J$ and $\tilde{\e}_1$ in
\begin{align}
u_{J}&=-{e^{\f{2\pi}{\b}(t_0+x_0)}(e^{\f{2\pi}{\b}(x+t)+i\e_J}+e^{\f{2\pi}{\b}(x_s+t_s)+i\tilde{\e}_J})(e^{i\e_A}-e^{i\e_B})\ov(e^{\f{2\pi}{\b}(x+t)+i\e_J}-e^{\f{2\pi}{\b}(x_0+t_0)+i\e_A})(e^{\f{2\pi}{\b}(x_s+t_s)+i\tilde{\e}_J}+e^{\f{2\pi}{\b}(x_0+t_0)+i\e_B})} \label{uab}\\
\bar{u}_{J}&=-{e^{\f{2\pi}{\b}(x_0-t_0)}(e^{\f{2\pi}{\b}(x-t)-i\e_J}+e^{\f{2\pi}{\b}(x_s-t_s)-i\tilde{\e}_J})(e^{-i\e_A}-e^{-i\e_B})\ov(e^{\f{2\pi}{\b}(x-t)-i\e_J}-e^{\f{2\pi}{\b}(x_0-t_0)-i\e_A})(e^{\f{2\pi}{\b}(x_s-t_s)-i\tilde{\e}_J}+e^{\f{2\pi}{\b}(x_0-t_0)-i\e_B})}\label{ubab}
\end{align}
Similar for $u_1$ and $\bar{u}_1$ by setting $x=x_s=x_1$ and $t=t_s=0$ with corresponding $\e$. In general, we need to take all $\e$'s to zero at the end of calculation, which seemingly leads to trivial $u_J$ and $\bar{u}_J$. However, we could smear it for some small range in time for $\g^L$ that equivalently sets $\e_{AB}$ small but finite. For $t_s-t_0-|x_s-x_0|,~t-t_0-|x-x_0|\gg \b$, it is clear from \eqref{uab} and \eqref{ubab} that 
\be\label{uabgamma-1}
u_{J}\ra -i(e^{-\f{2\pi}{\b}(x+t)}+e^{-\f{2\pi}{\b}(x_s+t_s)})e^{\f{2\pi}{\b}(t_0+x_0)}\e_{AB},\;
\bar{u}_{J}\ra -i(e^{\f{2\pi}{\b}(x-t)}+e^{\f{2\pi}{\b}(x_s-t_s)})e^{\f{2\pi}{\b}(t_0-x_0)}\e_{AB}
\ee
Similarly for $t_0-t_s-|x_s-x_0|,~t_0-t-|x-x_0|\gg \b$, we have 
\be\label{uabgamma-2}
u_{J}\ra i(e^{\f{2\pi}{\b}(x+t)}+e^{\f{2\pi}{\b}(x_s+t_s)})e^{-\f{2\pi}{\b}(t_0+x_0)}\e_{AB},\quad
\bar{u}_{J}\ra i(e^{\f{2\pi}{\b}(t-x)}+e^{\f{2\pi}{\b}(t_s-x_s)})e^{\f{2\pi}{\b}(x_0-t_0)}\e_{AB}
\ee
This shows that for finite $\e_{AB}$, $u_J$ and $\bar{u}_J$ are suppressed by $|t_0-t_s|$ and $|t_0-t|$ exponentially.

Set $t=t_s$ and $x=x_s$, and check the contour of $u_{J}$ and $\bar{u}_J$. For simplicity we will take $x_0-x>0$  from now on. We find that $u_J$ is on $-1$-th sheet when $t-t_0>x_0-x$, and on first sheet when $t-t_0<x_0-x$. On the other hand, $\bar{u}_J$ is on second sheet when $t_0-t>x_0-x$, and on first sheet when $t_0-t<x_0-x$. The value of $\mU_\eta$ on either sheet is given by 
\begin{align}
\mU_{\eta,-1}(u)&=\left( \frac { \eta ( - u ) } { \sqrt { 1- u } } \frac { 1} { ( 1- u ) ^ { - \eta / 2} e ^ { -i \pi \eta } - ( 1- u ) ^ { \eta / 2} e ^ {  i \pi \eta } } \right) ^ { 2h_J}\\
\mU_{\eta,1}(u)&=\left( \frac { \eta ( - u ) } { \sqrt { 1- u } } \frac { 1} { -( 1- u ) ^ { - \eta / 2} + ( 1- u ) ^ { \eta / 2} } \right) ^ { 2h_J }\\
\mU_{\eta,2}(u)&=\left( \frac { \eta ( - u ) } { \sqrt { 1- u } } \frac { 1} { ( 1- u ) ^ { - \eta / 2} e ^ { i \pi \eta } - ( 1- u ) ^ { \eta / 2} e ^ {  -i \pi \eta } } \right) ^ { 2h_J}
\end{align}
Taking large $c$ but $cu_J, c\bar{u}_J$ fixed limit and using \eqref{uabgamma-1} and \eqref{uabgamma-2}, we see that
\be \label{Ueta}
\mU_{\eta}(u_J)\mU_{\eta}(\bar{u}_J)\ra 
\begin{cases}
\left(1-\f{2 h_\g }{
\e_{AB}}Q_J \right)^{-2h_J},\quad &\text{for } |t_0-t|-|x_0-x|\gg\b\\
1,\quad &\text{for } |x_0-x|-|t_0-t|\gg\b
\end{cases}
\ee
where 
\be
Q_J=e^{\f{2\pi}{\b}(|t-t_0|-t_*-|x_0-x|)}
\ee
Similar results applies to $u_1$ and $\bar{u}_1$ in $\mV_\eta$, which in the same limit gives
\be \label{Veta}
\mV_{\eta}(u_1)\mV_{\eta}(\bar{u}_1)\ra 
\begin{cases}
\left(1-\f{2 h_\g }{
\e_{AB}}Q_1 \right)^{-2h_\mO},\quad &\text{for } |t_0|-|x_0-x_1|\gg\b\\
1,\quad &\text{for } |x_0-x_1|-|t_0|\gg\b
\end{cases}
\ee
where 
\be
Q_1=e^{\f{2\pi}{\b}(|t_0|-t_*-|x_0-x_1|)}
\ee
Since $\e_{AB}<0$, this shows that \eqref{Ueta}
and \eqref{Veta} are monotonically decreasing real function of $t$. 

For the contour of $r_1$ and $\bar{r}_1$, one should first track 
the contour of various $z$ coordinates in $v$ plane. In the following we will only consider $x_1<x<x_0$ \footnote{For the case $x_0$ is in between $x_1$ and $x$ (e.g. $x_1<x_0<x$), one will get trivial $A_\a$ that is not continuous in $x_0$ to the case $x_1<x<x_0$. This means that our $s$-channel approximation does not hold in such a case.}. Since the $z\ra v(z)$ map has singularity at $z_A$, we set the ordering of $\e$'s to be $\e_A<\e_1<\e_J<\e_B$ and $\tilde{\e}_J<\tilde{\e}_1$ for $\mV^+_\a$, and switch the ordering between $\e_J$ and $\e_1$ for $\mV^-_\a$. Start with $t=t_0=0$ and send $t$ and $t_0$ to various value. In this process, for both orderings, we find that $v(z_a)$ and $v(z_1)$ moves to $-1$-th sheet when $t-t_0>x_0-x$ and $-t_0>x_0-x_1$ respectively, and $v(\bar{z}_a)$ and $v(\bar{z}_1)$ moves to second sheet when $t_0-t>x_0-x$ and $t_0>x_0-x_1$ respectively, in which
\be 
v_{-1}=1-e^{2\pi i\eta}\left(1-\f{z_{AB}z}{z_{B}(z_{A}-z)}\right)^{\eta},\quad v_{2}=1-e^{-2\pi i\eta}\left(1-\f{z_{AB}z}{z_{B}(z_{A}-z)}\right)^{\eta}
\ee
while $v(z_b)$, $v(z_{\tilde{1}})$ and their antiholomorphic parts are all on the first sheet. 

Take these values of $v$ in $r$ plane, we find that for $\mV^+$ both $r_1$ and $\bar{r}_1$ stay in the first sheet, whereas for $\mV^-$ $r_1$ stays in first sheet but $\bar{r}_1$ moves to second sheet when $t>x-x_1$  just like ordinary case. This leads to
\be\label{Aa-1}
A_\a(r_1,\bar{r}_1)=\mV_{\a,1}(r_1)(\mV_{\a,1}(\bar{r}_1)-\mV_{\a,2}(\bar{r}_1)), \quad \text{for } t>x-x_1
\ee
Taking large $c$ and $e^t/c$ fixed limit for $|t_0|\ll t_*$ and $|t-t_0|\ll t_*$ case, one can show that
\be 
A_\a\sim 1- \le(1 + i Q \ri)^{-2 h_\sO}
\ee
where $Q$ is the same as \eqref{eq:df-Q}. For general range of $t_0$, the calculation of $A_\a$ becomes tricky and we are not completely sure how to calculate it reliably in the $s$-channel approximation. Physically, we should expect $|A_\a|$ is bounded as a $O(1)$ function.

\bibliography{references}

\begin{thebibliography}{35}%
\makeatletter
\providecommand \@ifxundefined [1]{%
 \@ifx{#1\undefined}
}%
\providecommand \@ifnum [1]{%
 \ifnum #1\expandafter \@firstoftwo
 \else \expandafter \@secondoftwo
 \fi
}%
\providecommand \@ifx [1]{%
 \ifx #1\expandafter \@firstoftwo
 \else \expandafter \@secondoftwo
 \fi
}%
\providecommand \natexlab [1]{#1}%
\providecommand \enquote  [1]{``#1''}%
\providecommand \bibnamefont  [1]{#1}%
\providecommand \bibfnamefont [1]{#1}%
\providecommand \citenamefont [1]{#1}%
\providecommand \href@noop [0]{\@secondoftwo}%
\providecommand \href [0]{\begingroup \@sanitize@url \@href}%
\providecommand \@href[1]{\@@startlink{#1}\@@href}%
\providecommand \@@href[1]{\endgroup#1\@@endlink}%
\providecommand \@sanitize@url [0]{\catcode `\\12\catcode `\$12\catcode
  `\&12\catcode `\#12\catcode `\^12\catcode `\_12\catcode `\%12\relax}%
\providecommand \@@startlink[1]{}%
\providecommand \@@endlink[0]{}%
\providecommand \url  [0]{\begingroup\@sanitize@url \@url }%
\providecommand \@url [1]{\endgroup\@href {#1}{\urlprefix }}%
\providecommand \urlprefix  [0]{URL }%
\providecommand \Eprint [0]{\href }%
\providecommand \doibase [0]{http://dx.doi.org/}%
\providecommand \selectlanguage [0]{\@gobble}%
\providecommand \bibinfo  [0]{\@secondoftwo}%
\providecommand \bibfield  [0]{\@secondoftwo}%
\providecommand \translation [1]{[#1]}%
\providecommand \BibitemOpen [0]{}%
\providecommand \bibitemStop [0]{}%
\providecommand \bibitemNoStop [0]{.\EOS\space}%
\providecommand \EOS [0]{\spacefactor3000\relax}%
\providecommand \BibitemShut  [1]{\csname bibitem#1\endcsname}%
\let\auto@bib@innerbib\@empty
\bibitem [{\citenamefont {Gao}\ \emph {et~al.}(2017)\citenamefont {Gao},
  \citenamefont {Jafferis},\ and\ \citenamefont {Wall}}]{Gao:2016bin}%
  \BibitemOpen
  \bibfield  {author} {\bibinfo {author} {\bibfnamefont {P.}~\bibnamefont
  {Gao}}, \bibinfo {author} {\bibfnamefont {D.~L.}\ \bibnamefont {Jafferis}}, \
  and\ \bibinfo {author} {\bibfnamefont {A.}~\bibnamefont {Wall}},\ }\href
  {\doibase 10.1007/JHEP12(2017)151} {\bibfield  {journal} {\bibinfo  {journal}
  {JHEP}\ }\textbf {\bibinfo {volume} {12}},\ \bibinfo {pages} {151} (\bibinfo
  {year} {2017})},\ \Eprint {http://arxiv.org/abs/1608.05687} {arXiv:1608.05687
  [hep-th]} \BibitemShut {NoStop}%
\bibitem [{\citenamefont {Maldacena}\ \emph {et~al.}(2017)\citenamefont
  {Maldacena}, \citenamefont {Stanford},\ and\ \citenamefont
  {Yang}}]{Maldacena:2017axo}%
  \BibitemOpen
  \bibfield  {author} {\bibinfo {author} {\bibfnamefont {J.}~\bibnamefont
  {Maldacena}}, \bibinfo {author} {\bibfnamefont {D.}~\bibnamefont {Stanford}},
  \ and\ \bibinfo {author} {\bibfnamefont {Z.}~\bibnamefont {Yang}},\ }\href
  {\doibase 10.1002/prop.201700034} {\bibfield  {journal} {\bibinfo  {journal}
  {Fortsch. Phys.}\ }\textbf {\bibinfo {volume} {65}},\ \bibinfo {pages}
  {1700034} (\bibinfo {year} {2017})},\ \Eprint
  {http://arxiv.org/abs/1704.05333} {arXiv:1704.05333 [hep-th]} \BibitemShut
  {NoStop}%
\bibitem [{\citenamefont {Susskind}(2016{\natexlab{a}})}]{Susskind:2014yaa}%
  \BibitemOpen
  \bibfield  {author} {\bibinfo {author} {\bibfnamefont {L.}~\bibnamefont
  {Susskind}},\ }\href {\doibase 10.1002/prop.201500094} {\bibfield  {journal}
  {\bibinfo  {journal} {Fortsch. Phys.}\ }\textbf {\bibinfo {volume} {64}},\
  \bibinfo {pages} {72} (\bibinfo {year} {2016}{\natexlab{a}})},\ \Eprint
  {http://arxiv.org/abs/1412.8483} {arXiv:1412.8483 [hep-th]} \BibitemShut
  {NoStop}%
\bibitem [{\citenamefont {Susskind}(2016{\natexlab{b}})}]{Susskind:2016jjb}%
  \BibitemOpen
  \bibfield  {author} {\bibinfo {author} {\bibfnamefont {L.}~\bibnamefont
  {Susskind}},\ }\href {\doibase 10.1002/prop.201600036} {\bibfield  {journal}
  {\bibinfo  {journal} {Fortsch. Phys.}\ }\textbf {\bibinfo {volume} {64}},\
  \bibinfo {pages} {551} (\bibinfo {year} {2016}{\natexlab{b}})},\ \Eprint
  {http://arxiv.org/abs/1604.02589} {arXiv:1604.02589 [hep-th]} \BibitemShut
  {NoStop}%
\bibitem [{\citenamefont {Susskind}\ and\ \citenamefont
  {Zhao}(2018)}]{Susskind:2017nto}%
  \BibitemOpen
  \bibfield  {author} {\bibinfo {author} {\bibfnamefont {L.}~\bibnamefont
  {Susskind}}\ and\ \bibinfo {author} {\bibfnamefont {Y.}~\bibnamefont
  {Zhao}},\ }\href {\doibase 10.1103/PhysRevD.98.046016} {\bibfield  {journal}
  {\bibinfo  {journal} {Phys. Rev.}\ }\textbf {\bibinfo {volume} {D98}},\
  \bibinfo {pages} {046016} (\bibinfo {year} {2018})},\ \Eprint
  {http://arxiv.org/abs/1707.04354} {arXiv:1707.04354 [hep-th]} \BibitemShut
  {NoStop}%
\bibitem [{\citenamefont {Takahasi}\ and\ \citenamefont
  {Umezawa}(1975)}]{Takahasi:1974zn}%
  \BibitemOpen
  \bibfield  {author} {\bibinfo {author} {\bibfnamefont {Y.}~\bibnamefont
  {Takahasi}}\ and\ \bibinfo {author} {\bibfnamefont {H.}~\bibnamefont
  {Umezawa}},\ }\href@noop {} {\bibfield  {journal} {\bibinfo  {journal}
  {Collect. Phenom.}\ }\textbf {\bibinfo {volume} {2}},\ \bibinfo {pages} {55}
  (\bibinfo {year} {1975})},\ \bibinfo {note} {[,303(1974)]}\BibitemShut
  {NoStop}%
\bibitem [{\citenamefont {Shenker}\ and\ \citenamefont
  {Stanford}(2014)}]{Shenker:2013pqa}%
  \BibitemOpen
  \bibfield  {author} {\bibinfo {author} {\bibfnamefont {S.~H.}\ \bibnamefont
  {Shenker}}\ and\ \bibinfo {author} {\bibfnamefont {D.}~\bibnamefont
  {Stanford}},\ }\href {\doibase 10.1007/JHEP03(2014)067} {\bibfield  {journal}
  {\bibinfo  {journal} {JHEP}\ }\textbf {\bibinfo {volume} {03}},\ \bibinfo
  {pages} {067} (\bibinfo {year} {2014})},\ \Eprint
  {http://arxiv.org/abs/1306.0622} {arXiv:1306.0622 [hep-th]} \BibitemShut
  {NoStop}%
\bibitem [{\citenamefont {Maldacena}\ \emph {et~al.}(2016)\citenamefont
  {Maldacena}, \citenamefont {Shenker},\ and\ \citenamefont
  {Stanford}}]{Maldacena:2015waa}%
  \BibitemOpen
  \bibfield  {author} {\bibinfo {author} {\bibfnamefont {J.}~\bibnamefont
  {Maldacena}}, \bibinfo {author} {\bibfnamefont {S.~H.}\ \bibnamefont
  {Shenker}}, \ and\ \bibinfo {author} {\bibfnamefont {D.}~\bibnamefont
  {Stanford}},\ }\href {\doibase 10.1007/JHEP08(2016)106} {\bibfield  {journal}
  {\bibinfo  {journal} {JHEP}\ }\textbf {\bibinfo {volume} {08}},\ \bibinfo
  {pages} {106} (\bibinfo {year} {2016})},\ \Eprint
  {http://arxiv.org/abs/1503.01409} {arXiv:1503.01409 [hep-th]} \BibitemShut
  {NoStop}%
\bibitem [{\citenamefont {Roberts}\ and\ \citenamefont
  {Stanford}(2015)}]{Roberts:2014ifa}%
  \BibitemOpen
  \bibfield  {author} {\bibinfo {author} {\bibfnamefont {D.~A.}\ \bibnamefont
  {Roberts}}\ and\ \bibinfo {author} {\bibfnamefont {D.}~\bibnamefont
  {Stanford}},\ }\href {\doibase 10.1103/PhysRevLett.115.131603} {\bibfield
  {journal} {\bibinfo  {journal} {Phys. Rev. Lett.}\ }\textbf {\bibinfo
  {volume} {115}},\ \bibinfo {pages} {131603} (\bibinfo {year} {2015})},\
  \Eprint {http://arxiv.org/abs/1412.5123} {arXiv:1412.5123 [hep-th]}
  \BibitemShut {NoStop}%
\bibitem [{\citenamefont {Fitzpatrick}\ \emph {et~al.}(2014)\citenamefont
  {Fitzpatrick}, \citenamefont {Kaplan},\ and\ \citenamefont
  {Walters}}]{Fitzpatrick:2014vua}%
  \BibitemOpen
  \bibfield  {author} {\bibinfo {author} {\bibfnamefont {A.~L.}\ \bibnamefont
  {Fitzpatrick}}, \bibinfo {author} {\bibfnamefont {J.}~\bibnamefont {Kaplan}},
  \ and\ \bibinfo {author} {\bibfnamefont {M.~T.}\ \bibnamefont {Walters}},\
  }\href {\doibase 10.1007/JHEP08(2014)145} {\bibfield  {journal} {\bibinfo
  {journal} {JHEP}\ }\textbf {\bibinfo {volume} {08}},\ \bibinfo {pages} {145}
  (\bibinfo {year} {2014})},\ \Eprint {http://arxiv.org/abs/1403.6829}
  {arXiv:1403.6829 [hep-th]} \BibitemShut {NoStop}%
\bibitem [{\citenamefont {Fitzpatrick}\ \emph
  {et~al.}(2015{\natexlab{a}})\citenamefont {Fitzpatrick}, \citenamefont
  {Kaplan},\ and\ \citenamefont {Walters}}]{fitzpatrick2015virasoro}%
  \BibitemOpen
  \bibfield  {author} {\bibinfo {author} {\bibfnamefont {A.~L.}\ \bibnamefont
  {Fitzpatrick}}, \bibinfo {author} {\bibfnamefont {J.}~\bibnamefont {Kaplan}},
  \ and\ \bibinfo {author} {\bibfnamefont {M.~T.}\ \bibnamefont {Walters}},\
  }\href@noop {} {\bibfield  {journal} {\bibinfo  {journal} {Journal of High
  Energy Physics}\ }\textbf {\bibinfo {volume} {2015}},\ \bibinfo {pages} {200}
  (\bibinfo {year} {2015}{\natexlab{a}})}\BibitemShut {NoStop}%
\bibitem [{\citenamefont {Fitzpatrick}\ \emph
  {et~al.}(2015{\natexlab{b}})\citenamefont {Fitzpatrick}, \citenamefont
  {Kaplan}, \citenamefont {Walters},\ and\ \citenamefont
  {Wang}}]{Fitzpatrick:2015qma}%
  \BibitemOpen
  \bibfield  {author} {\bibinfo {author} {\bibfnamefont {A.~L.}\ \bibnamefont
  {Fitzpatrick}}, \bibinfo {author} {\bibfnamefont {J.}~\bibnamefont {Kaplan}},
  \bibinfo {author} {\bibfnamefont {M.~T.}\ \bibnamefont {Walters}}, \ and\
  \bibinfo {author} {\bibfnamefont {J.}~\bibnamefont {Wang}},\ }\href {\doibase
  10.1007/JHEP09(2015)019} {\bibfield  {journal} {\bibinfo  {journal} {JHEP}\
  }\textbf {\bibinfo {volume} {09}},\ \bibinfo {pages} {019} (\bibinfo {year}
  {2015}{\natexlab{b}})},\ \Eprint {http://arxiv.org/abs/1504.01737}
  {arXiv:1504.01737 [hep-th]} \BibitemShut {NoStop}%
\bibitem [{\citenamefont {Streater}\ and\ \citenamefont
  {Wightman}(2016)}]{streater2016pct}%
  \BibitemOpen
  \bibfield  {author} {\bibinfo {author} {\bibfnamefont {R.~F.}\ \bibnamefont
  {Streater}}\ and\ \bibinfo {author} {\bibfnamefont {A.~S.}\ \bibnamefont
  {Wightman}},\ }\href@noop {} {\emph {\bibinfo {title} {PCT, spin and
  statistics, and all that}}}\ (\bibinfo  {publisher} {Princeton University
  Press},\ \bibinfo {year} {2016})\BibitemShut {NoStop}%
\bibitem [{\citenamefont {Cornalba}\ \emph
  {et~al.}(2007{\natexlab{a}})\citenamefont {Cornalba}, \citenamefont {Costa},
  \citenamefont {Penedones},\ and\ \citenamefont
  {Schiappa}}]{cornalba2007eikonal1}%
  \BibitemOpen
  \bibfield  {author} {\bibinfo {author} {\bibfnamefont {L.}~\bibnamefont
  {Cornalba}}, \bibinfo {author} {\bibfnamefont {M.~S.}\ \bibnamefont {Costa}},
  \bibinfo {author} {\bibfnamefont {J.}~\bibnamefont {Penedones}}, \ and\
  \bibinfo {author} {\bibfnamefont {R.}~\bibnamefont {Schiappa}},\ }\href@noop
  {} {\bibfield  {journal} {\bibinfo  {journal} {Journal of High Energy
  Physics}\ }\textbf {\bibinfo {volume} {2007}},\ \bibinfo {pages} {019}
  (\bibinfo {year} {2007}{\natexlab{a}})}\BibitemShut {NoStop}%
\bibitem [{\citenamefont {Cornalba}\ \emph
  {et~al.}(2007{\natexlab{b}})\citenamefont {Cornalba}, \citenamefont {Costa},
  \citenamefont {Penedones},\ and\ \citenamefont
  {Schiappa}}]{cornalba2007eikonal2}%
  \BibitemOpen
  \bibfield  {author} {\bibinfo {author} {\bibfnamefont {L.}~\bibnamefont
  {Cornalba}}, \bibinfo {author} {\bibfnamefont {M.~S.}\ \bibnamefont {Costa}},
  \bibinfo {author} {\bibfnamefont {J.}~\bibnamefont {Penedones}}, \ and\
  \bibinfo {author} {\bibfnamefont {R.}~\bibnamefont {Schiappa}},\ }\href@noop
  {} {\bibfield  {journal} {\bibinfo  {journal} {Nuclear Physics B}\ }\textbf
  {\bibinfo {volume} {767}},\ \bibinfo {pages} {327} (\bibinfo {year}
  {2007}{\natexlab{b}})}\BibitemShut {NoStop}%
\bibitem [{\citenamefont {Cornalba}\ \emph
  {et~al.}(2007{\natexlab{c}})\citenamefont {Cornalba}, \citenamefont {Costa},\
  and\ \citenamefont {Penedones}}]{cornalba2007eikonal}%
  \BibitemOpen
  \bibfield  {author} {\bibinfo {author} {\bibfnamefont {L.}~\bibnamefont
  {Cornalba}}, \bibinfo {author} {\bibfnamefont {M.~S.}\ \bibnamefont {Costa}},
  \ and\ \bibinfo {author} {\bibfnamefont {J.}~\bibnamefont {Penedones}},\
  }\href@noop {} {\bibfield  {journal} {\bibinfo  {journal} {Journal of High
  Energy Physics}\ }\textbf {\bibinfo {volume} {2007}},\ \bibinfo {pages} {037}
  (\bibinfo {year} {2007}{\natexlab{c}})}\BibitemShut {NoStop}%
\bibitem [{\citenamefont {Shenker}\ and\ \citenamefont
  {Stanford}(2015)}]{shenker2015stringy}%
  \BibitemOpen
  \bibfield  {author} {\bibinfo {author} {\bibfnamefont {S.~H.}\ \bibnamefont
  {Shenker}}\ and\ \bibinfo {author} {\bibfnamefont {D.}~\bibnamefont
  {Stanford}},\ }\href@noop {} {\bibfield  {journal} {\bibinfo  {journal}
  {Journal of High Energy Physics}\ }\textbf {\bibinfo {volume} {2015}},\
  \bibinfo {pages} {132} (\bibinfo {year} {2015})}\BibitemShut {NoStop}%
\bibitem [{\citenamefont {Camanho}\ \emph {et~al.}(2016)\citenamefont
  {Camanho}, \citenamefont {Edelstein}, \citenamefont {Maldacena},\ and\
  \citenamefont {Zhiboedov}}]{camanho2016causality}%
  \BibitemOpen
  \bibfield  {author} {\bibinfo {author} {\bibfnamefont {X.~O.}\ \bibnamefont
  {Camanho}}, \bibinfo {author} {\bibfnamefont {J.~D.}\ \bibnamefont
  {Edelstein}}, \bibinfo {author} {\bibfnamefont {J.}~\bibnamefont
  {Maldacena}}, \ and\ \bibinfo {author} {\bibfnamefont {A.}~\bibnamefont
  {Zhiboedov}},\ }\href@noop {} {\bibfield  {journal} {\bibinfo  {journal}
  {Journal of High Energy Physics}\ }\textbf {\bibinfo {volume} {2016}},\
  \bibinfo {pages} {20} (\bibinfo {year} {2016})}\BibitemShut {NoStop}%
\bibitem [{\citenamefont {Perlmutter}(2016)}]{perlmutter2016bounding}%
  \BibitemOpen
  \bibfield  {author} {\bibinfo {author} {\bibfnamefont {E.}~\bibnamefont
  {Perlmutter}},\ }\href@noop {} {\bibfield  {journal} {\bibinfo  {journal}
  {Journal of High Energy Physics}\ }\textbf {\bibinfo {volume} {2016}},\
  \bibinfo {pages} {69} (\bibinfo {year} {2016})}\BibitemShut {NoStop}%
\bibitem [{\citenamefont {Maldacena}\ and\ \citenamefont
  {Qi}(2018)}]{Maldacena:2018lmt}%
  \BibitemOpen
  \bibfield  {author} {\bibinfo {author} {\bibfnamefont {J.}~\bibnamefont
  {Maldacena}}\ and\ \bibinfo {author} {\bibfnamefont {X.-L.}\ \bibnamefont
  {Qi}},\ }\href@noop {} {\  (\bibinfo {year} {2018})},\ \Eprint
  {http://arxiv.org/abs/1804.00491} {arXiv:1804.00491 [hep-th]} \BibitemShut
  {NoStop}%
\bibitem [{\citenamefont {Bak}\ \emph {et~al.}(2018)\citenamefont {Bak},
  \citenamefont {Kim},\ and\ \citenamefont {Yi}}]{Bak:2018txn}%
  \BibitemOpen
  \bibfield  {author} {\bibinfo {author} {\bibfnamefont {D.}~\bibnamefont
  {Bak}}, \bibinfo {author} {\bibfnamefont {C.}~\bibnamefont {Kim}}, \ and\
  \bibinfo {author} {\bibfnamefont {S.-H.}\ \bibnamefont {Yi}},\ }\href
  {\doibase 10.1007/JHEP08(2018)140} {\bibfield  {journal} {\bibinfo  {journal}
  {JHEP}\ }\textbf {\bibinfo {volume} {08}},\ \bibinfo {pages} {140} (\bibinfo
  {year} {2018})},\ \Eprint {http://arxiv.org/abs/1805.12349} {arXiv:1805.12349
  [hep-th]} \BibitemShut {NoStop}%
\bibitem [{\citenamefont {Maldacena}\ \emph {et~al.}(2018)\citenamefont
  {Maldacena}, \citenamefont {Milekhin},\ and\ \citenamefont
  {Popov}}]{Maldacena:2018gjk}%
  \BibitemOpen
  \bibfield  {author} {\bibinfo {author} {\bibfnamefont {J.}~\bibnamefont
  {Maldacena}}, \bibinfo {author} {\bibfnamefont {A.}~\bibnamefont {Milekhin}},
  \ and\ \bibinfo {author} {\bibfnamefont {F.}~\bibnamefont {Popov}},\
  }\href@noop {} {\  (\bibinfo {year} {2018})},\ \Eprint
  {http://arxiv.org/abs/1807.04726} {arXiv:1807.04726 [hep-th]} \BibitemShut
  {NoStop}%
\bibitem [{\citenamefont {Caceres}\ \emph {et~al.}(2018)\citenamefont
  {Caceres}, \citenamefont {Misobuchi},\ and\ \citenamefont
  {Xiao}}]{Caceres:2018ehr}%
  \BibitemOpen
  \bibfield  {author} {\bibinfo {author} {\bibfnamefont {E.}~\bibnamefont
  {Caceres}}, \bibinfo {author} {\bibfnamefont {A.~S.}\ \bibnamefont
  {Misobuchi}}, \ and\ \bibinfo {author} {\bibfnamefont {M.-L.}\ \bibnamefont
  {Xiao}},\ }\href@noop {} {\  (\bibinfo {year} {2018})},\ \Eprint
  {http://arxiv.org/abs/1807.07239} {arXiv:1807.07239 [hep-th]} \BibitemShut
  {NoStop}%
\bibitem [{\citenamefont {Fu}\ \emph {et~al.}(2018)\citenamefont {Fu},
  \citenamefont {Grado-White},\ and\ \citenamefont {Marolf}}]{Fu:2018oaq}%
  \BibitemOpen
  \bibfield  {author} {\bibinfo {author} {\bibfnamefont {Z.}~\bibnamefont
  {Fu}}, \bibinfo {author} {\bibfnamefont {B.}~\bibnamefont {Grado-White}}, \
  and\ \bibinfo {author} {\bibfnamefont {D.}~\bibnamefont {Marolf}},\
  }\href@noop {} {\  (\bibinfo {year} {2018})},\ \Eprint
  {http://arxiv.org/abs/1807.07917} {arXiv:1807.07917 [hep-th]} \BibitemShut
  {NoStop}%
\bibitem [{\citenamefont {Hayden}\ and\ \citenamefont
  {Preskill}(2007)}]{Hayden:2007cs}%
  \BibitemOpen
  \bibfield  {author} {\bibinfo {author} {\bibfnamefont {P.}~\bibnamefont
  {Hayden}}\ and\ \bibinfo {author} {\bibfnamefont {J.}~\bibnamefont
  {Preskill}},\ }\href {\doibase 10.1088/1126-6708/2007/09/120} {\bibfield
  {journal} {\bibinfo  {journal} {JHEP}\ }\textbf {\bibinfo {volume} {09}},\
  \bibinfo {pages} {120} (\bibinfo {year} {2007})},\ \Eprint
  {http://arxiv.org/abs/0708.4025} {arXiv:0708.4025 [hep-th]} \BibitemShut
  {NoStop}%
\bibitem [{\citenamefont {Yoshida}\ and\ \citenamefont
  {Kitaev}(2017)}]{Yoshida:2017non}%
  \BibitemOpen
  \bibfield  {author} {\bibinfo {author} {\bibfnamefont {B.}~\bibnamefont
  {Yoshida}}\ and\ \bibinfo {author} {\bibfnamefont {A.}~\bibnamefont
  {Kitaev}},\ }\href@noop {} {\  (\bibinfo {year} {2017})},\ \Eprint
  {http://arxiv.org/abs/1710.03363} {arXiv:1710.03363 [hep-th]} \BibitemShut
  {NoStop}%
\bibitem [{\citenamefont {Bao}\ \emph {et~al.}(2018)\citenamefont {Bao},
  \citenamefont {Chatwin-Davies}, \citenamefont {Pollack},\ and\ \citenamefont
  {Remmen}}]{Bao:2018msr}%
  \BibitemOpen
  \bibfield  {author} {\bibinfo {author} {\bibfnamefont {N.}~\bibnamefont
  {Bao}}, \bibinfo {author} {\bibfnamefont {A.}~\bibnamefont {Chatwin-Davies}},
  \bibinfo {author} {\bibfnamefont {J.}~\bibnamefont {Pollack}}, \ and\
  \bibinfo {author} {\bibfnamefont {G.~N.}\ \bibnamefont {Remmen}},\
  }\href@noop {} {\  (\bibinfo {year} {2018})},\ \Eprint
  {http://arxiv.org/abs/1808.05963} {arXiv:1808.05963 [hep-th]} \BibitemShut
  {NoStop}%
\bibitem [{\citenamefont {Nahum}\ \emph {et~al.}(2018)\citenamefont {Nahum},
  \citenamefont {Vijay},\ and\ \citenamefont {Haah}}]{Nahum:2017yvy}%
  \BibitemOpen
  \bibfield  {author} {\bibinfo {author} {\bibfnamefont {A.}~\bibnamefont
  {Nahum}}, \bibinfo {author} {\bibfnamefont {S.}~\bibnamefont {Vijay}}, \ and\
  \bibinfo {author} {\bibfnamefont {J.}~\bibnamefont {Haah}},\ }\href {\doibase
  10.1103/PhysRevX.8.021014} {\bibfield  {journal} {\bibinfo  {journal} {Phys.
  Rev.}\ }\textbf {\bibinfo {volume} {X8}},\ \bibinfo {pages} {021014}
  (\bibinfo {year} {2018})},\ \Eprint {http://arxiv.org/abs/1705.08975}
  {arXiv:1705.08975 [cond-mat.str-el]} \BibitemShut {NoStop}%
\bibitem [{\citenamefont {Blake}\ \emph {et~al.}(2017)\citenamefont {Blake},
  \citenamefont {Lee},\ and\ \citenamefont {Liu}}]{Blake:2017ris}%
  \BibitemOpen
  \bibfield  {author} {\bibinfo {author} {\bibfnamefont {M.}~\bibnamefont
  {Blake}}, \bibinfo {author} {\bibfnamefont {H.}~\bibnamefont {Lee}}, \ and\
  \bibinfo {author} {\bibfnamefont {H.}~\bibnamefont {Liu}},\ }\href@noop {} {\
   (\bibinfo {year} {2017})},\ \Eprint {http://arxiv.org/abs/1801.00010}
  {arXiv:1801.00010 [hep-th]} \BibitemShut {NoStop}%
\bibitem [{\citenamefont {Cotler}\ and\ \citenamefont
  {Jensen}(2018)}]{Cotler:2018zff}%
  \BibitemOpen
  \bibfield  {author} {\bibinfo {author} {\bibfnamefont {J.}~\bibnamefont
  {Cotler}}\ and\ \bibinfo {author} {\bibfnamefont {K.}~\bibnamefont
  {Jensen}},\ }\href@noop {} {\  (\bibinfo {year} {2018})},\ \Eprint
  {http://arxiv.org/abs/1808.03263} {arXiv:1808.03263 [hep-th]} \BibitemShut
  {NoStop}%
\bibitem [{\citenamefont {Haehl}\ and\ \citenamefont {Rozali}(2018)}]{moshe}%
  \BibitemOpen
  \bibfield  {author} {\bibinfo {author} {\bibfnamefont {F.~M.}\ \bibnamefont
  {Haehl}}\ and\ \bibinfo {author} {\bibfnamefont {M.}~\bibnamefont {Rozali}},\
  }\href@noop {} {\  (\bibinfo {year} {2018})},\ \Eprint
  {http://arxiv.org/abs/1808.02898} {arXiv:1808.02898 [hep-th]} \BibitemShut
  {NoStop}%
\bibitem [{\citenamefont {{Levitov}}\ \emph {et~al.}(1995)\citenamefont
  {{Levitov}}, \citenamefont {{Shytov}},\ and\ \citenamefont
  {{Yakovets}}}]{levitov}%
  \BibitemOpen
  \bibfield  {author} {\bibinfo {author} {\bibfnamefont {L.~S.}\ \bibnamefont
  {{Levitov}}}, \bibinfo {author} {\bibfnamefont {A.~V.}\ \bibnamefont
  {{Shytov}}}, \ and\ \bibinfo {author} {\bibfnamefont {A.~Y.}\ \bibnamefont
  {{Yakovets}}},\ }\href {\doibase 10.1103/PhysRevLett.75.370} {\bibfield
  {journal} {\bibinfo  {journal} {Physical Review Letters}\ }\textbf {\bibinfo
  {volume} {75}},\ \bibinfo {pages} {370} (\bibinfo {year} {1995})},\ \Eprint
  {http://arxiv.org/abs/cond-mat/9406117} {cond-mat/9406117} \BibitemShut
  {NoStop}%
\bibitem [{\citenamefont {G{\"a}rttner}\ \emph {et~al.}(2017)\citenamefont
  {G{\"a}rttner}, \citenamefont {Bohnet}, \citenamefont {Safavi-Naini},
  \citenamefont {Wall}, \citenamefont {Bollinger},\ and\ \citenamefont
  {Rey}}]{garttner2017measuring}%
  \BibitemOpen
  \bibfield  {author} {\bibinfo {author} {\bibfnamefont {M.}~\bibnamefont
  {G{\"a}rttner}}, \bibinfo {author} {\bibfnamefont {J.~G.}\ \bibnamefont
  {Bohnet}}, \bibinfo {author} {\bibfnamefont {A.}~\bibnamefont
  {Safavi-Naini}}, \bibinfo {author} {\bibfnamefont {M.~L.}\ \bibnamefont
  {Wall}}, \bibinfo {author} {\bibfnamefont {J.~J.}\ \bibnamefont {Bollinger}},
  \ and\ \bibinfo {author} {\bibfnamefont {A.~M.}\ \bibnamefont {Rey}},\
  }\href@noop {} {\bibfield  {journal} {\bibinfo  {journal} {Nature Physics}\
  }\textbf {\bibinfo {volume} {13}},\ \bibinfo {pages} {781} (\bibinfo {year}
  {2017})}\BibitemShut {NoStop}%
\bibitem [{\citenamefont {Maldacena}\ and\ \citenamefont
  {Susskind}(2013)}]{Maldacena:2013xja}%
  \BibitemOpen
  \bibfield  {author} {\bibinfo {author} {\bibfnamefont {J.}~\bibnamefont
  {Maldacena}}\ and\ \bibinfo {author} {\bibfnamefont {L.}~\bibnamefont
  {Susskind}},\ }\href {\doibase 10.1002/prop.201300020} {\bibfield  {journal}
  {\bibinfo  {journal} {Fortsch. Phys.}\ }\textbf {\bibinfo {volume} {61}},\
  \bibinfo {pages} {781} (\bibinfo {year} {2013})},\ \Eprint
  {http://arxiv.org/abs/1306.0533} {arXiv:1306.0533 [hep-th]} \BibitemShut
  {NoStop}%
\bibitem [{\citenamefont {Prudnikov}\ \emph {et~al.}(1992)\citenamefont
  {Prudnikov}, \citenamefont {Brychkov},\ and\ \citenamefont
  {Marichev}}]{Prudnikov1992}%
  \BibitemOpen
  \bibfield  {author} {\bibinfo {author} {\bibfnamefont {A.}~\bibnamefont
  {Prudnikov}}, \bibinfo {author} {\bibfnamefont {Y.~A.}\ \bibnamefont
  {Brychkov}}, \ and\ \bibinfo {author} {\bibfnamefont {O.}~\bibnamefont
  {Marichev}},\ }\href@noop {} {\emph {\bibinfo {title} {Integrals and Series,
  Volume 3: More Special Functions}}}\ (\bibinfo  {publisher} {Gordon and
  Breach, New York},\ \bibinfo {year} {1992})\BibitemShut {NoStop}%
\end{thebibliography}%

\end{document}